\documentclass[11pt]{article}
\usepackage[margin=1in]{geometry}
\usepackage{setspace}
\renewcommand{\baselinestretch}{1.15} 

\usepackage{graphicx}
\graphicspath{ {figures/}}
\usepackage{natbib}

\usepackage{bm}
\usepackage{bbm} 
\usepackage[color,notref,notcite]{showkeys}
\usepackage{bigfoot}
\usepackage{amsmath,amsfonts,amssymb,amsthm,booktabs,color,epsfig,graphicx,hyperref,url}

\allowdisplaybreaks

\theoremstyle{plain}
\newtheorem{theorem}{Theorem}

\newtheorem{lemma}{Lemma}
\newtheorem{corollary}{Corollary}

\theoremstyle{definition}
\newtheorem{assumption}{Assumption}
\newtheorem{assumptions}[assumption]{Assumptions}

\newtheorem{remark}{Remark}

\usepackage{setspace}
\usepackage{rotating}
\usepackage{hvfloat}

\def\Prob{\mathbb{P}}
\def\E{\mathbb{E}}
\def\Var{\mathbb{V}\mathrm{ar}}
\def\Cov{\mathbb{C}\mathrm{ov}}

\def\ik{(\mathbf{i},k)}

\makeatletter
\newcommand{\oset}[3][.075ex]{%
  \mathrel{\mathop{#3}\limits^{
    \vbox to#1{\kern-2\ex@
    \hbox{$\scriptstyle#2$}\vss}}}}
\makeatother

\newcommand{\ccirc}{\vcenter{\hbox{$\scriptscriptstyle\circ$}}}

\newcommand\lowprime{\mkern-7.5mu%
                     \raise1.15ex\hbox{$\scriptstyle\prime$}
                     \mkern-2.5mu}
\newcommand\lowpprime{\mkern-7.5mu%
                    \raise1.15ex\hbox{$\scriptscriptstyle\prime\prime$}
                     \mkern-2.5mu}
\newcommand\lowk{\mkern-7.5mu%
                     \raise1.15ex\hbox{$\scriptstyle (k)$}
                     \mkern-2.5mu}
\newcommand*{\gensymb}[4]{%
  {\oset{\hspace{#1}#2}{#3}#4}}
\newcommand*{\gensymbsub}[6]{%
  {\oset{\hspace{#1}#2}{#3}#4}_{\hspace{#5} #6}}

\newcommand{\Pk}[1]{%
  \gensymbsub{0ex}{}{\Psi}{}{-.3ex}{#1 \vphantom{\hat{H}}}}
\newcommand{\Pkc}[1]{%
  \gensymbsub{0ex}{\ccirc}{\Psi}{}{-.3ex}{#1 \vphantom{\hat{H}}}}
\newcommand{\Pkcc}[1]{%
  \gensymbsub{0ex}{\ccirc \ccirc}{\Psi}{}{-.3ex}{#1 \vphantom{\hat{H}}}}
\newcommand{\Pkcr}[2]{%
  \gensymbsub{0ex}{\ccirc \scriptsize{#2}}{\Psi}{}{-.3ex}{#1 \vphantom{\hat{H}}}}
\newcommand{\Pkr}[2]{%
  \gensymbsub{0ex}{\scriptsize{#2}}{\Psi}{}{-.3ex}{#1 \vphantom{\hat{H}}}}
\newcommand{\Pkrs}[3]{%
  \gensymbsub{0ex}{\scriptsize{#2} \scriptsize{#3}}{\Psi}{}{-.3ex}{#1 \vphantom{\hat{H}}}}

\newcommand{\btsk}[2]{%
  \gensymbsub{.5ex}{\scriptsize{#1}}{B}{}{-0ex}{#2 \vphantom{\hat{H}}}}
\newcommand{\btssk}[3]{%
  \gensymbsub{.5ex}{\scriptsize{#1} \hspace{-0ex} \scriptsize{#2}}{B}{}{-.2ex}{#3 \vphantom{\hat{H}}}}

\DeclareMathOperator*{\argmin}{\arg\,\min}

\title{Inference for overparametrized\\ hierarchical Archimedean copulas}
\author{Samuel Perreault$^\dag$\footnote{Corresponding author: samuel.perreault@utoronto.ca} \quad Yanbo Tang$^\ddag$  \quad Ruyi Pan$^\dag$ \quad Nancy Reid$^\dag$\\[2ex]
$^\dag$University of Toronto, Canada \qquad $^\ddag$Imperial College London, United Kingdom}

\date{November 15, 2024}

\begin{document}

\maketitle

\begin{abstract}
Hierarchical Archimedean copulas (HACs) are multivariate uniform distributions constructed by nesting Archimedean copulas into one another, and provide a flexible approach to modeling non-exchangeable data.
However, this flexibility in the model structure may lead to over-fitting when the model estimation procedure is not performed properly.
In this paper, we examine the problem of structure estimation and more generally on the selection of a parsimonious model from the hypothesis testing perspective.
Formal tests for structural hypotheses concerning HACs have been lacking so far, most likely due to the restrictions on their associated parameter space which hinders the use of standard inference methodology.
Building on previously developed asymptotic methods for these non-standard parameter spaces, we provide an asymptotic stochastic representation for the maximum likelihood estimators of (potentially) overparametrized HACs, which we then use to formulate a likelihood ratio test for certain common structural hypotheses.
Additionally, we also derive analytical expressions for the first- and second-order partial derivatives of two-level HACs based on Clayton and Gumbel generators, as well as general numerical approximation schemes for the Fisher information matrix. 
\end{abstract}

\section{Introduction} \label{sec:intro}

Multivariate hierarchical modeling is an appealing and effective framework for capturing complex dependency structures in data.
In particular, copula-based approaches have become increasingly popular, and many interesting and flexible copulas have been introduced reflect for a wide array of dependence structures.
We focus on a sub-class of multivariate hierarchical models proposed in \cite{Joe:1997} called hierarchical Archimedean copulas (HACs), which are widely used in finance, insurance and risk management \citep{Savu/Trede:2010,Hofert/Scherer:2011,Abdallah/Boucher/Cossette:2015,Cossette/Marceau/Mtalai:2019,Li/Balasooriya/Liu:2021}.
Our main purpose is to provide insights into the asymptotic behaviour of the maximum likelihood estimator (MLE) and likelihood ratio statistic within the framework of overparametrized HACs. We call a HAC overparametrized if it contains more parameters than needed to specify the true distribution.

As with many hierarchical models, a key barrier to the practical usage of HACs is in finding an appropriate tree structure (hierarchy) for the data at hand, one that is complex enough to capture the complexity of the data, but simple enough to prevent over-fitting.
We provide a method for testing some important structural hypotheses for possibly overparametrized HACs through likelihood ratio statistic. 
In order to characterize the asymptotic distribution of the likelihood ratio, we need to consider parameters on the boundary of the parameter space, a setting which does not allow the use of standard inference methodology.

Our main result, Theorem~\ref{thm:mle-law}, provides an asymptotic stochastic representation for the likelihood ratio statistic in these non-standard conditions, and thus partially extends Theorem~1 of \cite{Hofert/Machler/McNeil:2012} for Archimedean copulas, and Theorem~1.b of \cite{Okhrin/Okhrin/Schmid:2013}, which deals with non-overparametrized HACs with parameters in the interior of the parameter space, and a multi-stage maximum likelihood estimation procedure.

\subsection{Hierarchical Archimedean Copulas}

As the name suggests, HACs are multivariate distributions with uniform margins whose building blocks are Archimedean copulas. A $p$-variate Archimedean copula $C_{\theta}$ with parameter $\theta$ is fully characterized by its generator $\psi_{\theta}$ via
\begin{equation}
    C_{\theta}(\bm{u}) = \psi_{\theta}\{\phi_{\theta}(u_{1}) +\ldots+\phi_{\theta}(u_{p})\} , \qquad \text{for all } \bm{u} \in [0,1]^p,
\end{equation}
where $\phi_{\theta}(s) := \inf \{t \in [0,\infty] : \psi_{\theta}(t) = s\}$ is the generalized inverse of $\psi_{\theta}$.
It was shown by \cite{McNeil/Neslehova:2009} that an Archimedean copula $C_{\theta}$ is proper, in that it is a multivariate distribution with $U(0,1)$ univariate margins, if and only if $\psi_{\theta}$ is $d$-monotone: a continuous function on $[0,\infty]$ admitting derivatives $\psi^{(k)}$ up to the order $k=d-2$ satisfying $(-1)^k \psi^{(k)}(t) \geqslant 0$ for all $k \in \{ 0, \ldots, d-2 \}$ and $t \in (0,\infty)$, and $(-1)^{d-2} \psi^{(d-2)}$ is decreasing and convex on $(0,\infty)$.
Prime examples of $d$-monotone generators are the Clayton and Gumbel generators, respectively given by
\begin{align} \label{eq:generators}
\psi_{\theta}(t) = (1+t)^{-1/\theta} \qquad \text{and} \qquad \psi_{\theta}(t) = \exp(-t^{1/\theta}) ,
\end{align}
where $\theta \in (0, \infty)$ for the Clayton generator and $\theta \in [1,\infty)$ for the Gumbel generator.
These are in fact examples of completely monotone generators: infinitely differentiable generators whose derivatives satisfy $(-1)^k \psi^{(k)}(t) \geqslant 0$ for all $k \in \{ 0, 1, \ldots \}$ and $t \in (0,\infty)$.

\begin{figure}
    \centering
    \includegraphics[width=.25\textwidth]{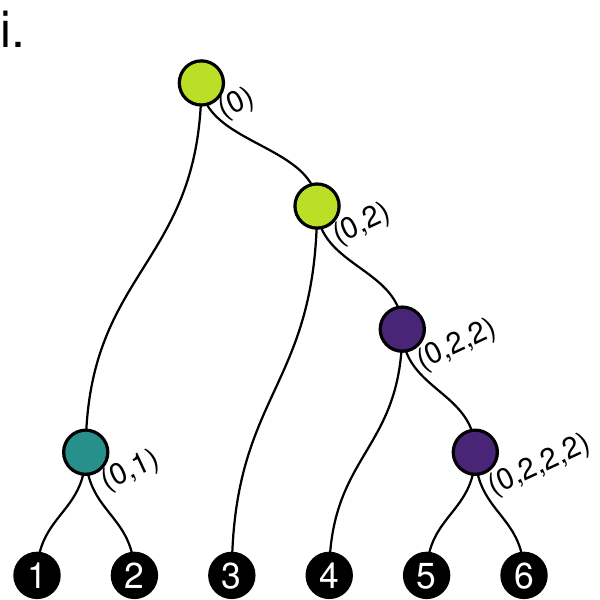}
    \hspace{.1\textwidth}
    \includegraphics[width=.25\textwidth]{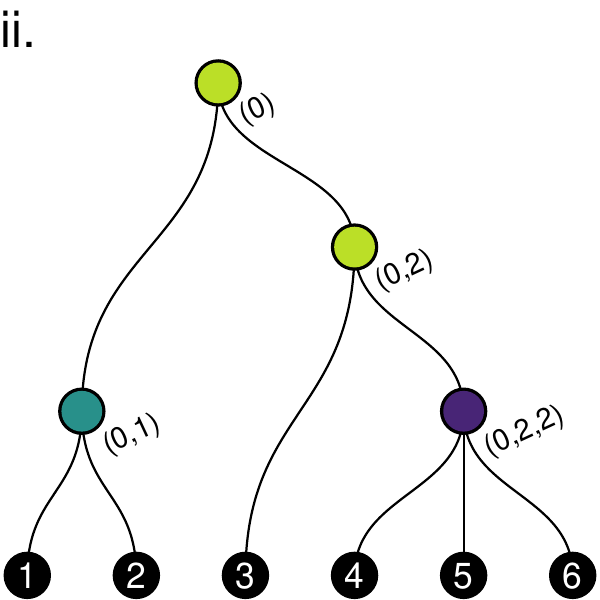}
    \hspace{.1\textwidth}
    \includegraphics[width=.25\textwidth]{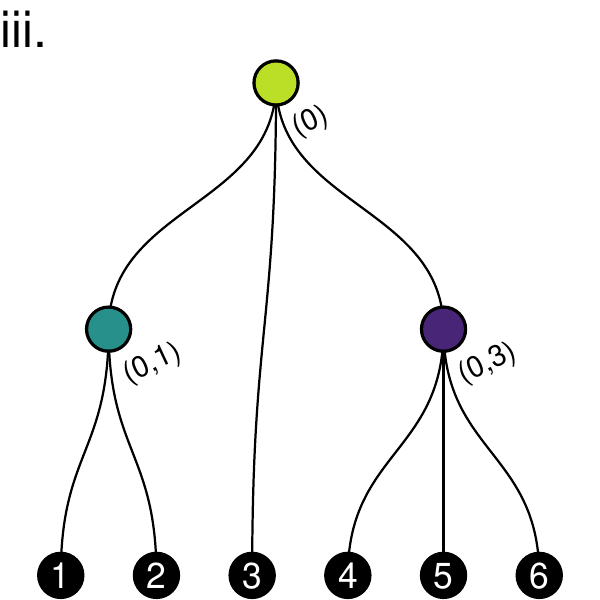}
    \caption{Some tree structures partially characterizing HACs. 
    The indexing of the nodes is as described around \eqref{eq:hac-recursive}.
    Each node corresponds to an Archimedean copula, which we assume is described by one parameter.
    When the Archimedean copulas associated to a parent-child pair of nodes are of the same type and have equal parameters, the child node can be collapsed into its parent. The tree in panel ii.\ is obtained by collapsing node $(0,2,2,2)$ in panel i.\ and similarly, the tree in panel iii.\ is obtained by collapsing node $(0,2)$ in panel ii.}
    \label{fig:trees}
\end{figure}

A HAC is constructed by nesting Archimedean copulas into one another, and is partially characterized by a tree whose leaf nodes each represent a unique variable.
As illustrated in Figure~\ref{fig:trees}, we index the nodes of this tree using vectors $\mathbf{i}$, which encodes the path to the corresponding node from the root, whose index is always $(0)$.
We denote the set of all indices corresponding to non-leaf nodes by $\mathcal{I}$; for all $\mathbf{i} \in \mathcal{I}$, $K_{\mathbf{i}}$ denotes the number of children of node $\mathbf{i}$ (the nodes directly below $\mathbf{i}$) and $d_{\mathbf{i}}$ denotes the number of leaf nodes below it (the nodes below $\mathbf{i}$ which has no child nodes).
Using this notation, a HAC $C$ can be represented recursively as
{\small 
\begin{equation} \label{eq:hac-recursive}
        C^{\mathbf{i}}(\bm{u}) = C_{\mathbf{i}}\{C^{(\mathbf{i},1)}(\bm{u}_1), \dots, C^{(\mathbf{i},K_{\mathbf{i}})}(\bm{u}_{K_{\mathbf{i}}})\} , \quad 
        \text{for all } \mathbf{i} \in \mathcal{I} \text{ and } \bm{u} = (\bm{u}_k \in [0,1]^{d_{\ik}}: k=1,\dots,K_{\mathbf{i}}),
\end{equation}}
where $C_{\mathbf{i}}$ is an Archimedean copula and $C^{\ik}$, $k=1,\dots,K_{\mathbf{i}}$, are themselves HACs of lower-dimensions; $C^{(0)} = C$ and $C^{\ik}$ is the identity function when $\ik$ is a leaf node.
For each $\mathbf{i} \in \mathcal{I}$, we define $\psi_{\mathbf{i}}$ and $\phi_{\mathbf{i}}$ as the generator and inverse generator of $C_{\mathbf{i}}$ respectively, and $\theta_{\mathbf{i}}$ as the corresponding scalar parameter; $\bm{\theta} := (\theta_{\mathbf{i}})_{\mathbf{i} \in \mathcal{I}}$ is the full parameter vector.

\begin{figure}[t]
    \centering
    \includegraphics[width = .75\textwidth]{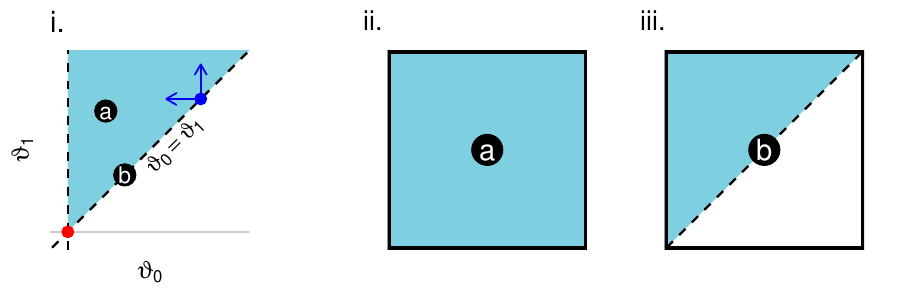}
    \caption{(i) Depiction of the parameter space $\bm{\Theta}$ for a two-level HAC with two parameters, which is a cone; its origin is indicated by a red point.
    The blue arrows show the directions used to compute numerical derivatives via finite differences at a given point (in blue): ``backward'' on the $\theta_0$-dimension and ``forward'' on the $\theta_1$-dimension; see Section~\ref{sec:sigma-estimation}. 
    (ii, iii) Depiction of the asymptotic local parameter space $\mathcal{A}$, which equals $\mathbb{R}^2$ when $\bm{\vartheta}$ is in the interior of $\bm{\Theta}$ (ii), but is a half-space with boundary given by the line $\theta_{0}=\theta_{1}$ when $\vartheta_{0}=\vartheta_{1}$ (iii).}
    \label{fig:sets1}
\end{figure}

In contrast with Archimedean copulas, non-trivial HACs are not exchangeable, which makes them much more attractive in real data settings, where exchangeability is often an unreasonable assumption.
This increased flexibility, comes at the cost of additional theoretical considerations however;
using Archimedean copulas as building blocks for a HAC does not necessarily yield a proper copula.
To guarantee that a given HAC $C = C_{\bm{\theta}}$ is proper, one may verify the sufficient, but not necessary condition that, for any $\mathbf{i},\ik \in \mathcal{I}$, the parent-child composition $\phi_{\mathbf{i}} \circ \psi_{\ik}$ defines a completely monotone generator.
For some families of completely monotone generators such as the Gumbel or Clayton, it suffices to check whether $\theta_{\mathbf{i}} \leqslant \theta_{\ik}$ (\cite{Holena/Bajer/Scavnicky:2015}, see also Table~2.3 in \cite{Hofert:2010}).
This suggests restricting the parameter space $\bm{\Theta} \subset \mathbb{R}^{|\mathcal{I}|}$ to 
\begin{equation} \label{eq:Theta}
	\bm{\Theta} = \Big\{ \bm{\theta} = (\theta_{\mathbf{i}})_{\mathbf{i} \in \mathcal{I}} \in \Theta^{|\mathcal{I}|} : \theta_{\mathbf{i}} \leqslant \theta_{\ik} \text{ for all } \mathbf{i},\ik \in \mathcal{I}, \text{ and } k \in \{1,\dots,K_{\mathbf{i}}\} \Big\} ,
\end{equation}
where $\Theta$ encodes further requirements on individual parameters imposed by specific generator families; for example, $\Theta = (0,\infty)$ and $\Theta = [1,\infty)$ for the Clayton and Gumbel families, respectively.
See Figure~\ref{fig:sets1}.i for a depiction of $\bm{\Theta}$ for two-parameter HACs.
Although advances have been made on the topic, verifying that a HAC mixing many types of generators is a proper copula is difficult in general; see \cite{Rezapour:2015} and \cite{Holena/Bajer/Scavnicky:2015} for sufficient conditions for constructing proper HACs through nesting, and \cite{Hering/al:2010}, \cite{Zhu/Wang/Tan:2016} and \cite{Cossette/Gadoury/Marceau/Mtalai:2017} for construction methods that necessarily yields proper copulas.

The main issue we focus on arises when two or more parameters corresponding to neighbouring nodes, say $\mathbf{i}$ and $(\mathbf{i},1)$, are equal, meaning that the parameter vector lies on the boundary of $\bm{\Theta}$.
These cases are of particular interest, as $\phi_{\mathbf{i}} \circ \psi_{(\mathbf{i},k)}$ implicitly appearing in 
 \eqref{eq:hac-recursive} then reduces to the identity map, thus collapsing the child node into its parent node:
$$
    C^{\mathbf{i}} = C_{\mathbf{i}} \circ (C^{(\mathbf{i},1)},\dots,C^{(\mathbf{i},K_{\mathbf{i}})}) = C_{\mathbf{i}} \circ (C^{(\mathbf{i},1,1)},\dots,C^{(\mathbf{i},1,K_{(\mathbf{i},1)})} , C^{(\mathbf{i},2)} ,\dots,C^{(\mathbf{i},K_{\mathbf{i}})}) ,
$$
where, on the right-hand side of the equation, $C_{\mathbf{i}}$ is now a $(K_{\mathbf{i}} + K_{(\mathbf{i},1)} - 1)$-variate function.
This is depicted in Figure~\ref{fig:trees}.
In view of this, structural hypotheses about HACs can be formulated in terms of equalities between parameters, 
but these equalities (partially) characterize the boundary of the parameter space, which violates the standard assumptions on the parameter space for inference.

This last concern is intimately related with the more general, and extensively studied problem of determining an appropriate HAC structure from data alone; see \cite{Okhrin/Okhrin/Schmid:2013, Segers/Uyttendaele:2014, Matsypura/al:2016, Gorecki/Hofert/Holena:2017-JSCS, Uyttendaele:2018, Okhrin/Ristig:2024} for specific algorithms, and \cite{Gorecki/Hofert/Holena:2017-DM} for a review of methods based on Kendall's tau.
Hypothesis testing offers an attractive framework for collapsing nodes, and it can play an important part in structure learning methods.
This is made explicit in \cite{Segers/Uyttendaele:2014}, which uses hypothesis tests based on Kendall distributions \citep{Genest/Rivest:1993,Nelsen/al:2003} as part of their structure learning algorithm, and to some extent by \cite[Chapter~3]{Perreault:2020}, which discusses structure learning for hierarchical models based on structural hypothesis tests for rank correlations developed in \cite{Perreault/Neslehova/Duchesne:2023}.

In addition to these theoretical considerations, the constrained parameter space $\bm{\Theta}$ and the difficulty of working with HAC densities give rise to some computational and inferential challenges for the practical usage of HACs (discussed for example in Section~5.1 of \cite{Savu/Trede:2010} and Section~4 of \cite{Okhrin/Okhrin/Schmid:2013}).
One such challenge is that common numerical optimization techniques for maximizing the log-likelihood may yield a local maximum rather than a global one.
To alleviate this problem, \cite{Okhrin/Okhrin/Schmid:2013} proposed a multi-stage estimation procedure, in which parameters are estimated sequentially, starting from the lower levels then upward and treating previously estimated parameters as fixed.
We explore the asymptotic behaviour of the MLEs in fixed dimensions where local maxima are less of a concern and we therefore assume we have access to the MLEs.

\subsection{Objectives and paper organization}
To provide insight into the uncertainty related to node collapse for overparametrized HACs, we consider the node collapse problem from the perspective of hypothesis testing.
Formally, let $\mathcal{U}_n$ be a sample of $n$ observations from a HAC $C = C_{\bm{\vartheta}}$ given by \eqref{eq:hac-recursive} with one-parameter $\infty$-monotone generators of a unique type and $\bm{\Theta}$ of the form \eqref{eq:Theta}.
We consider hypotheses of the kind
\begin{equation} \label{eq:H}
    H_{\circ} : \bm{\vartheta} \in \bm{\Theta}_{\circ} \quad \text{against} \quad H_{\bullet} : \bm{\vartheta} \in \bm{\Theta} ,
\end{equation}
where $\bm{\Theta}_{\circ}$ is created by taking unions and intersections of subspaces defined by $\bm{\Theta}_{\ik} := \{ \bm{\theta} \in \bm{\Theta} : \theta_{(\mathbf{i},k)} = \theta_{\mathbf{i}}\}$.
Of particular interest are cases when $\bm{\Theta}_{\circ} = \bm{\Theta}_{\mathbf{i}}$ for some $\mathbf{i}$ (single node collapse), $\bm{\Theta}_{\circ} = \cap_{k=1}^{K_{\mathbf{i}}}\bm{\Theta}_{\ik}$ (collapsing all children of node $\mathbf{i}$), and $\bm{\Theta}_{\circ} = \cup_{k=1}^{K_{\mathbf{i}}}\bm{\Theta}_{\ik}$ (collapsing at least one child of node $\mathbf{i}$).

Our main result provides approximate $p$-values for the likelihood ratio test of $H_{\circ}$ against $H_{\bullet}$. 
Specifically, we derive the asymptotic null distribution as $n\to\infty$ of the test statistic given by
\begin{equation} \label{eq:stat}
	L_n := 2 \{\ell_n(\bm{\hat\theta}_{\bullet}) - \ell_n(\bm{\hat\theta}_{\circ})\} , \quad \text{where  } \ell_n(\bm{\theta}) = \sum\nolimits_{\bm{u} \in \mathcal{U}_n} \ln c_{\bm{\theta}}(\bm{u}) ,
\end{equation}
where $c_{\bm{\vartheta}}$ is the copula density associated with $C_{\bm{\vartheta}}$,
and $\bm{\hat\theta}_{\circ}$ and $\bm{\hat\theta}_{\bullet}$ are the (constrained) maximum likelihood estimators of $\bm{\vartheta}$ under $H_{\circ}$ and $H_{\bullet}$, respectively.
Due to overparametrization, the limiting distribution is not the usual $\chi^2$, it is now mixture distribution whose parameters depend heavily on the underlying space.
In practice, it may be difficult to find the asymptotic null distribution, as it involves weights which may not be known apriori.
Our focus is to give intuition on behavior of the type-I error control and the power of the test under different configurations of the parameter space, which relates to the tree structure of the overparametrized model.
In addition to tests based on the asymptotic distribution of $L_n$, we consider so-called \emph{conditional} tests, as discussed by \citep{Susko:2013}.
When applicable, these are usually much easier to perform than the unconditional tests, although they are in general conservative, see Section 3.4.

As a secondary contribution, we derive analytical formulas for computing the score and the Hessian (or Fisher information) matrix for two-level Clayton and Gumbel HACs, and we investigate analogous numerical methods for approximating these quantities for general HACs. 
While the analytical formulas may prove useful for future computational or theoretical works, as the score function and the Hessian are often needed for optimization purposes or for deriving the properties of maximum likelihood estimators, their primary purpose here is to help quantifying the effect of numerical error on $p$-value approximations.

The paper is organized as follows.
In \S\ref{sec:basic-assumptions}, we discuss our theoretical framework.
In \S\ref{sec:tests}, we present our main result and derive tests for $H_{\circ}$ from it.
We further discuss power under local alternatives, nuisance parameters and a conditional versions of the test.
In \S\ref{sec:sigma-estimation}, provide numerical methods for estimating the Fisher information matrix required for these tests; we briefly explain how to derive analytical estimates for two-level hierarchical Clayton or Gumbel copulas in Appendix~\ref{app:two-level}.
We investigate the performance of the proposed test in simulations in \S\ref{sec:sim-study}. Specifically, we assess the power and size of the tests in finite sample settings as well as under local alternatives.
We conclude the paper in \S\ref{sec:discussion} by discussing the implications of our work for structure learning, focusing on the popular algorithm of \cite{Okhrin/Okhrin/Schmid:2013}, as well as the newly proposed, and promising algorithm of \cite{Okhrin/Ristig:2024} based on penalized estimation. 
The proofs of our results are presented in Appendix~\ref{app:proofs} of the main document, while the additional appendices referred to throughout (Appendices~\ref{app:two-level}--\ref{app:sim-study}) are available as online supplementary material.

Throughout, we use $\mathcal{N}(\bm{\mu},\bm{S})$  to denote a Gaussian distribution with mean $\bm{\mu}$ and covariance $\bm{S}$, $\mathcal{M}(m,\bm{\gamma})$ to denote a multinomial distribution for $m$ trials with $|\bm{\gamma}|$ possible results whose probabilities are given by $\bm{\gamma}$, and $\mathcal{B}(\gamma)$ to denote a Bernoulli distribution with parameter $\gamma$.

\section{Assumptions} \label{sec:basic-assumptions} %
Throughout the paper, we use the following set of assumptions.
Assumptions~\ref{ass:basic-framework} reformulate and extend the theoretical framework discussed in \S\ref{sec:intro}. It ensures that the nesting of the Archimedean copulas produces a valid HAC.
\begin{assumptions} \label{ass:basic-framework}
Let $p$ and $d$ be integers and $\mathcal{C}_{\bm{\Theta}} = \{ C_{\bm{\theta}} : \bm{\theta} \in \bm{\Theta} \}$ be a family of $d$-variate HACs defined recursively by \eqref{eq:hac-recursive} with $\bm{\Theta} \subset \mathbb{R}^p$ as in \eqref{eq:Theta}, and generated by one-parameter, completely monotone generators of a unique type.
Also let $\mathcal{U}_n$ be a dataset of $n$ independent observations from the copula $C_{\bm{\vartheta}} \in \mathcal{C}_{\bm{\Theta}}$ and further assume that $\bm{\vartheta}$ is in the interior of $\Theta^{p}$, where $\Theta \subseteq [0,\infty)$ is the Archimedean parameter space defined below \eqref{eq:Theta}.
We use $\bm{\vartheta}$ for a fixed true value of the parameter, while we use $\bm{\theta}$ to mean a value of the parameter which may be distinct from $\bm{\vartheta}$.
\end{assumptions}

The requirement that the HAC be generated from a single family could be relaxed, although, as discussed in the introduction, it is in general difficult to derive the parameter constraints ensuring that \ref{eq:hac-recursive} defines a proper copula. Additionally some general formulations may prevent the possibility of node collapsing.
The assumption $\Theta \subseteq [0,\infty)$, which restricts us to positive dependence, covers most popular HACs.
For example, the parameter space of all ten generators in Table~2.2 of \cite{Hofert:2010} is either $[0,1)$, $(0,\infty)$ or $[1,\infty)$.

In addition to Assumptions~\ref{ass:basic-framework}, we make the following usual regularity assumptions on the data generating model.
They are in general mild, although often difficult to check. 
\begin{assumptions} \label{ass:stochastic}
Assume that under the null, as $n \rightarrow \infty$, the MLEs $\bm{\hat\theta}_{\circ}$ and $\bm{\hat\theta}_{\bullet}$ are consistent estimators of $\bm{\vartheta}$, and that the local parameter spaces $\mathcal{A}_{n} := \sqrt{n}(\bm{\Theta} - \bm{\vartheta})$ and $\mathcal{A}_{\circ n} := \sqrt{n}(\bm{\Theta}_\circ - \bm{\vartheta})$ converges to sets $\mathcal{A}$ and $\mathcal{A}_{\circ}$, respectively.
Also assume that the log-likelihood function $\ell_n(\theta)$ is differentiable in quadratic mean in a neighbourhood of $\bm{\vartheta}$ with non-singular Fisher information matrix, and that for every $\bm{\theta}_1$ and $\bm{\theta}_2$ in a neighbourhood of $\bm{\vartheta}$, 
\begin{equation*} \label{eq:Lipschitz}
   | \ln c_{\bm{\theta}_1}(u) - \ln c_{\bm{\theta}_2}(u)  | \leqslant l'(u)\  \| \bm{\theta}_1 - \bm{\theta}_2 \|_2    
\end{equation*}
for some function $l': [0,1]^d \to \mathbb{R}$ such that $\mathbb{E}_{\bm{\vartheta}}\{ l'(\bm{U})^2\} < \infty$, where $\| \cdot \|_2$ denotes the Euclidean norm.
\end{assumptions}
\noindent The local parameter spaces are obtained by centering and rescaling around $\bm{\vartheta}$ in $\bm{\Theta}$ and $\bm{\Theta}_{\circ}$.
Consider for example a two-level HAC with parameter $\bm{\vartheta} = (\vartheta_0,\vartheta_1)$ and the hypothesis $H_{\circ}:\vartheta_0=\vartheta_1$.
Figure~\ref{fig:sets1}.i shows $\bm{\Theta}$ (in turquoise) and $\bm{\Theta}_{\circ}$ (the line $\vartheta_0=\vartheta_1$).
When $\bm{\vartheta}$ is in the interior of $\bm{\Theta}$, $\mathcal{A} = \mathbb{R}^2$, but when $\bm{\vartheta} \in \bm{\Theta}_{\circ}$ (Figure~\ref{fig:sets1}.ii), then $\mathcal{A}$ is a half plane (Figure~\ref{fig:sets1}.iii).
Similarly, $\mathcal{A}_{\circ} = \{ (x,x) : x \in \mathbb{R} \}$, provided that $\bm{\vartheta} \in \bm{\Theta}_{\circ}$.

Concerning the consistency of the constrained maximum likelihood estimators, recall that we assume (in Assumptions~\ref{ass:basic-framework}) that each component of the true parameter $\bm{\vartheta}$ lies in the interior of $\Theta$.
This condition, along with others, is often use to ensure consistency; see, \emph{e.g.}, Assumption~1 (ii) in \cite{Okhrin/Ristig:2024}.
For estimation purposes, we use the (full maximum likelihood estimation) algorithm available in the \texttt{HAC} package \citep{Okhrin/Ristig:2014} in \texttt{R}, which is based numerical optimization.
While such method does not guarantee that the estimators are indeed global maximizers, in practice one can repeat the estimation procedure with several different starting values to maximize the probability of finding the global maximizers.
Thus, to limit the scope of the paper, we assume that our MLEs are indeed global maxima, and we point to the works of \cite{Geyer:1994} and \cite{Shapiro:2000} for extensions to potentially local MLEs.

The good asymptotic behaviour of the local parameter spaces is perhaps the easiest condition to check. As we will see in \S\ref{sec:tests}, all the specific structural hypotheses we consider lead to local spaces converging to cones: sets $\mathcal{C} \subseteq \mathbb{R}^p$ such that $\bm{x} \in \mathcal{C}$ if and only if $\lambda \bm{x} \in \mathcal{C}$ for all $\lambda \geqslant 0$.
The existence of such cones is often referred to as Chernoff regularity, due to their use by the latter in his seminal work \cite{Chernoff:1954} on inference under non-standard conditions. 
The exact shape of these cones has a significant impact on the limiting null distributions of our tests.

The differentiability in quadratic mean (DQM) of the log-likelihood and the non-singularity of the Fisher information matrix at the true parameter are standard conditions in the asymptotics literature; the latter condition is difficult to check due to generally cumbersome (even for very shallow HACs) differentiation formulas.
Fortunately, the invertibility condition can, to some extent, be checked numerically.
In particular, non-invertability of the Fisher information matrix is usually flagged during estimation steps, as it generally leads to computational issues during numerical optimization.
The DQM condition is satisfied if the likelihood is second-order continuously differentiable, but we note the DQM condition alone may not be sufficient to guarantee the good performance of numerical procedures needed to estimate the Fisher information matrix.
To guarantee numerical convergence of finite difference methods which we employ, it is sufficient to have an additional order of smoothness, for example to numerically estimate the second-order derivative it is sufficient to require smoothness on the third-order derivative.
Luckily, most models used in practice are very smooth with most being infinitely differentiable functions.
These assumptions are the analogous versions of Assumptions 2 (i) and 3 in \cite{Okhrin/Ristig:2024}.

Finally the Lipschitz assumption on the log-likelihood function is standard in the analysis of parametric models, see \cite[Chapter 5]{vanderVaart:1998}, and may be further weakened to an assumption bracketing number on the log-likelihood \cite[Theorem 19.4]{vanderVaart:1998}.
This particular assumption is not as often used the HAC literature, so we verify this for a two-level Clayton HAC, showing that these assumptions can be satisfied in practice; indeed most smooth parametric families of Archimedean copulas should satisfy these requirements. 
This Assumption is similar to Assumption 2 (ii) in \cite{Okhrin/Ristig:2024}, as they are both used to invoke a uniform strong law of large numbers needed to prove the asymptotic normality of the MLE.

\begin{figure}
    \centering
    \includegraphics[width = .99\textwidth]{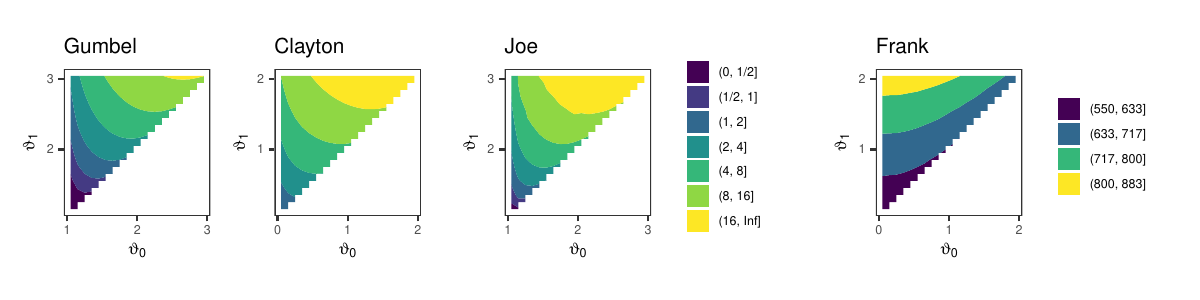}
    \vspace*{-5mm}
    \caption{Numerical approximation of the determinant of $\bm{\Sigma}$ for Clayton and Gumbel trivariate HACs with trivial structure $\mathcal{G} = \{\{1,2\}, 3\}$, as a function of their parameters $\bm{\vartheta} = (\vartheta_0, \vartheta_1)$. It is computed on a grid in a neighborhood of the origin $\bm{o} = (o, o)$ of the cone $\bm{\Theta}$; $o = 1$ for Gumbel and Joe generators and $o = 0$ for Clayton and Frank generators. Note that the vertical boundary $\{o\} \times \Theta$ is part of $\bm{\Theta}$ only if $o \in \Theta$, but the oblique boundary $\{(\vartheta,\vartheta) : \vartheta \in \Theta\}$ always is. All values on this grid are positive, meaning that $\bm{\Sigma}$ is invertible.}
    \label{fig:determinant}
\end{figure}

For an example of how Assumptions~B can be verified in practice for a simple Clayton HAC, see Appendix~\ref{app:Clayton}. The only condition not verified is the invertibility of the Fisher information matrix which we provide some numerical evidence for in Figure~\ref{fig:determinant}.

\section{Likelihood ratio tests} \label{sec:tests} %

We are now ready to investigate the asymptotic behavior of the MLE and the likelihood ratio $L_n$ for testing hypotheses $H_{\circ}$ against $H_{\bullet}$, as defined by \eqref{eq:Theta} and \eqref{eq:H}.
Specifically, we provide an asymptotic stochastic representation for $L_n$, which we then use to derive explicit distributions in some special cases.
We also discuss handling nuisance parameters as well as an alternative method for approximating the null distribution of $L_n$ that is computationally simpler, but which might be overly conservative in certain cases.
Note that for practical purposes, one could exploit available sampling algorithms for HACs \citep{Whelan:2004, McNeil:2008, Hofert:2010, Hofert:2011, Hofert/Machler:2011, Okhrin/Ristig:2014, Grothe/Hofert:2015, Gorecki/Hofert/Okhrin:2021, copula:2023} to approximate the null distribution of $L_n$ by generating new datasets from the HAC with parameter $\bm{\hat\theta}_{\circ}$ and computing the associated value of $L_n$ for each of them. 
However this requires repeated numerical optimizations and is computationally demanding even for moderately large parameter spaces.

\subsection{Test via the asymptotic null distribution of $L_n$} \label{sec:Ln-asymptotic}

As it will be useful later on, we consider local alternatives characterized by a deviation parameter $\bm{h} \in \mathbb{R}^p$; results under the null are obtained by setting $\bm{h} = \bm{0}$.
Our main result involves a $p$-dimensional random vector $\bm{Z} \sim \mathcal{N}(\bm{h},\bm{\Sigma})$, with $\bm{\Sigma}$ as in Assumptions~\ref{ass:stochastic}, and its projections onto the sets $\mathcal{A}$ and $\mathcal{A}_{\circ}$, based on the squared $\bm{\Sigma}$-Mahalanobis distance:
\begin{equation} \label{eq:q-projections}
	\bm{Z}_{\bullet} := \inf\nolimits_{\bm{z} \in \mathcal{A}} q_{\bm{Z}}(\bm{z})
    \quad \text{and} \quad
    \bm{Z}_{\circ} := \inf\nolimits_{\bm{z} \in \mathcal{A}_{\circ}} q_{\bm{Z}}(\bm{z}) ,
    \qquad 
    q_{\bm{Z}}(\bm{z}) = (\bm{Z} - \bm{z})^\top \bm{\Sigma}^{-1} (\bm{Z} - \bm{z}) .
\end{equation}
In practice, when $\bm{Z} \not\in \bm{\Theta}$, we compute $\bm{Z}_{\bullet}$ by projecting $\bm{Z}$ onto all faces of $\mathcal{A}$ and selecting the optimal one; $\bm{Z}_{\circ}$ is computed similarly.
The following theorem, which provides an asymptotic stochastic representation for $L_n$, is a consequence of Theorem~16.7 of \cite{vanderVaart:1998}.
\begin{theorem} \label{thm:mle-law}
Under Assumptions~\ref{ass:basic-framework} and \ref{ass:stochastic}, let $\bm{\hat\theta}_{\circ}$ and $\bm{\hat\theta}_{\bullet}$ be the maximum likelihood estimators of $\bm{\vartheta}_n' = \bm{\vartheta} + \bm{h}_n/\sqrt{n}$ ($\bm{h}_n \to \bm{h} \in \mathbb{R}^p$ as $n \to \infty$) over $\bm{\Theta}_{\circ}$ and $\bm{\Theta}$, respectively, and for $\bm{Z} \sim \mathcal{N}(\bm{h},\bm{\Sigma})$, let $\bm{Z}_{\bullet}$, $\bm{Z}_{\circ}$, and $q_{\bm{Z}}$ be as in \eqref{eq:q-projections}. Then $\bm{Z}_{\circ}$ and $\bm{Z}_{\bullet}$ are unique for almost all $\bm{Z}$ and
\begin{equation} \label{eq:mle-law}
\sqrt{n}(\bm{\hat\theta}_{\bullet} - \bm{\hat\theta}_{\circ}) \rightsquigarrow \bm{Z}_{\bullet}-\bm{Z}_{\circ}
	\qquad \text{and} \qquad
L_n \rightsquigarrow L_{\infty} := q_{\bm{Z}}(\bm{Z}_{\circ})-q_{\bm{Z}}(\bm{Z}_{\bullet}) \qquad \text{as } n \to \infty ,
\end{equation}
where $\rightsquigarrow$ denotes convergence in distribution.
\end{theorem}
\noindent Theorem~\ref{thm:mle-law} provides asymptotic stochastic representations not only for $L_n$, but for many other popular test statistics via the continuous mapping Theorem, most notably for Wald-type statistics.
When $\bm{\Sigma}$ is unknown, as is often the case, Theorem~\ref{thm:mle-law} remains valid when $\bm{\Sigma}$ is replaced with a consistent estimator $\bm{\hat\Sigma}$, by Slutsky's Lemma.
We provide such an estimator in \S\ref{sec:sigma-estimation}.

Given a consistent estimate $\bm{\hat\Sigma}$ of $\bm{\Sigma}$, a $p$-value for testing $H_{\circ}$ against $H_{\bullet}$ can be obtained by approximating the distribution of $L_n$ using Monte Carlo replicates based on $\bm{Z} \sim \mathcal{N}(\bm{0}, \bm{\hat\Sigma})$, following \eqref{eq:q-projections} and \eqref{eq:mle-law}.
Alternatively, one can try to refine Theorem~\ref{thm:mle-law}, following for example \cite{Bartholomew:1959a, Bartholomew:1959b, Bartholomew:1961, Kudo:1963, Moran:1971b, Chant:1974, Shapiro:1985, Self/Liang:1987}.
In particular, it can be shown that, under certain conditions, the distribution of $L_\infty$ is a mixture of independent chi-squared distributions, often referred to as a $\bar{\chi}^2$ distribution.
To derive such refinements, we make use of the following lemma, adapted from \citet[Lemma~3.1]{Shapiro:1985}.
\begin{lemma} \label{lem:chi}
Let $\bm{Z} \sim \mathcal{N}(\bm{0},\bm{\Sigma})$ for some positive definite matrix $\bm{\Sigma}$ and $\mathcal{C}$ be the cone defined by $\mathcal{C} = \{\bm{z} \in \mathbb{R}^p : \bm{x}_k^\top \bm{z} \leqslant 0,\ k=1,\dots,p\}$ for some fixed vectors $\{\bm{x}_k\}_{k=1}^p$. 
Further suppose that there exists $\bm{P}_{\circ}$ and $\bm{P}_{\bullet}$ such that the following holds: $\bm{Z}_{\circ} = \bm{P}_{\circ} \bm{Z}$ and $\bm{Z}_{\bullet} = \bm{P}_{\bullet} \bm{Z}$ whenever $\bm{Z} \in \mathcal{C}$; $\bm{P} := \bm{P}_{\bullet} - \bm{P}_{\circ}$ is a matrix of rank $\nu$ such that $\bm{\Sigma}^{1/2} \bm{P}^\top \bm{\Sigma}^{-1} \bm{P} \bm{\Sigma}^{1/2}$ is idempotent; and $\bm{P} \bm{x}_k \in \{\bm{0}, \bm{x}_k\}$ for each $k \in\{1,\dots,p\}$.
Then, $(L_\infty | \bm{Z} \in \mathcal{C}) \sim \chi_{\nu}^2$.
\end{lemma}

To give a general idea of what the asymptotic distribution of the likelihood ratio test look like in practice we now consider some simple HACs. Generally the mixture weights will depend on $\bm\Sigma$, but we begin with an example in which $L_{\infty}$ follows a $\bar{\chi}^2$ distribution with mixture weights that do not depend on $\bm{\Sigma}$.
It concerns two-level HACs with only two non-leaf nodes, which, in the notation of \eqref{eq:hac-recursive}, corresponds to the case $K_{(0)} \geqslant 2$, $d_{(0,1)} \geqslant 2$ and $d_{(0,k)} = 1$, for $2 \leqslant k \leqslant K_{(0)}$.

\begin{corollary} \label{cor:simple}
Suppose that $C_{\bm{\vartheta}}$ is a two-level HAC with parameter $\bm{\vartheta} = (\vartheta_0, \vartheta_1)$, then under the assumptions of Theorem~\ref{thm:mle-law} and $H_{\circ}: \vartheta_0 = \vartheta_1$ ($\bm{\Theta}_{\circ} = \bm{\Theta}_{(0,1)}$), $L_{\infty} \sim W_1 \chi_1^2$, with $W_1 \sim  \mathcal{B}(1/2)$.
\end{corollary}

The following corollaries treat a similar case, the only difference being that the root node contains two twin non-leaf children; $K_{(0)} \geqslant 2$, $d_{(0,1)}=d_{(0,2)} \geqslant 2$ and $d_{(0,k)} = 1$ ($2 < k \leqslant K_{(0)}$) in \eqref{eq:hac-recursive}.
They respectively concern intersection and union hypotheses.
Note that the mixture weights depends on $\bm{\Sigma}$ this time.

\begin{corollary} \label{cor:simple-2}
Suppose that $C_{\bm{\vartheta}}$ is a two-level HAC with parameter $\bm{\vartheta} = (\vartheta_0, \vartheta_1, \vartheta_2)$,
let $\bm{Z}$ be the Gaussian vector given in Theorem~\ref{thm:mle-law} with $\sigma_k^2 = \Var(Z_k)$, and $\sigma_{k,k+1} = \Cov(Z_k,Z_{k+1})$ for $k=0,1$. 
Then, under the assumptions of Theorem~\ref{thm:mle-law} and $H_{\circ}: \vartheta_0 = \vartheta_1 = \vartheta_2$ ($\bm{\Theta}_{\circ} = \bm{\Theta}_{(0,1)} \cap \bm{\Theta}_{(0,2)}$),
$$
    L_{\infty} \sim  W_1 \chi_1^2 + W_2 \chi_2^2 , \quad \text{for some } (W_0, W_1, W_2) \sim  \mathcal{M}\{1, (\gamma_0,\gamma_1,\gamma_2)\} ,
$$
where $\gamma_0=\cos^{-1}(\beta)/(2\pi)$, $\gamma_1 = 1/2$ and $\gamma_2 = 1/2 - \gamma_0$ with $\beta=(\sigma_0^2 - 2\sigma_{01} + \sigma_{12})/(\sigma_0^2 - 2\sigma_{01} + \sigma_{1}^2)$.
\end{corollary}

\begin{corollary} \label{cor:simple-3}
Suppose that $C_{\bm{\vartheta}}$ is a two-level HAC with parameter $\bm{\vartheta} = (\vartheta_0, \vartheta_1, \vartheta_2)$ and let $W_1 \sim  \mathcal{B}(1/2)$,
then under the assumptions of Theorem~\ref{thm:mle-law} and $H_{\circ}: \vartheta_0 \in \{\vartheta_1, \vartheta_2\}$ ($\bm{\Theta}_{\circ} = \bm{\Theta}_{(0,1)} \cup \bm{\Theta}_{(0,2)}$),  $L_\infty \preceq W_1 \chi_1^2$ if $\vartheta_1 = \vartheta_2$, where $\preceq$ denotes stochastic dominance, and $L_\infty \sim W_1 \chi_1^2$ otherwise.
\end{corollary}
In this latter case, it is not specified by the null whether only one of or both $\vartheta_1$ and $\vartheta_2$ equal $\vartheta_0$. The distribution of $L_{\infty}$ is thus unknown, but one can ensure that the corresponding test is asymptotically conservative, at worst, by using $W_1 \chi_1^2$ as the reference distribution for $L_n$.

\subsection{Power under local alternatives} \label{sec:power-local} %
Fix an Archimedean generator family and a suppose that $\bm{\vartheta}$ satisfies some hypothesis $H_{\circ}$ of interest.
To investigate the asymptotic properties of the likelihood ratio test of \S\ref{sec:Ln-asymptotic}, one can define local departures from $H_{\circ}$ by setting some entries of $\bm{h}$ to non-zero values in Theorem~\ref{thm:mle-law} and studying the asymptotic distribution of $L_n$ in these cases.
Specifically, when $\bm{h}$ depends only on a scalar $h$, one can produce so-called power curves by recording the power of the test at some given nominal level $\alpha$ for each value of $h$.

As an example, consider testing $H_{\circ}:\vartheta_0 = \vartheta_1$ against $H_{\bullet}$ of \eqref{eq:H} in the HAC of Corollary~\ref{cor:simple}, when the parameter generating the data is given by $\bm{\vartheta} + \bm{h}_n/\sqrt{n}$, where $\bm{\vartheta} = (\vartheta, \vartheta)$ for some $\vartheta \in \Theta$ and $\bm{h}_n \to \bm{h}$ as $n \to \infty$. 
To allow for a meaningful comparison across families we set the local parameter $\bm{h}_n$ so that it corresponds to a specific discrepancy in Kendall's $\tau$; this is desirable as Kendall's $\tau$ has a universal interpretation not linked to any specific parametrization of the copula. We can do this by exploiting the bijective relationship between the parameter $\vartheta$ and Kendall's $\tau$ in bivariate Archimedean copulas.
More precisely, for a given generator $\psi$ and a given discrepancy $h'$ on the $\tau$ scale, we set
$$
\bm{h}_n := \sqrt{n}\{ \tau_{\psi}^{-1}(\tau_{\psi}(\bm{\vartheta}) + h' \bm{e}/\sqrt{n}) -  \bm{\vartheta}\} , \qquad \text{with } \bm{e} := (e_0, e_1) \text{ such that } e_1 - e_0 = 1 ,
$$
where $\tau_{\psi}:\Theta \to [0,1]$ is the family-specific function that returns the Kendall correlation corresponding to $\vartheta$; see Table~2.2 in \cite{Hofert:2010} for analytical expressions of $\tau_\psi$ for a variety of families and the \texttt{copula} package in \texttt{R} \cite{copula:2023} for implementations of $\tau_{\psi}$ and $\tau_{\psi}^{-1}$.
One can check that for Gumbel, Clayton, Frank and Joe generators, $\bm{h} = h \bm{e}$ for some $h$; in particular, $h = h'/(1-\tau)^2$ for the Gumbel family and $h = 2h'/(1-\tau)^2$ for the Clayton, where $\tau = \tau_{\psi}(\vartheta)$.
\begin{remark}
The local power of the test is invariant to the choice of $\bm{e}$, as the constraint $e_1-e_0 = 1$ ensures that $\bm{h}$ lies on a line parallel to the local null space $\mathcal{A}_{\circ} = \{(z_0,z_1) : z_0 = z_1\}$.
\end{remark}

For a given nominal level $\alpha$ and a given generator family, the resulting power curve takes the form $\beta_{\alpha}(h') := \Prob\{L > c_{\alpha}\}$, where $c_\alpha$ is the $(1-2\alpha)$th quantile of the $\chi_1^2$ distribution.
We compute this latter using Monte Carlo replicates of $\bm{Z} \sim \mathcal{N}(\bm{h},\bm{\tilde\Sigma})$, where $\bm{\tilde\Sigma}$ is an estimator of $\bm{\Sigma}$ based on a very large dataset generated from the null HAC model.
Note that Monte Carlo sampling of $\bm{Z}$ is not required to compute $\Prob(L = 0)$, since $\Prob(L = 0) = \Phi(0)$, where $\Phi$ is the normal cumulative distribution function with mean parameter $h_1 - h_0$ ($\bm{h} = (h_0,h_1)$) and variance parameter $\sigma_0^2 + \sigma_1^2 - 2\sigma_{01}$ (in the notation of Corollary~\ref{cor:simple-2}).

\begin{figure}[t]
    \centering
    \includegraphics[width=.9\textwidth]{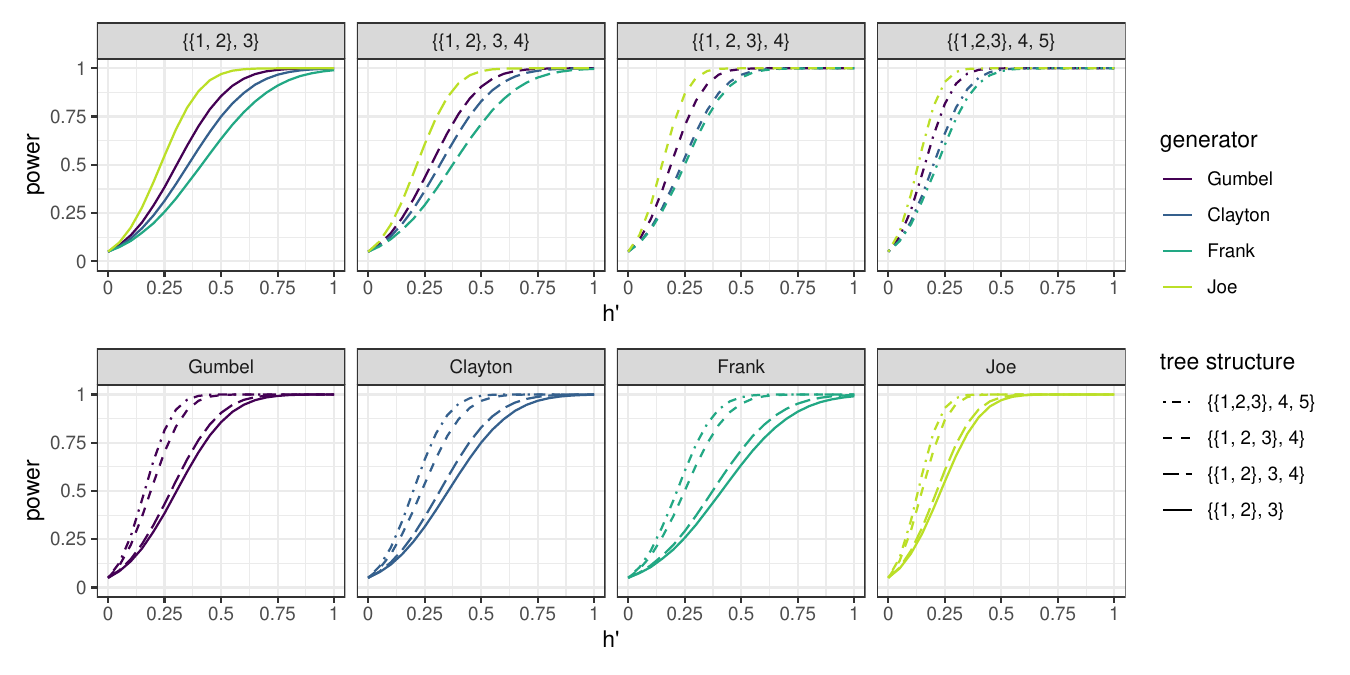}
    \caption{Power curves at nominal level $\alpha = 0.05$ for tests of $H_{\circ}: \vartheta_0 = \vartheta_1$ against $H_{\bullet} : (\vartheta,\vartheta) + \bm{h}_n/\sqrt{n}$, when $\vartheta = \tau_{\psi}^{-1}(1/3)$.   
    Four generator families (Gumbel, Clayton, Frank and Joe) and four two-parameter structures are considered, for a total of sixteen distinct distributions. 
    The top and bottom panels show the same results from a different perspective:
    the top panels compare the local power across the families for a given tree structure, while the bottom panels compare the local power across tree structures for a given family.}
    \label{fig:power-curves}
\end{figure}

Figure~\ref{fig:power-curves} shows the power curves, computed at nominal level $\alpha = 0.05$, for Gumbel, Clayton, Frank and Joe HACs and four distinct structures involving two parameters.
The top panels highlight the difference in performance across the families.
For each structure considered, the test performed best for the Joe case, followed by the Gumbel, Clayton, and then Frank cases.
The bottom panels, which show the same results from a different perspective, highlight the fact that power increases with the number of variables.

\subsection{Nuisance parameters} \label{sec:nuisance}
We now consider cases where the parameter vector of interest, constrained by $H_{\circ}$, is a sub-vector of $\bm{\vartheta}$, so that the latter contains so-called nuisance parameters.
For convenience, we assume that the parameters of interest, say $(\vartheta_{\mathbf{i}})_{\mathbf{i} \in \mathcal{I}'}$ for some $\mathcal{I}' \subset \mathcal{I}$, are such that the nodes indexed by $\mathcal{I}'$ corresponds to a contiguous sub-tree of the HAC.
Without loss of generality, we further assume that $\mathcal{I}'$ contains the root node; otherwise one could get rid of some nuisance parameters by working directly with the margin $C^{\bm{j}}$ for which $\bm{j} = \argmin_{\mathbf{i} \in \mathcal{I}'} |\mathbf{i}|$, as defined in \ref{eq:hac-recursive}.

We first discuss cases assuming all nuisance parameters, say $\vartheta_{(\mathbf{i},k)}$, are distinct from their respective neighboring parameters, i.e., $\vartheta_{\mathbf{i}} < \vartheta_{(\mathbf{i},k)} <  \vartheta_{(\mathbf{i},k,r)}$ for all $r \in \{1,\dots,K_{(\mathbf{i},k)}\}$.
Aside from the extra variation they might induce in the estimation procedure, such parameters do not affect the application of Theorem~\ref{thm:mle-law}, as the dimensions in $\mathcal{A}_{\bullet}$ and $\mathcal{A}_{\circ}$ to which they correspond then all coincide with $\mathbb{R}$.
This is exemplified in the following corollary dealing with a case similar to Corollary~\ref{cor:simple} but with an additional nuisance parameter, and for which we recover the same limiting distribution for $L_n$; its proof follows from that of Corollary~\ref{cor:simple-3}.
\begin{corollary} \label{cor:nuisance}
Suppose that $C_{\bm{\vartheta}}$ is a two-level HAC with parameter $\bm{\vartheta} = (\vartheta_0, \vartheta_1, \vartheta_2)$ such that $\vartheta_1 < \vartheta_2$,
then under the assumptions of Theorem~\ref{thm:mle-law} and $H_{\circ}: \vartheta_0 = \vartheta_1$ ($\bm{\Theta}_{\circ} = \bm{\Theta}_{(0,1)}$), $L_{\infty} \sim W_1 \chi_1^2$, where $W_1 \sim  \mathcal{B}(1/2)$.
\end{corollary}

A more serious challenge arises when it is not clear whether the nuisance parameters are distinct from their neighboring parameters or not, thus creating some uncertainty as to exactly which alternative hypothesis $H_{\bullet}: \bm{\vartheta} \in \bm{\Theta}_{\bullet}$ (with $\bm{\Theta}_{\bullet} \subseteq \bm{\Theta}$) to use.
Since different choices of alternative hypothesis lead to a different null space $\bm{\Theta}_{\circ}$ and in turn different local spaces $\mathcal{A}$ and $\mathcal{A}_{\circ}$, this creates the possibility of a mismatch between the limiting distribution used and the true distribution of $L_{\infty}$.
For example, we derive the distribution of $L_\infty$ in a case similar to Corollary~\ref{cor:nuisance}, this time with all parameters being equal.
\begin{corollary} \label{cor:nuisance-2}
Suppose that $C_{\bm{\vartheta}}$ is two-level HAC with parameter $\bm{\vartheta} = (\vartheta_0, \vartheta_1, \vartheta_2)$ such that $\vartheta_0=\vartheta_1=\vartheta_2$,
let $\bm{Z}$ be the Gaussian vector given in Theorem~\ref{thm:mle-law}, $\sigma_k^2 = \Var(Z_k)$ and $\sigma_{k,k+1} = \Cov(Z_k,Z_{k+1})$ for $k=0,1$.
Further let $\beta=(\sigma_0^2 - 2\sigma_{01} + \sigma_{12})/(\sigma_0^2 - 2\sigma_{01} + \sigma_{1}^2)$ and assume that $\beta \geqslant 0$.
Then, under the assumptions of Theorem~\ref{thm:mle-law} and $H_{\circ}: \vartheta_0 = \vartheta_1$ ($\bm{\Theta}_{\circ} = \bm{\Theta}_{(0,1)}$),
$$
    L_{\infty} \sim W_1 \chi_1^2 + W_2 \chi_2^2 , \quad \text{for some } (W_0, W_1, W_2) \sim  \mathcal{M}\{1, (\gamma_0'',\gamma_1'',\gamma_2'')\} ,
$$
where $\gamma_0'' = 1/4 + \cos^{-1}(\beta)/(2\pi)$, $\gamma_1'' = 1/2$ and $\gamma_2'' = 1/2 - \gamma_0''$.
\end{corollary}

Naturally, if one knew that $\vartheta_0 = \vartheta_2$, one would eliminate $\vartheta_2$ by collapsing the HAC and then relying on Corollary~\ref{cor:simple} for inference.
Corollary~\ref{cor:nuisance-2} shows that the cost of keeping $\vartheta_2$ in the model is extra noise in the form of an additional $\chi_2^2$ component. 
On the other hand, we see by comparing Corollaries~\ref{cor:nuisance} and \ref{cor:nuisance-2} that
for such two-level HACs ($\beta > 0$) with $\vartheta_0 = \vartheta_2$, keeping both parameters in the model leads to a conservative test.
Furthermore, the two tests are equivalent when $\beta = 0$.

Extensions to two-level structures with more nuisance parameters sharing the same value should be relatively simple.
In more complex cases involving, say, $q$ nuisance parameters with unclear status, a simple strategy is to repeat the test for all $2^{q}$ candidates sets $\bm{\Theta}_{\bullet}$ obtained by setting each of them either equal to their closest neighbour (and collapsing the HAC) or not, and then combine or summarize the observed $p$-values.
In particular, using the largest of the p-values thus obtained guarantees that the test is at worst conservative, although for large $q$ we expect this to perform poorly.

\begin{remark} \label{rem:hybrid}
    When $q$ is large, one might instead compute a single p-value using a hybrid null distribution.
To illustrate the idea, consider again testing $H_{\circ}: \vartheta_0 = \vartheta_1$ against $H_{\bullet}: \vartheta_0 < \vartheta_1$ in the presence of a nuisance parameter $\vartheta_2$, as in Corollaries~\ref{cor:nuisance} and \ref{cor:nuisance-2}, and let $h_n = \sqrt{n}(\hat\theta_{\circ 2} - \hat\theta_{\circ 0})$ and $\bm{h}_n := (0,0,h_n)$; in general, $\bm{h}_n$ would have $q$ non-zero entries.
To test $H_{\circ}$, it seems reasonable to use a null distribution similar to that in Corollary~\ref{cor:nuisance} when $h_n$ is large, but similar to that in Corollary~\ref{cor:nuisance-2} when $h_n$ is small.
This can be done by setting the mean of $\bm{Z}$ in Theorem~\ref{thm:mle-law} to $\bm{h}_n$.
If $\vartheta_0 < \vartheta_2$ is true, then $h_n \to \infty$ as $n \to \infty$ and one eventually recovers the asymptotic null distribution in Corollary~\ref{cor:nuisance}.
If however $\vartheta_0 = \vartheta_2$ is true, then $h_n'$ does not converge to a constant, although $h_n' = 0$ with probability ${\sim}\gamma_0''$ (as in Corollary~\ref{cor:nuisance-2}) for large $n$.
Thus, in this case, one does not exactly recover the asymptotic null distribution in Corollary~\ref{cor:nuisance-2}.
Yet, it seems reasonable to believe that the distribution thus obtained is a sensible approximation of the unknown true null distribution.
See Appendix Figure~\ref{fig:sets2} for a visual description of the hybrid null distribution.
We investigate the specific case discussed above in simulations in \S\ref{sec:sim-study}, and find that the approximation is very accurate in the cases considered. 
\end{remark}

\subsection{A conditional test based on the asymptotic null distribution of $L_n$} \label{sec:conditional}
To circumvent the estimation of $\bm{\Sigma}$ when testing $H_\circ$, one may consider the method discussed by \cite{Susko:2013} which is based on \citep{Bartholomew:1961}.
The main idea is to partition the space $\bm{\Theta}$ (and, accordingly, $\mathcal{A}$) into regions (say $\bm{\Theta}^\nu$ and $\mathcal{A}^\nu$) for which the distribution of $(L_{\infty} | \bm{Z}_{\bullet} \in \mathcal{A}^\nu)$ is known, and when $\bm{\hat\theta}_{\bullet} \in \bm{\Theta}^\nu$ we base our test on the latter distribution rather than on that of $L_\infty$.
This method is thus particularly useful when $L_\infty$ is a $\bar\chi^2$ distribution, as in the case developed in Corollaries~\ref{cor:simple}--\ref{cor:nuisance-2}, where $(L_{\infty} | \bm{Z}_{\bullet} \in \mathcal{A}^\nu) \sim \chi_\nu^2$ and $\nu$ is the number of extra constraints introduced by passing from $\bm{\hat\theta}_{\bullet}$ to $\bm{\hat\theta}_{\circ}$.
In doing so, however, one should expect generally conservative sizes and some loss in power, although in some special cases it might lead to an increase in power \citep[Section~5]{Susko:2013}.

\begin{remark} \label{rem:cond}
As an example, consider the HAC in Corollary~\ref{cor:simple}, where $\bm{\vartheta} = (\vartheta_0,\vartheta_1)$ and $H_{\circ}:\vartheta_0=\vartheta_1$.
In this case, $\bm{P}_{\circ}$ is always a matrix of rank 2, and $\bm{P}_{\bullet} = \bm{I}$, when $Z_0 < Z_1$, or $\bm{P}_{\bullet} = \bm{P}_{\circ}$, when $Z_0 > Z_1$.
The event $Z_0 = Z_1$ has zero probability.
Thus, $(L_n| \bm{\hat\theta}_{\bullet} \in \bm{\Theta}^{\nu}) \rightsquigarrow \chi_{\nu}^2$, where $\bm{\Theta}^{0} = \bm{\Theta}_{\circ}$ and $\bm{\Theta}^{1} = \{(\theta_0,\theta_1) : \theta_0 < \theta_1\}$.
The test rejects $H_\circ$ only when $\nu=1$ and $\Prob(\chi_1^2 > L_n) < \alpha$, where $\alpha$ is the nominal test level.
In particular, it has an asymptotic size of $\alpha/2$.
\end{remark}

For comparison, consider the HAC in Corollary~\ref{cor:simple-2}, where $\bm{\vartheta} = (\vartheta_0,\vartheta_1,\vartheta_2)$, $H_{\circ}:\vartheta_0=\vartheta_1=\vartheta_2$ and $H_{\bullet}:\vartheta_0 \leqslant \vartheta_1,\vartheta_2$.
In this case, $\bm{\Theta}^{1} = \cup_{k=0,1} \{\bm{\theta} : \theta_0 = \theta_{1+k} < \theta_{2-k}\}$ and 
$\bm{\Theta}^{2} = \{\bm{\theta} : \theta_0 < \theta_{1}, \theta_{2}\}$.
The test rejects $H_\circ$ only when $\nu \in \{1,2\}$ and $\Prob(\chi_\nu^2 > L_n) < \alpha$, and therefore, in the notation of Corollary~\ref{cor:simple-2}, the test has an asymptotic size of $(\gamma_1 + \gamma_2) \alpha > \alpha/2$.
This suggests more generally that the asymptotic size of the conditional test in similar situations is $(1-\gamma_0) \alpha$, where $\gamma_0 = \Prob(\bm{\hat\theta}_{\bullet} \in \bm{\Theta}_{\circ})$.
For hypotheses $H_{\circ}$ involving many intersections ($\bm{\Theta}_{\circ} = \cap_{\mathcal{I}_{\circ}} \bm{\Theta}_{\mathbf{i}}$ with $|\mathcal{I}_{\circ}|$ large), we expect that $\gamma_0$ be close to zero and hence that the loss in size be negligible.

\begin{figure}
    \centering
    \includegraphics{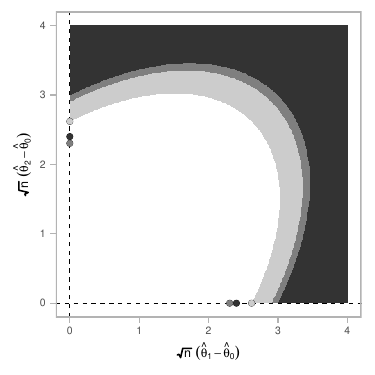}
    \vspace*{-5mm}
    \caption{Depiction of the rejection regions associated with the unconditional test based on Theorem~\ref{thm:mle-law} (all shades of gray) and the exact (two darkest shade of gray) and conservative (the darkest shade of gray) conditional tests discussed in \S\ref{sec:conditional} under the setup of Corollary~\ref{cor:simple-2} ($H_{\circ}: \vartheta_0 = \vartheta_1 = \vartheta_2$) with $\bm{\Sigma} = \bm{I}$.
    To ease the interpretation, the regions are plotted in terms of the vector $\sqrt{n}(\hat\theta_{\bullet 1}- \hat\theta_{\bullet 0},\hat\theta_{\bullet 2}- \hat\theta_{\bullet 0})$ as opposed to $\sqrt{n}(\bm{\theta} - \bm{\vartheta})$; the spaces $\mathcal{A}$ and $\mathcal{A}_{\circ}$ are mapped to $[0,\infty) \times [0,\infty)$ and the origin $(0,0)$, respectively.
    The points on the $x$ and $y$ axes correspond to the critical values used by the tests when $\hat\theta_{\bullet 0} = \hat\theta_{\bullet 2}$ and $\hat\theta_{\bullet 0} = \hat\theta_{\bullet 1}$, respectively.}
    \label{fig:simple-2-rejection}
\end{figure}

The above discussion also suggests that the conditional test can be made asymptotically exact by replacing the desired nominal level $\alpha$ by $\alpha/(1-\gamma_0)$, although this might require estimating $\gamma_0$.
Even in this case, however, the conditional and unconditional tests are still intrinsically different, as they do not use the same critical regions.
The conditional test uses a critical value that varies depending on the location of $\bm{\hat\theta}_{\bullet}$ in $\bm{\Theta}$, while the unconditional test uses a fixed critical value.
This is depicted in Figure~\ref{fig:simple-2-rejection}, which shows the rejection regions associated with the unconditional and the two conditional tests (the conservative test and the asymptotically exact test) under the setup of Corollary~\ref{cor:simple-2} for $\bm{\Sigma} = \bm{I}$.
Analogously, one can think of the difference between the conditional and the unconditional tests as coming from their implied long term sampling schemes.
The unconditional test proposes that we resample from the data generating distribution, while the conditional tests propose that we sample from data which produces an MLE in the same the region of $\bm{\hat\theta}_{\bullet}$.
The loss in size for the conditional method stems from the fact that the null is never rejected when $\bm{\hat\theta}_{\bullet}$ is contained in $\bm{\Theta}_{\circ}$.

\section{Estimation of the Fisher information matrix} \label{sec:sigma-estimation}
We now turn to the estimation of $\bm{\Sigma}$, which is required in most situations to apply Theorem~\ref{thm:mle-law}.
By the second Bartlett identity, $\bm{\Sigma}$ is the inverse of the Fisher Information matrix $\bm{I}_{\bm{\vartheta}} := -\E\{ \bm{J}_{ \bm{\vartheta}}(\bm{U})\}$, where $\bm{J}_{ \bm{\vartheta}}(\bm{U})=\partial^2 \ln c_{\bm{\theta}}(\bm{U})/(\partial \bm{\theta}\ \partial \bm{\theta}^\top)$ evaluated at $\bm{\theta} = \bm{\vartheta}$.
Since $\bm{I}_{\bm{\vartheta}}$ is often difficult to express explicitly and $\bm{\vartheta}$ is unknown, $\bm{\Sigma}$ is usually estimated from the data $\mathcal{U}_n := \{\bm{U}_{r}\}_{r=1}^n$ as the inverse of $(1/n) \sum_{r=1}^n \bm{J}_{\bm{\hat\theta}}(\bm{U}_r)$ where $\bm{\hat\theta} \in \{\bm{\hat\theta}_{\circ}, \bm{\hat\theta}_{\bullet}\}$.
Under weak conditions such as those in Assumptions~\ref{ass:basic-framework}--\ref{ass:stochastic}, this estimator is consistent.
In Appendix~\ref{app:two-level}, we provide the necessary formulas to compute this estimator for two-level Clayton and Gumbel HACs.
But even in these simple cases the complex nature of $c_{\bm{\theta}}$ makes the derivations cumbersome. Therefore we expect that for more complex HACs an additional numerical procedure for estimating $\Sigma$ would be beneficial, we explore this below.

We consider approximating the second-order partial derivatives by their finite difference analogues;
this requires a bit of care to ensure that the log-likelihood is evaluated at points inside the parameter space.
For example, consider two parameters of interest (constrained by $H_{\circ}$) which lie on two consecutive levels and let $\ell_r^*(\bm{\theta}) := \ln c_{\bm{\theta}} (\bm{U}_r)$ for all $\bm{U}_r \in \mathcal{U}_n$.
In this case, we use backward differences for the parameter from the higher level, and forward differences for the parameter from the lower level in order to remain in the parameter space; this is depicted Figure~\ref{fig:sets1}.i (blue arrows).
Concretely, we approximate the off-diagonal entry of $\bm{I}_{\bm{\theta}}$ associated with two nodes $\mathbf{i}$ and $\ik$ with
\begin{equation} \label{eq:finite-difference}
    \frac{\partial^2 \ell_r^*(\bm{\theta})}{\partial \theta_{\mathbf{i}}\ \partial \theta_{\ik}}  \approx D(\bm{\theta}|\bm{U}_r) := \frac{\ell_r^*(\bm{\theta} + \bm{\delta}_{\ik}) - \ell_r^*(\bm{\theta}) - \ell_r^*(\bm{\theta}  - \bm{\delta}_{\mathbf{i}} + \bm{\delta}_{\ik}) + \ell_r^*(\bm{\theta} - \bm{\delta}_{\mathbf{i}})}{\delta_{\mathbf{i}}^* \delta_{\ik}^*} ,    
\end{equation}
where for node $\bm{j} \in \{\mathbf{i}, \ik\}$, $\bm{\delta}_{\bm{j}}$ is a $p$-dimensional vector with a single non-zero entry, and $\delta_{\bm{j}}^*$ is the (node-specific) value of this latter.
The numerical error of this scheme will be negligible in the limit if the third order derivative exists and is bounded in a neighbourhood of the evaluation point; this can be easily argued through a Taylor expansion.

For a given node $\mathbf{i}$, it is possible that $\ell_r^*(\bm{\theta} - \bm{\delta}_{\mathbf{i}})$  varies considerably with respect to small changes in $\theta_{\mathbf{i}}$, leading to unstable behaviour when $\ell_r^*(\bm{\theta})$ and $\ell_r^*(\bm{\theta} - \bm{\delta}_{\mathbf{i}})$ are numerically indistinguishable.
To alleviate this problem, we follow a reasoning similar to that of \S\ref{sec:power-local} and define $\delta_{\mathbf{i}}^*$ so that it corresponds to a small discrepancy of $\delta^\tau$ in the associated Kendall correlation value; we use $\delta^\tau = 0.005$.
Specifically, we let $\delta_{\mathbf{i}}^* = \tau_{\psi}^{-1}\{\tau_{\psi}(\theta_{\mathbf{i}}) - \delta^\tau\} - \theta_{\mathbf{i}}$, where $\tau_{\psi}(\theta)$ is the family-specific function that returns the Kendall correlation corresponding to the Archimedean parameter $\theta$, and we proceed similarly for $\delta_{\ik}^*$.

More general hypotheses for equality of parameters from adjacent nodes on three or more levels, say $H_{\circ}: \vartheta_{(0)}=\vartheta_{(0,1)}=\vartheta_{(0,1,1)}$, is more complex.
This is because the estimate $\hat\theta_{\circ (0,1)}$ must satisfy $\hat\theta_{\circ (0)} \leqslant \hat\theta_{\circ (0,1)} \leqslant \hat\theta_{\circ (0,1,1)}$, leaving no space around it to compute numerical derivatives.
In such cases, one can further exploit Theorem~\ref{thm:mle-law} and replace the subvector $(\hat\theta_{\circ (0)},\hat\theta_{\circ (0,1)},\hat\theta_{\circ (0,1,1)})$ of the null-constrained MLE by $(\hat\theta_{\circ (0)},\hat\theta_{\circ (0,1)},\hat\theta_{\circ (0,1,1)}) + (-\delta_{(0)}^*, 0, \delta_{(0,1,1)}^*) g(n)$, where $g(n)$ is $O(1/\sqrt{n})$.
The hypothesized parameter, which satisfies a specific local alternative to $H_{\circ}$, now lies in the interior of $\bm{\Theta}$, allowing the use of central finite differences to approximate the partial derivatives, while leaving $\bm{\Sigma}$ invariant.
Other cases can be handled similarly by choosing a suitable local deviation from the null.

\begin{figure}[t!]
    \centering
    \includegraphics[scale=1]{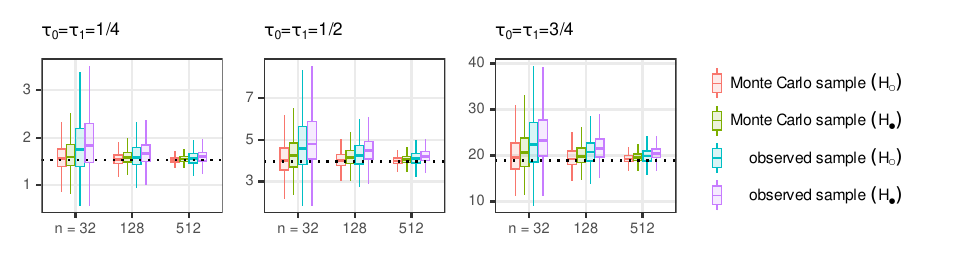}
    \caption{Boxplots comparing four estimators of $\Sigma_{00}$ based on the methodology of \S\ref{sec:sigma-estimation} for a Clayton HAC with structure $\{\{1,2\},3\}$ and three pairs of identical parameters: $(\vartheta_0, \vartheta_1) = (\vartheta, \vartheta)$ with $\vartheta \in \{\theta : \tau_{\psi}(\theta) = 1/4,1/2,3/4\}$; the null hypothesis is $H_\circ : \vartheta_0 = \vartheta_1$. The four estimators are obtained by considering either $\bm{\hat\theta}_{\circ}$ ($H_\circ$) or $\bm{\hat\theta}_{\bullet}$ ($H_\bullet$), and either the observed sample (of size $n$) or a Monte Carlo sample (of size $N = 10^5$) generated from $C_{\bm{\hat\theta}}$. The dashed lines indicate the true value of $\Sigma_{00}$, approximated using Monte Carlo sampling and the explicit formulas given in Appendix~\ref{app:two-level}.}
    \label{fig:sigma-clayton}
\end{figure}

In addition to the estimator of $\bm{I}_{\bm{\vartheta}}$ just described, that is, $(1/n) \sum_{i=1}^n D(\bm{\hat\theta}|\bm{U}_i)$ with $D$ a finite difference scheme such as that in \eqref{eq:finite-difference} and $\bm{\hat\theta} \in \{\bm{\hat\theta}_{\circ}, \bm{\hat\theta}_{\bullet}\}$, we investigate a second approach based on Monte Carlo replicates.
Specifically, we propose to exploit known sampling algorithms for HACs (e.g., from the \texttt{HAC} package \cite{Okhrin/Ristig:2014}) to approximate the expectation $\E D(\bm{\hat\theta}|\bm{U}) \approx \bm{I}_{\bm{\vartheta}}$ with $(1/N) \sum_{i=1}^N D(\bm{\hat\theta}|\bm{U}_i')$, where $N$ is a large integer and $\{\bm{U}_i'\}_{i=1}^N$ are Monte Carlo replicates from $C_{\bm{\hat\theta}}$.
Simulations involving minimal two-parameter Gumbel and Clayton HACs show that the use of Monte Carlo sampling can significantly reduce the estimation noise for small sample sizes.
This can be seen in Figure~\ref{fig:sigma-clayton}, which provides boxplots of the results for the estimation of $\Sigma_{00} := \Var(Z_0)$ for Clayton HACs in a simple scenario, and in Appendix Figure~\ref{fig:sigma-full}, which shows the results for all entries of $\bm{\Sigma}$ and for both families in the same scenario.
The results also suggest that the finite difference approximation is very accurate and that, when the null is true, the use of $\bm{\hat\theta} = \bm{\hat\theta}_{\bullet}$ introduces a bias for small sample sizes.
For large sample sizes, the estimators based on the observed and Monte Carlo data both perform well.

\section{Simulation study} \label{sec:sim-study} %
We now assess the finite sample performance of the tests described in \S\ref{sec:tests}, using the estimation methods for $\bm{\Sigma}$ described in \S\ref{sec:sigma-estimation} when necessary.
To this end, we consider four types of hypotheses, described below, for which we generated datasets of sizes $n \in \{32,128,512\}$ using Gumbel, Clayton and Frank HACs involving two or three parameters.
Recall the definition of $\tau_\psi^{-1}$ from \S\ref{sec:power-local}.
The parameters $(\vartheta_k)_{k=0}^K$, $K \in \{1,2\}$ are chosen such that $\vartheta_k = \tau_\psi^{-1}(\tau_k) + \delta_k$ with $\tau_k \in \{ 1/4, 1/2, 3/4\}$ and $\delta_k \in \{0,1/10\}$; non-zero $\delta_k$ values are used to create departures from the null.
To further evaluate the impact of model misspecification, we model each dataset generated using Gumbel, Clayton and Frank HACs.
For each scenario considered, empirical rejection rates given a nominal level of $\alpha = 0.05$ are computed from 1000 replications.
The main conclusions are described below, while the full results are reported in Tables~\ref{tab:simple-uncond}--\ref{tab:nuisance-hybrid-2} in Appendix~\ref{app:sim-study}.\footnote{For estimators $\bm{\hat\Sigma}$ of $\bm{\Sigma}$ based on Monte Carlo sampling, we use $10^5$ replicates.
When the distribution of $L$ in Theorem~\ref{thm:mle-law} is not covered by one of the corollaries, we use $5 \times 10^3$ Monte Carlo replicates based on $\bm{Z} \sim \mathcal{N}(\bm{h}, \bm{\hat\Sigma})$ to approximate it.}

\subsection{Scenario I: simple hypotheses}
The first scenario treats the simple hypothesis $H_{\circ}: \vartheta_0 = \vartheta_1$ for a trivariate HAC with structure $\mathcal{G} = \{\{1,2\}, 3\}$ and parameters $\bm{\vartheta} = (\vartheta_0,\vartheta_1)$, $\vartheta_0 \leqslant \vartheta_1$.
Six combinations of parameters identified as Cases a--f are considered, of which only the first three are under the null.
Tests were performed using both unconditional and conditional approaches; the mixture distribution of $W_1\chi_1^2$ given in Corollary~\ref{cor:simple} was used as the unconditional approach and $\chi_1^2$ was used when $L_n > 0$ for the conditional approach as in Remark~\ref{rem:cond}.
The results are reported in Appendix Tables~\ref{tab:simple-uncond}--\ref{tab:simple-cond}.

We begin with the results for the unconditional test.
When the null hypothesis is true and the fitted model coincides with that used to generate the data, the rejection rates agree with the nominal level, even for smaller sample sizes.
However, when the fitted model does not coincide with that used to generate the data rejection rates are often dangerously larger than the nominal level.
Surprisingly, this tendency reverses when the null hypothesis is false, as model misspecification leads to lower rejection rates in these cases.
This highlights the importance of thoroughly assessing the fit of the chosen family prior to performing the test, and perhaps of selecting the best fitting family among a large variety of candidates.
The results for Cases~d--f also show that the test's ability to identify departures from the null increases with the sample size and with the strength of the dependence, as expected.

The results for the conditional test follow a similar pattern than those of the unconditional test, the main difference being that the rejection rates are generally smaller.
When the null is true and the model is correctly specified, the rejection rates are approximately half of those of the unconditional model, in accordance with Remark~\ref{rem:cond}.

\subsection{Scenario II: intersection hypotheses}
In the second scenario, we consider the intersection hypothesis $H_{\circ}: \vartheta_0 = \vartheta_1 = \vartheta_2$ for four-variate HACs with structure $\mathcal{G} = \{\{1,2\}, \{3,4\}\}$.
Nine combinations of parameters identified as Cases a--i are considered: $\vartheta_0=\vartheta_1=\vartheta_2$ for Cases a--c; $\vartheta_0=\vartheta_1<\vartheta_2$ for Cases d--f; and $\vartheta_0 < \vartheta_1=\vartheta_2$ for Cases g--i.
We test $H_{\circ}$ using the unconditional test based on asymptotic null distribution of $L_n$ given in Corollary~\ref{cor:simple-2} using the four different estimators of $\bm{\Sigma}$ appearing in Figure~\ref{fig:sigma-clayton}.
All four estimators are obtained by inverting the observed Fisher information. 
They differ in the choice of estimate $\bm{\hat\theta} \in \{\bm{\hat\theta}_{\bullet},\bm{\hat\theta}_{\circ}\}$ at which the observed information is computed and the dataset used, either the observed sample or a Monte Carlo sample based on $\bm{\hat\theta}$; see \S\ref{sec:sigma-estimation} for more details.
The results are reported in Appendix Tables~\ref{tab:int-0-mc}--\ref{tab:int-1-obs}.

The conclusions are broadly the same as for Scenario~I.
Unsurprisingly, the tests are more powerful against alternatives for which both $\vartheta_0 \neq \vartheta_1$ and $\vartheta_0 \neq \vartheta_2$, as in Cases~g--i.
More importantly, the results strongly suggest that the performance of the test is not robust to the choice of estimator $\bm{\hat\Sigma}$.
In particular, the estimators based on $\bm{\hat\theta}_{\bullet}$ yields a liberal test under the null, while this is not the case for those based on $\bm{\hat\theta}_{\circ}$.\footnote{Note that some of the estimates $\bm{\hat\Sigma}$ based on $\bm{\hat\theta}_{\bullet}$ were not positive definite, making it impossible to compute p-values in these cases. 
This almost never occurred for cases when the null is satisfied, and occurred 164 times at of 1000 replications in the worst case.
Replacing all the missing p-values by zero (or by one) has little effect on the results.}
In contrast, the estimation method for $\bm{\Sigma}$ had little impact on the performance of these tests.
The superior performance of the estimator based on $\bm{\hat\theta}_{\circ}$ is likely caused by the additional structure imposed on them; recall that some entries of $\bm{\Sigma}$ are equal under the null, and that the corresponding equalities are enforced in $\bm{\hat\Sigma}$ when computed under $H_{\circ}$.
This was also observed in a similar context in \cite{Perreault/Neslehova/Duchesne:2023}.

\subsection{Scenario III: union hypotheses}
In the third scenario, we generate the data as in Cases~a--i of Scenario~II, but we instead consider the union hypothesis $H_{\circ}: \vartheta_0 \in \{\vartheta_1,\vartheta_2\}$.
To compute $p$-values, we use the one-component mixture consisting of $\chi_1^2$ with probability $1/2$ and zero otherwise, as suggested by Corollary~\ref{cor:simple-3}.
The results are reported in Appendix Table~\ref{tab:union}.

Again, the results broadly agree the with those for Scenarios~I and II.
However, the rejection rates are approximately equal to the nominal level only when either $\vartheta_0 = \vartheta_1$ or $\vartheta_0 = \vartheta_2$, but not both, which occurs only in Cases~d--f.
This highlights the fact by using the reference distribution described above we implicitly assume that at least one of $\vartheta_1$ and $\vartheta_2$ is different from $\vartheta_0$.
Consequently, when this is not the case ($\vartheta_0 \not\in \{\vartheta_1,\vartheta_2\}$), we obtain rejection rates that are conservative, and hence a lower than expected type I error at no cost in terms of type II error.

\subsection{Scenario IV: simple hypotheses with nuisance}
In the last scenario, we consider the simple hypothesis $H_{\circ}: \vartheta_0 = \vartheta_1$ of Scenario I, but this time with four-variate HACs as in Scenarios~II and III.
This introduces a nuisance parameter, which we deal with using the hybrid null distribution described in Remark~\ref{rem:hybrid}.
For comparison, we also test $H_{\circ}$ assuming $\vartheta_0 = \vartheta_2$, in which case we follow Corollary~\ref{cor:simple} and compare $L_n$ to the $(1-2\alpha)$th quantile of a $\chi_1^2$.
The results are reported in Appendix Tables~\ref{tab:nuisance-simplified-1}--\ref{tab:nuisance-hybrid-2}.

The results suggest that the hybrid method performs well in all six cases where the null is satisfied (Cases~a--f).
It thus compares favourably to the simplified method on that aspect, as the latter holds its level only when $\vartheta_0 = \vartheta_2$ (Cases~a--c), and not when $\vartheta_0 \neq \vartheta_2$ (Cases~d--f); this is more pronounced with large sample sizes $n$.
Despite this, the empirical rejections rates associated with the hybrid method are only very slightly below those associated with the simplified method, whether or not $\vartheta_0 = \vartheta_2$.

\section{Discussion} \label{sec:discussion} %

\begin{figure}[t]
    \centering
    \includegraphics[width=0.2\linewidth]{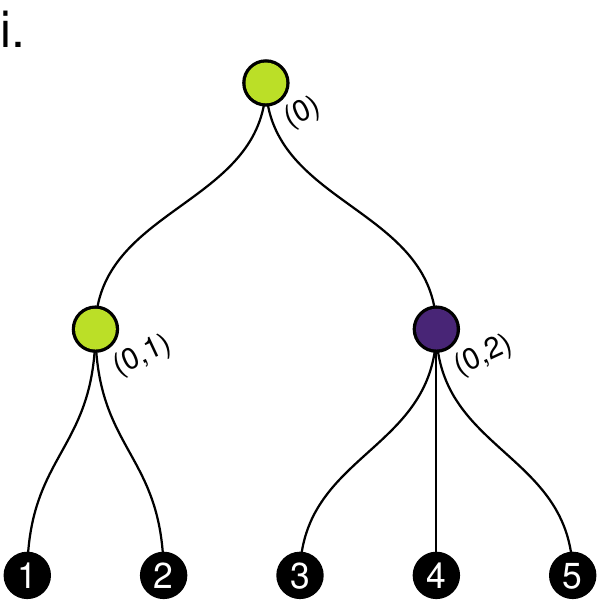}
    \hspace{.05\textwidth}
    \includegraphics[width=0.2\linewidth]{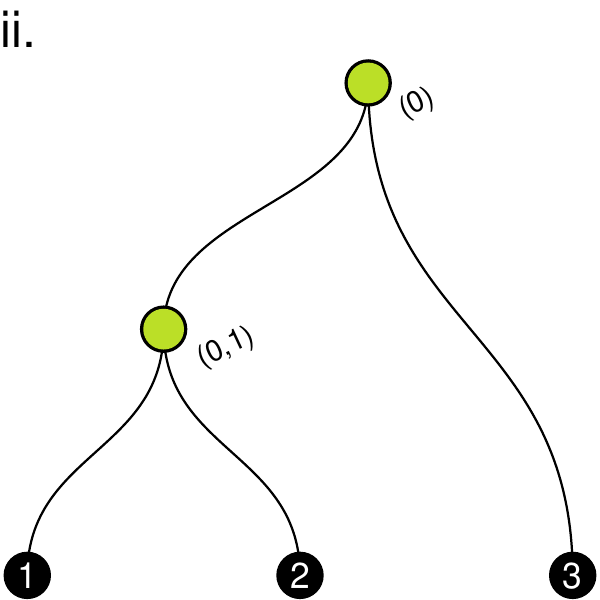}
    \hspace{.05\textwidth}
    \includegraphics[width=0.2\linewidth]{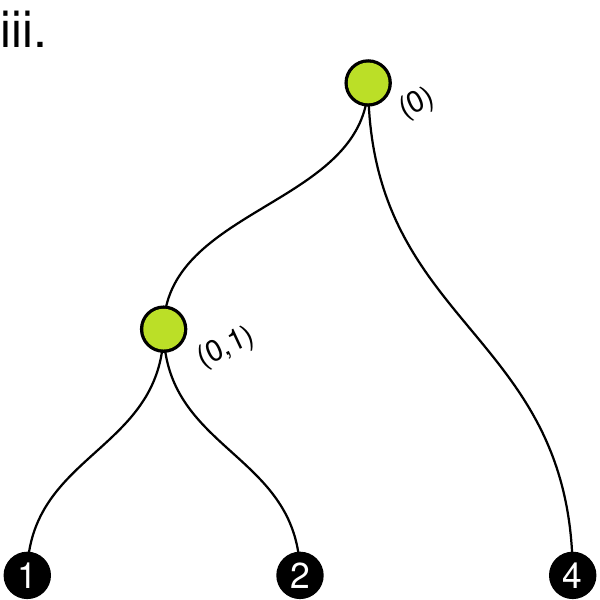}
    \hspace{.05\textwidth}
    \includegraphics[width=0.2\linewidth]{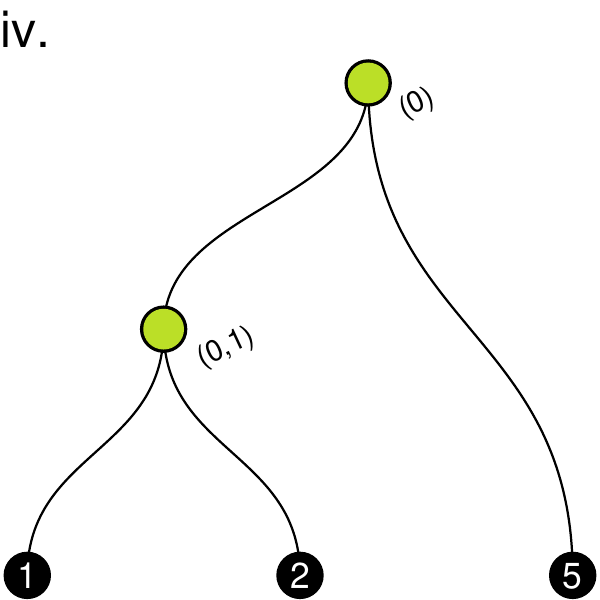}

\caption{Simple example illustrating the construction of a HAC composite likelihood that is free of nuisance parameters. 
Consider a random vector $(U_1,\dots,U_5)$ with a HAC distribution $C_{\bm{\vartheta}}$ with corresponding tree structure as illustrated in Panel (i), where $\bm{\vartheta} = (\vartheta_0,\vartheta_1,\vartheta_2)$ and $\vartheta_2$ is a nuisance parameter.
Panels (ii--iv) show the tree structures underlying the margins corresponding to $(U_1,U_2,U_3)$, $(U_1,U_2,U_4)$ and $(U_1,U_2,U_5)$, respectively. Since none of them depend on $\vartheta_2$, they can be used to construct a composite likelihood free of $\vartheta_2$.}
    \label{fig:composite}
\end{figure}

In this paper, we studied asymptotic behavior of the maximum likelihood estimate and the likelihood ratio statistic testing for overparametrized HACs.
As in \cite{Hofert/Machler/McNeil:2012}, who studied likelihood inference for Archimedean copulas, we have assumed that the marginal distributions are known.
Yet, we can draw conclusions that are likely to apply also in cases when the margins are estimated and when variants of maximum likelihood are used; e.g., two-stage \citep{Genest/Ghoudi/Rivest:1995, Joe:1997}, multistage \citep{Okhrin/Okhrin/Schmid:2013}, or multistage penalized \citep{Okhrin/Ristig:2024} maximum likelihood.

Our theoretical results show that the asymptotic behavior of the maximum likelihood estimate and the likelihood ratio statistics can be very complex, depending on the geometry of the local parameter spaces under the null and alternative.
In order to provide insight into how these quantities behave in practice we examined several shallow HACs with a single, one-parameter family of generators more carefully.
As the asymptotic distribution of the likelihood ratio is not always available, we also discussed alternative approaches which condition on the region in which the maximum likelihood estimate lies; the drawback of this approach is that, without further modification, it is conservative.
Some details on the approximation of the Fisher information matrix were provided, the discussion being specialized to cases where standard finite difference schemes cannot be used.

From Theorem~\ref{thm:mle-law}, it is clear that when the generators are correctly specified and a reasonable initial (possibly overfitting) tree structure is set, maximum likelihood estimation has a positive probability of recovering the true tree structure.
This has some implications on existing structure learning algorithms such as the one considered in \cite{Okhrin/Okhrin/Schmid:2013}, where the collapse of a child node $(\mathbf{i},k)$ into its parent $\mathbf{i}$ is triggered when $\hat{\theta}_{(\mathbf{i},k)} - \hat{\theta}_{\mathbf{i}} < \epsilon$ for some small threshold parameter $\epsilon$.
From the same result it follows that the probability of recovering the true structure converges to one asymptotically.
However, our results suggest that it might be beneficial to use distinct thresholds to collapse child-parent nodes, as the Fisher information matrix varies with the generator considered and the location of the true parameter.
More generally they suggest that structure learning methods can benefit from using the information contained in the curvature of the log-likelihood as opposed to only relying on the difference between parameters; one such method, based on penalized estimation, was recently proposed by \cite{Okhrin/Ristig:2024}.

Our work also raises some concerns about the practical usage of maximum likelihood inference and its variants.
Our simulation study highlights that misspecifying the generators may lead to dramatically inflated rejection rates for structural likelihood ratio tests, which in turn suggests that structure learning algorithms are at risk of returning simplistic tree structures in most practical situations.
A deeper analysis of our simulation results reveals that for a HAC with parameter $\bm{\vartheta}$ in the interior of the parameter space, the best estimator given misspecified generators (e.g., $\bm{\eta}^*$ in Theorem~1 of \cite{Okhrin/Okhrin/Schmid:2013}) can lie on the boundary of the parameter space.
This highlights the importance of considering a large family of generators, and further motivates the development of user-friendly estimation methods that allow mixing generator families within HACs.
It also suggests that in cases when the object of inferential interest is the structure itself and not the parameters, nonparametric methods \citep{Segers/Uyttendaele:2014, Perreault/Duchesne/Neslehova:2019, Zhang/Jin/Bai:2021} might be more appropriate.

Finally, one of the main challenges underlying the implementation of the proposed tests is handling a large number of nuisance parameters.
As a potential approach, we suggest investigating composite likelihood methods \citep{Lindsay:1988}, which have been used by \cite{Cossette/Gadoury/Marceau/Robert:2019}, \cite{Chaboui/al:2021} and \cite{Gorecki/Hofert:2023} in the context of HAC modeling.
Specifically, one could construct an objective function by multiplying marginal likelihoods, as asymptotic guarantees are available for these constructions under some conditions; see, e.g., \cite{Varin:2008} and \cite{Varin/Reid/Firth:2011}.
We illustrate this idea for a simple example in Figure~\ref{fig:composite}, where one can eliminate nuisance parameters at the cost of some loss in efficiency.
Extending our results to composite likelihood inference under boundary conditions, following ideas in, e.g., \cite{Huang/Ning/Cai/al:2020} and \cite{Azadbakhsh/al:2021}, could possibly provide likelihood ratio statistics whose distribution is free of nuisance parameters.

\section*{Acknowledgments}
S.P. was supported by the Data Science Institute at the University of Toronto and N.R. was supported by the Natural Sciences and Engineering Research Council of Canada [grant number RGPIN-2020-05897].

\singlespacing
\bibliographystyle{apalike}

\newpage
\appendix
\gdef\thesection{\Alph{section}} 

\numberwithin{equation}{section}
\numberwithin{table}{section}
\numberwithin{figure}{section}

\renewcommand{\baselinestretch}{1.3} 

\section{Proofs} \label{app:proofs}

\subsubsection*{Proof of Corollary~\ref{cor:simple}}
In this case, $\mathcal{A} = \{\bm{z} : z_1 \leqslant z_2 \}$ is a half space whose only boundary is $\mathcal{A}_{\circ} = \{\bm{z} : z_1 = z_2\}$.
The Gaussian vector $\bm{Z} = (Z_0,Z_1)$ in Theorem~\ref{thm:mle-law}, centered at $(0,0) \in \mathcal{A}_{\circ}$, is thus equally likely to fall on either side $\mathcal{A}_{\circ}$, regardless of $\bm{\Sigma}$.
When $Z_0 \geq Z_1$, then $\bm{Z}_{\bullet}$ lies on $\mathcal{A}_{\circ}$, implying $\bm{Z}_{\circ} = \bm{Z}_{\bullet}$ and therefore $L_n = L_{\infty} = 0$ with probability one.
In the other case when $Z_0 < Z_1$, then $\bm{Z}_{\bullet} = \bm{Z}$ and $\bm{Z}_{\circ} = \bm{P}_{\circ}\bm{Z}$, for some matrix $\bm{P}_{\circ}$ of rank 2. One can check that Lemma~\ref{lem:chi} holds with $\bm{P} = \bm{I} - \bm{P}_{\circ}$, which implies $(L_{\infty} | \bm{Z} \in \mathcal{A}) \sim \chi_1^2$, from which the result follows.

\bigskip

\begin{figure}[t]
    \centering
    \includegraphics[width=.75\textwidth]{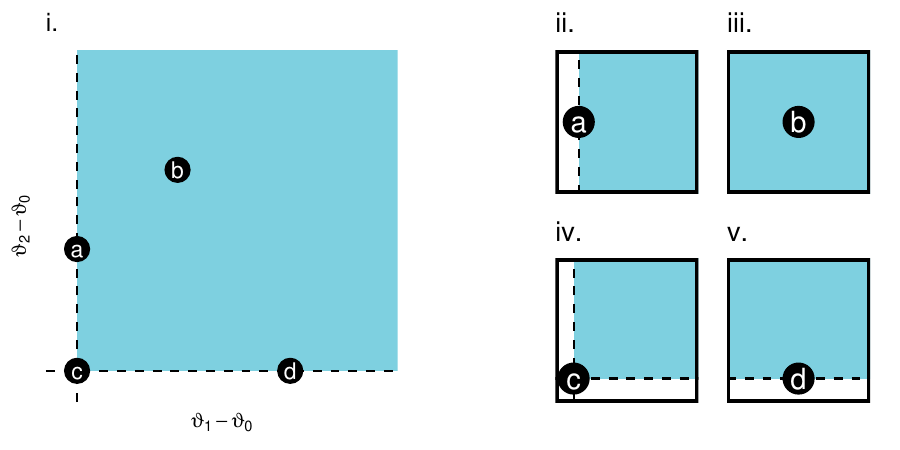}
    \caption{(i) Depiction, in the two-dimensional plane, of the three-dimensional parameter space $\bm{\Theta}$ for a two-level HAC with parameters $\vartheta_0$, $\vartheta_1$ and $\vartheta_2$.
    (ii--v) Depiction of the asymptotic local parameter space $\mathcal{A}^*$ associated with $\bm{Z}^*$ as defined in the proofs of Corollaries~\ref{cor:simple-2}, \ref{cor:simple-3}, and \ref{cor:nuisance-2}. More specifically, (ii) arises when $\vartheta_0 = \vartheta_1 < \vartheta_2$; (iii) arises when $\vartheta_0 < \vartheta_1, \vartheta_2$; (iv) arises when $\vartheta_0 = \vartheta_1 = \vartheta_2$; and (v) arises when $\vartheta_0 = \vartheta_2 < \vartheta_2$.
    \vspace{.1cm}\\
    \textbf{Complement to Remark~\ref{rem:hybrid}}: Consider testing $H_\circ : \vartheta_0 = \vartheta_1$ against $H_\bullet : \vartheta_0 < \vartheta_1$ in the presence of a nuisance parameter $\vartheta_2$ with the hybrid null distribution discussed in Remark~\ref{rem:hybrid}.
    The null distribution in question is that of normal random vector $\bm{Z}_n = (Z_{0n},Z_{1n},Z_{2n})$ centered at $\sqrt{n}(0,0,h_n)$, where $h_n = \hat\theta_2 - \hat\theta_0$. To get some intuition, one may instead focus on $\bm{Z}_n^* = (Z_{1n}^*,Z_{2n}^*) = (Z_{1n}-Z_{0n}, Z_{2n}-Z_{0n})$, whose distribution is normal and centered at $(0,h_n)$.
    Regardless of the value of $\vartheta_2 - \vartheta_0$, the local parameter space $\mathcal{A}^*$ associated with $\bm{Z}_n^*$ is that given in panel (iv), and the associated local null space $\mathcal{A}_\circ^*$ is the non-negative y-axis defined by $Z_{2n}^* = 0$, that is, $Z_{0n} = Z_{2n}$.
    However, when $\vartheta_0 < \vartheta_2$, then $h_n \to \infty$ as $n\to \infty$ and the probability that $Z_{1n}^* = 0$ vanishes.
    Therefore, it does not make a difference to assume that $\mathcal{A}^*$ is as in panel (ii), which leads to a null distribution for $\bm{Z}$ as in the setup of Corollary~\ref{cor:simple-3} ($\vartheta_0 = \vartheta_1 < \vartheta_2$) and Corollary~\ref{cor:nuisance}.
    If instead $\vartheta_0 = \vartheta_2$, then $h_n$ does not converge to zero as $n \to \infty$. 
    In this case the distribution of $\bm{Z}_n^*$ is centered somewhere on the non-negative part of the y-axis (possibly at $(0,0)$, but not necessarily), but the probability that $\bm{Z}_{2n} = 0$ does not vanish as $n \to \infty$ and $\mathcal{A}^*$ may not coincide with (ii).} \label{fig:sets2}
\end{figure}

\textbf{Additional notation for the proofs of Corollaries~\ref{cor:simple-2}, \ref{cor:simple-3} and \ref{cor:nuisance-2}, below.}
Instead of $\bm{Z}$, we consider $\bm{Z}^* := (Z_1, Z_2) - Z_0$, whose distribution is given by $\bm{Z}^* \sim \mathcal{N}(\bm{0}, \bm{\Sigma}_*)$, where $\bm{\Sigma}_* = \sigma_0^2 - 2\sigma_{01} + \sigma_1^2 \bm{I} + (1-\bm{I})\sigma_{12}$.
Denote the two-dimensional analogues of $\mathcal{A}$ and $\mathcal{A}_{\circ}$ associated with $\bm{Z}^*$ by $\mathcal{A}^*$ and $\mathcal{A}_{\circ}^*$ (see Figure~\ref{fig:sets2} for depictions of $\mathcal{A}^*$ in specific cases), and the projection of $\bm{Z}^*$ onto $\mathcal{A}^*$ and $\mathcal{A}_{\circ}^*$ by $\bm{Z}_\bullet^*$ and $\bm{Z}_\circ^*$, respectively.
Also define $\beta := (\sigma_0^2 - 2\sigma_{01} + \sigma_{12})/(\sigma_0^2 - 2\sigma_{01} + \sigma_{1}^2)$.

\subsubsection*{Proof of Corollary~\ref{cor:simple-2}}
One can verify that $L_{\infty} = q_{\bm{Z}^*}(\bm{Z}_{\circ}^*) - q_{\bm{Z}^*}(\bm{Z}_{\bullet}^*)$ and that $\bm{\Sigma}_*$ has eigenvalues $\lambda_{\pm} = (1 \pm 1)(\sigma_0^2 - 2\sigma_{01}) + \sigma_{1}^2 \pm \sigma_{12}$ with corresponding eigenvectors $\bm{v}_{\pm} = (1, \pm 1)/\sqrt{2}$.
Also note that $\mathcal{A}^* = \{\bm{z} : \bm{z} \geqslant \bm{0}\}$ and $\mathcal{A}_{\circ}^* = \{(0,0)\}$ in this case, and let $\mathcal{L}_y$ and $\mathcal{L}_x$ be the vertical and horizontal boundaries of $\mathcal{A}^*$.

In what follows, we partition $\mathbb{R}^2$ into four disjoint (up to zero-probability sets) cones based on the matrix $\bm{P} = \bm{P}_{\bullet} - \bm{P}_{\circ}$ as given in Lemma~\ref{lem:chi}.
Specifically, we write $\mathbb{R}^2 = \mathcal{C}_0 \cup \mathcal{C}_1^x \cup \mathcal{C}_1^y \cup \mathcal{C}_2$, with 
$\mathcal{C}_0 := P_{\bullet}^{-1} \mathcal{A}_{\circ}^*$, 
$\mathcal{C}_1^x := P_{\bullet}^{-1} \mathcal{L}_x$,
$\mathcal{C}_1^y := P_{\bullet}^{-1} \mathcal{L}_y$,
and $\mathcal{C}_2 := P_{\bullet}^{-1} \mathbb{R}_{+}^2$, respectively, where $P_{\bullet}$ and $P_{\circ}$ are the operators projecting $\bm{z} \in \mathbb{R}^2$ onto $\mathcal{A}^*$ and $\mathcal{A}_{\circ}^*$ based on the Mahalanobis distance characterized by $\bm{\Sigma}_*^{-1}$.
Further let $\mathcal{C}_1 := \mathcal{C}_1^y \cup \mathcal{C}_1^x$.
One can verify using Lemma~\ref{lem:chi} that $(L_{\infty} | \bm{Z}^* \in \mathcal{C}_k) \sim \chi_k^2$ for $k=0,1,2$; it remains to determine $\gamma_k := \Prob(\bm{Z}^* \in \mathcal{C}_k)$ for each $k$.

Following \cite[p.~140]{Shapiro:1985}, we compute $\gamma_0$ by finding the opening angle $\alpha_0$ of the cone $\bm{\Sigma}_*^{-1/2} \mathcal{C}_0$, whose upper and lower boundaries are the half lines $\mathcal{L}_{\pm} := \{\bm{z} : z_1 = a(\lambda_+ \pm \lambda_-),\ z_2 = a(\lambda_+ \mp \lambda_-),\ a \in \mathbb{R}_{-}\}$, respectively. 
More precisely, we have $\gamma_0=\alpha_0/(2\pi)$, where $\alpha_0 = \cos^{-1} ( \bm{w}_{+}^\top \bm{w}_- )$ with $\bm{w}_{\pm} \in \bm{\Sigma}_*^{-1/2} \mathcal{L}_{\pm}$ such that $\| \bm{w}_{\pm} \| = 1$.
Through algebraic manipulations we can show that $\bm{w}_{\pm} = (\sqrt{\lambda_+} \bm{v}_+ \pm \sqrt{\lambda_-} \bm{v_-})/\sqrt{\lambda_+ + \lambda_-}$,
and thus $\bm{w}_+^\top \bm{w}_- = (\lambda_+ - \lambda_-)/(\lambda_+ + \lambda_-) = (\sigma_0^2 - 2\sigma_{01} + \sigma_{12})/(\sigma_0^2 - 2\sigma_{01} + \sigma_{1}^2) = \beta$.

To compute $\gamma_1$, first note that $\Prob(\bm{Z}^* \in \mathcal{C}_1^y) = \Prob(\bm{Z}^* \in \mathcal{C}_1^x)$, and so $\gamma_1 = 2 \Prob(\bm{Z}^* \in \mathcal{C}_1^y)$.
Also, $\mathcal{C}_1^y$ has boundaries given by $\mathcal{L}_+$ and $\mathcal{L}_y$.
For any vector $\bm{z}_{+} \in \mathcal{L}_+$, we have $\bm{z}_+^\top \bm{\Sigma}_*^{-1} \in \mathcal{L}_x$, and hence $\bm{z}_+^\top \bm{\Sigma}_*^{-1} \bm{z} = 0$ for any $\bm{z} \in \mathcal{L}_y$.
The opening angle can thus be obtained as $\alpha_1^+ = \cos^{-1} ( 0 ) = \pi/2$, and so $\gamma_1 = 2\Prob(\bm{Z}^* \in \mathcal{C}_1^y) = 2 \alpha_1^+/(2\pi) = 1/2$ and $\gamma_2 = 1 - \gamma_1 - \gamma_0 = 1/2 - \gamma_0$.

\subsubsection*{Proof of Corollary~\ref{cor:simple-3}}

For the case when $\vartheta_0 = \vartheta_1 < \vartheta_2$, we have that $\mathcal{A}^*$ is the right half-plane $(\{0\}\cup\mathbb{R}_+) \times \mathbb{R}$ and $\mathcal{A}_{\circ}^*$ the vertical line $\{0\} \times \mathbb{R}$.
The result follows from the proof of Corollary~\ref{cor:simple}.
    
For the case, $\vartheta_0 = \vartheta_1 = \vartheta_2$, recall the definitions of $\mathcal{L}_x$, $\mathcal{L}_y$, $\mathcal{L}_-$ and $\mathcal{L}_+$ from the proof of Corollary~\ref{cor:simple-2}, and note that $\mathcal{A}^* = (\mathbb{R}_+ \cup \{0\})^2$ and $\mathcal{A}_\circ^* = \{(0,0)\} \cup \mathcal{L}_y \cup \mathcal{L}_x$.
Further denote by $\bm{Z}^x$ and $\bm{Z}^y$ the projections of $\bm{Z}^*$ onto $\mathcal{L}_x$ and $\mathcal{L}_y$, respectively.
We consider the two cases $\beta > 0$ and $\beta \leqslant 0$ seperately.

\begin{figure}[t]
    \centering
    \includegraphics[width=0.425\linewidth]{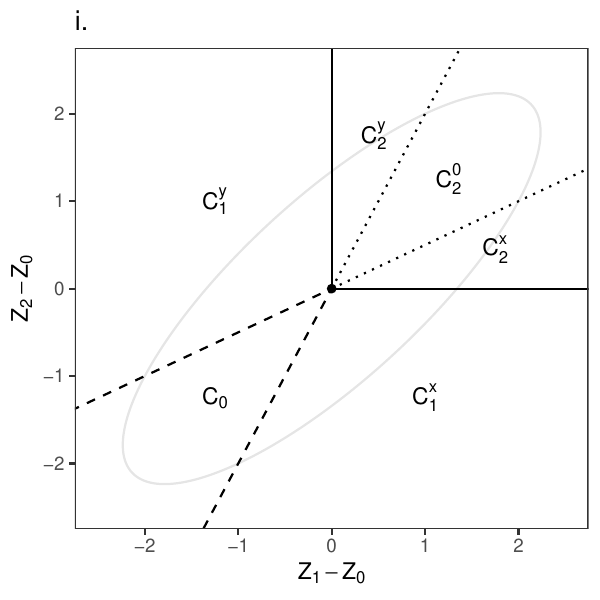}
    \hspace{.1\textwidth}
    \includegraphics[width=0.425\linewidth]{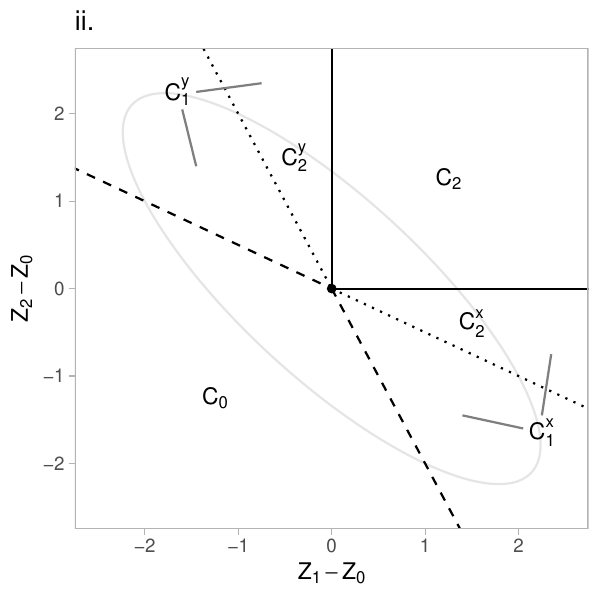}    
        \caption{Partition of $\mathbb{R}^2$ used in the proofs of Corollaries~2, 3 and 5. They are characterized by the half-lines $\mathcal{L}_x := \mathbb{R}_+ \times \{0\}$ and $\mathcal{L}_y := \{0\} \times \mathbb{R}_+$, the dashed half-lines $\mathcal{L}_-$ and $\mathcal{L}_+$ and their reflections, the dotted half-lines $-\mathcal{L}_-$ and $-\mathcal{L}_+$; $\mathcal{L}_-$ and $-\mathcal{L}_-$ are the ones with steepest slopes.
        Panel (i) depicts the case $\beta > 0$, for which $\mathcal{C}_2 = \mathcal{C}_2^x \cup \mathcal{C}_2^0 \cup \mathcal{C}_2^y$.
        Panel (ii) depicts the case $\beta < 0$, for which $\mathcal{C}_2^x \subset \mathcal{C}_1^x$ and $\mathcal{C}_2^y \subset \mathcal{C}_1^y$.
        For reference, a level set of the Gaussian density with covariance $\Cov(Z_1 - Z_0, Z_2 - Z_0)$ is shown in gray.}
    \label{fig:proofs}
\end{figure}

For the case $\beta > 0$, consider the sets $\mathcal{C}_0$, $\mathcal{C}_1^x$, $\mathcal{C}_1^y$ and $\mathcal{C}_2$, as defined in the proof of Corollary~\ref{cor:simple-2} (i.e., as in Figure~\ref{fig:proofs}.i), and
we partition $\mathcal{C}_2 = \mathbb{R}_+^2$ in three regions in a similar fashion such that $\mathcal{C}_0$, $(\mathcal{C}_1^x \cap \mathbb{R}_-^2)$ and $(\mathcal{C}_1^y \cap \mathbb{R}_-^2)$ partition $\mathbb{R}_-^2$.
Denote the resulting cones $\mathcal{C}_2^x$ (with boundaries $-\mathcal{L}_+$ and $\mathcal{L}_x$), $\mathcal{C}_2^y$ (with boundaries $-\mathcal{L}_-$ and $\mathcal{L}_y$) and $\mathcal{C}_2^0$ (with boundaries $-\mathcal{L}_+$ and $-\mathcal{L}_-$) and note that $L_\infty = \min\{q_{\bm{Z}}(\bm{Z}^x),q_{\bm{Z}}(\bm{Z}^y)\}$, which always equals $q_{\bm{Z}}(\bm{Z}^x)$ when $\bm{Z}^* \in \mathcal{C}_2^x$ and $q_{\bm{Z}}(\bm{Z}^y)$ when $\bm{Z}^* \in \mathcal{C}_2^y$.
Further note that $\Prob(\bm{Z}^* \in \mathcal{C}_2^0) = \gamma_0$, and thus
$$
    \Prob(\bm{Z}^* \in \mathcal{D}^x) = \Prob(\bm{Z}^* \in \mathcal{D}^y) = (\gamma_2-\gamma_0)/2 + \gamma_0 = 1/4 \qquad \mathcal{D}^x := \mathcal{C}_2^x \cup \mathcal{C}_2^0,\ \mathcal{D}^y := \mathcal{C}_2^y \cup \mathcal{C}_2^0 ,
$$
and by Lemma~\ref{lem:chi}, $(q_{\bm{Z}}(\bm{Z}^x)|\bm{Z}^* \in \mathcal{D}^x) \sim \chi_1^2$ and $(q_{\bm{Z}}(\bm{Z}^y)|\bm{Z}^* \in \mathcal{D}^y) \sim \chi_1^2$.
Combining all this, we get
\begin{align*} \label{eq:bound}
    \Prob(L_{\infty} > z)\quad  &=\quad  \Prob(L_{\infty} > z \cap \bm{Z}^* \in \mathcal{C}_2)\quad =\quad \Prob\{(L_{\infty} > z \cap \bm{Z}^* \in \mathcal{D}^x) \cup (L_{\infty} > z \cap \bm{Z}^* \in \mathcal{D}^y) \}\\
    \quad &\leqslant\quad \Prob(L_{\infty} > z \cap \bm{Z}^* \in \mathcal{D}^x) + \Prob(L_{\infty} > z \cap \bm{Z}^* \in \mathcal{D}^y)\quad =\quad 2 \Prob(L_{\infty} > z \cap \bm{Z}^* \in \mathcal{D}^x)\\
    \quad &=\quad 2 \Prob(\bm{Z}^* \in \mathcal{D}^x)\ \Prob(L_{\infty} > z\ | \bm{Z}^* \in \mathcal{D}^x)\quad <\quad (1/2) \Prob\{\chi_1^2 > z) ,
\end{align*}
since $L_\infty = \min\{q_{\bm{Z}}(\bm{Z}^x),q_{\bm{Z}}(\bm{Z}^y)\} \leqslant q_{\bm{Z}}(\bm{Z}^x)$ when $\bm{Z}^* \in \mathcal{D}^x$.
In other words, $L_{\infty} \preceq W \xi$ for $W \sim  \mathcal{B}(1/2)$ and some $\xi$ such that $\Prob(\xi > x) < \Prob(\chi_1^2 > x)$.

When $\beta \leqslant 0$, the cones $\mathcal{C}_2^x$ (with boundaries $-\mathcal{L}_+$ and $\mathcal{L}_x$) and $\mathcal{C}_2^y$ (with boundaries $-\mathcal{L}_-$ and $\mathcal{L}_y$) are no longer contained in $\mathcal{C}_2 = \mathbb{R}_+^2$.
In fact, $\mathcal{C}_2^x \subset \mathcal{C}_1^x \subseteq \mathbb{R}_+ \times \mathbb{R}_-$ and $\mathcal{C}_2^y \subset \mathcal{C}_1^y \subseteq \mathbb{R}_- \times \mathbb{R}_+$; see Figure~\ref{fig:proofs}.ii.
By design we still have that $(q_{\bm{Z}}(\bm{Z}^x)|\bm{Z}^* \in \mathcal{C}_2\cup\mathcal{E}^x) \sim \chi_1^2$, $(q_{\bm{Z}}(\bm{Z}^y)|\bm{Z}^* \in \mathcal{C}_2\cup\mathcal{E}^y) \sim \chi_1^2$, and 
$$
\Prob(\bm{Z}^* \in \mathcal{D}^x) = \Prob(\bm{Z}^* \in \mathcal{D}^y) = \Prob(\bm{Z}^* \in \mathcal{C}_2^x) + \Prob(\bm{Z}^* \in \mathcal{C}_2) = (\gamma_0 - \gamma_2)/2 + \gamma_2  = 1/4 .
$$
Hence,
\begin{align*}
    \Prob(L_{\infty} > z)\quad &=\quad \Prob\{L_{\infty} > z \cap (\bm{Z}^* \in \mathcal{E}^x \cup \mathcal{C}_2 \cup \mathcal{E}^y)\}\\
    &=\quad \Prob\{(L_{\infty} > z \cap \bm{Z}^* \in \mathcal{D}^x) \cup (L_{\infty} > z \cap \bm{Z}^* \in \mathcal{D}^y) \}\\
    \quad &\leqslant\quad \Prob(L_{\infty} > z \cap \bm{Z}^* \in \mathcal{D}^x) + \Prob(L_{\infty} > z \cap \bm{Z}^* \in \mathcal{D}^y)\\
    &= \quad 2 \Prob(L_{\infty} > z \cap \bm{Z}^* \in \mathcal{D}^x)\\
    \quad &=\quad 2 \Prob(\bm{Z}^* \in \mathcal{D}^x)\ \Prob(L_{\infty} > z\ | \bm{Z}^* \in \mathcal{D}^x)\\
    &< \quad (1/2) \Prob\{\chi_1^2 > z) ,
\end{align*}
since
$$
(L_{\infty}\ |\ \bm{Z}^* \in \mathcal{D}^x)\ =\ ( \mathbbm{1}\{\bm{Z}^* \in \mathcal{C}_2\} \min\{q_{\bm{Z}}(\bm{Z}^x),q_{\bm{Z}}(\bm{Z}^y)\}\ |\ \bm{Z}^* \in \mathcal{D}^x)\ \leqslant\ ( q_{\bm{Z}}(\bm{Z}^x)\ |\ \bm{Z}^* \in \mathcal{D}^x) .
$$
This concludes the proof.

\subsubsection*{Proof of Corollary~\ref{cor:nuisance-2}}

Use the notation of Corollary~\ref{cor:simple-2}, but this time define the null space as $\mathcal{A}_{\circ}^* = \{(0,0)\} \cup \mathcal{L}_y$. 
Now, if $\bm{Z}^* \in \mathcal{C}_0 \cup \mathcal{C}_1^y$ then $\bm{Z}_{\bullet}^* \in \mathcal{A}_{\circ}^*$, and so $L_\infty = 0$ with probability one.
If $\bm{Z}^* \in \mathcal{C}_1^x$, then $\bm{Z}_{\bullet}^* \in \mathcal{L}_{1}^-$ and $\bm{Z}_{\circ}^* = (0,0)$; an application of Lemma~\ref{lem:chi} shows that $(L_\infty | \bm{Z}^* \in \mathcal{C}_1^x) \sim \chi_1^2$.

To treat the case $\bm{Z}^* \in \mathcal{C}_2$, consider the partition $\mathcal{C}_2 = \mathcal{C}_2^x \cup \mathcal{C}_2^0 \cup \mathcal{C}_2^y$ as defined in the proof of Corollary~\ref{cor:simple-3}.
When $\bm{Z}^* \in \mathcal{C}_2^x \cup \mathcal{C}_2^0$, then $\bm{Z}_{\bullet}^* = \bm{Z}^*$ and $\bm{Z}_{\circ}^* \in \mathcal{L}_y$, meaning that $(L_\infty | \bm{Z}^* \in \mathcal{C}_2^x \cup \mathcal{C}_2^0) \sim \chi_1^2$.
In contrast, if $\bm{Z}^* \in \mathcal{C}_2^y$ then $\bm{Z}_{\bullet}^* = \bm{Z}^*$ and  $\bm{Z}_{\circ}^* = (0,0)$, and so $(L_\infty | \bm{Z}^* \in \mathcal{C}_2^y) \sim \chi_2^2$.
Note that, although it is not shown here, the application of Lemma~\ref{lem:chi} used to derive the last two conditional distributions is valid only when $\beta \geqslant 0$, hence the additional condition in this corollary's statement.

By the same reasoning as in the proof of Corollary~\ref{cor:simple-3}, we get $\gamma := \Prob(\bm{Z}^* \in \mathcal{C}_2^y) = \gamma_0 + (\gamma_2 - \gamma_0)/2 = 1/4 - \cos^{-1}(\beta)$, and hence $\gamma_1'' = \Prob(\bm{Z}^* \in \mathcal{C}_1^x) + \Prob(\bm{Z}^* \in \mathcal{C}_2) - \Prob(\bm{Z}^* \in \mathcal{C}_2^y) = \gamma_1 + \gamma_2 - \gamma = 1/2$ and $\gamma_2'' = \Prob(\bm{Z}^* \in \mathcal{C}_2^y) = \gamma$.

\clearpage
\newpage

\section{Two-level HACs} \label{app:two-level} %
In this section, we consider two-level HACs generated by a single one parameter family with $\infty$-monotone generators.
Specifically, we provide analytic formulas for the first- and second-order derivatives of commonly used families of copulas and explicit formulas for the Clayton and Gumbel generators.
We begin with some notation and formulas for HAC densities specific for two level HACs.

\subsection{Restricted framework, general notation and HAC density}
To ease the notation, we let $K := K_0$ and drop the vector-index notation by using $k$ instead of $(0,k)$ to index the elements associated with node $(0,k)$, for example $\psi_k := \psi_{(0,k)}$ and let $\mathcal{I} := \{0,\dots,K\}$. Also recall that $d := \sum_{i=1}^K d_i$ and let $\bm{d} := (d_i)_{i=1}^{K}$.
Additionally, we follow \cite{Hofert/Pham:2013} and define
\begin{align*} 
 t_i(\bm{u}_i) := \sum_{j=1}^{d_i} \phi_i(u_{ij}) , \qquad \bm{t}(\bm{u}) := \{t_i(\bm{u}_i)\}_{i=1}^{K} \qquad \text{and} \qquad
 t(\bm{u}) := \sum_{i=1}^{K}\phi_{0}\left(C_{i}(\bm{u}_i)\right),
\end{align*}
for $i=1,\dots,K$, $\bm{u}_i = (u_{ij})_{j=1}^{d_i} \in (0,1)^{d_i}$ and $\bm{u} = (\bm{u}_i)_{i=1}^{K}$.
Note that $t_i$ and $t$ are such that $C_i = \psi_i \circ t_i$ and $C = \psi_{0} \circ t$.

In what follows, as well as in Tables~\ref{tab:generators}--\ref{tab:Psi-function}, we use the \textit{prime} symbol to denote the first-order derivative of a function with respect to its argument, and parenthesized integers to denote higher-order derivatives.
When a function involves only one parameter, we use the dot and double dot symbols to denote its first- and second-order partial derivatives with respect to the parameter.
If the function involves multiple parameters, then we replace the dots by the indices of the parameters with respect to which the derivatives are taken (see, e.g., Table~\ref{tab:B-function}), which allows more compact formulas.
For example, we define the derivatives of the generator $\psi_{\theta}$ as
\begin{equation} \label{eq:notation-1}
\begin{split}
\psi_{\theta}'(t) := \frac{\partial}{\partial t}\psi_{\theta}(t) ,
\qquad
\psi_{\theta}^{(k)}(t) := \frac{\partial^k}{\partial^k t}\psi_{\theta}(t) ,
\qquad
\dot{\psi}_{\theta}(t) := \frac{\partial}{\partial \theta} \psi_{\theta}(t) ,\\
\ddot{\psi}_{\theta}(t) := \frac{\partial^2}{\partial \theta^2} \psi_{\theta}(t) ,
\qquad
\dot{\psi}_{\theta}'(t) := \frac{\partial^2}{\partial \theta \partial t} \psi_{\theta}(t).  \qquad   \qquad
\end{split}
\end{equation}
The expression for these derivatives for Clayton and Gumbel generators are provided in Table~\ref{tab:generators}. Note that the higher-order derivatives of $\psi_{\theta}(t)$ with respect to $t$ involve the falling factorial function $(x)_n := \sum_{j=1}^n x^j s(n,j)$ and the function $s_{nk}(x) := \sum_{j=k}^n x^j s(n,j) S(j,k)$, where $s(n,k)$ and $S(n,k)$ are Stirling numbers of the first and second kind \citep[Sections~24.1.3-4]{Goldberg/Newman/Haynsworth:1972}.
These functions and their derivatives are given explicitly in Table~\ref{tab:s-function}.
Derivatives of multi-parameter functions are defined similarly
\begin{equation} \label{eq:notation-2}
\begin{split}
\gensymb{0ex}{\ccirc}{t}{}(\bm{u}) := \frac{\partial}{\partial \theta_0} t(\bm{u}) ,
\qquad
\gensymb{0ex}{\scriptstyle{s}}{t}{}(\bm{u}) := \frac{\partial}{\partial \theta_s}  t(\bm{u}) ,
\qquad 
\gensymb{0ex}{\ccirc\hspace{-0.09ex}\ccirc}{t}{}(\bm{u}) := \frac{\partial^2}{\partial \theta_0 \partial \theta_0}  t(\bm{u}) ,\\
\gensymb{0ex}{\scriptstyle{s}\hspace{-0.125ex}\scriptstyle{r}}{t}{}(\bm{u}) := \frac{\partial}{\partial \theta_s \partial \theta_r}  t(\bm{u}) ,
\qquad
\gensymb{0ex}{\scriptstyle{\ccirc}\hspace{-0.1ex}\scriptstyle{s}}{t}{}(\bm{u}) := \frac{\partial}{\partial \theta_0 \partial \theta_s}  t(\bm{u}) .    \qquad \qquad
\end{split}
\end{equation}

Following Theorem~3.3 of \cite{Hofert/Pham:2013}, if $\phi_0 \circ \psi_i$ ($i=0,\dots,K$) have completely monotone first-order derivatives, the density $c$ associated to a HAC $C$ (whose generators are completely monotone) have the general form, 
\begin{align} \label{eq:density-hac}
c(\bm{u})  = \left\{\sum_{k=K}^d B_{k}(\bm{u})\Psi_{k}(\bm{u})\right\}\prod_{i=1}^{K}\prod_{j=1}^{d_i}\phi_i'(u_{ij}), \qquad B_{k}(\bm{u}) := b_{k}(\bm t(\bm{u})), \quad \Psi_{k}(\bm{u}) := \psi_{0}^{(k)}(t(\bm{u})),
\end{align}
for $\bm{u} \in (0,1)^d$, where for any $\bm{t} \in (0,\infty)^{K}$, $b_{k}(\bm{t})$ is given by
\begin{align} \label{eq:b-function}
b_{k}(\bm{t})=\sum_{\bm{q} \in \mathcal{Q}_{\bm{d},k}}\prod_{i=1}^{K} a_{iq_i}(t_i),
\end{align}
with $\mathcal{Q}_{\bm{d}, k} = \{\bm{q} \in \mathbb{N}^{K}:\sum_{i=1}^{K}q_i=k, q_i \le d_i, i \in \{1,\ldots,K\}\}$ and $a_{iq_i}$ as in \cite[][Eq.~(12)]{Hofert/Pham:2013}.
In general, the functions $a_{iq_i}$ ($i=1,\dots,d_0$) have complicated expressions in terms of Bell polynomials, but for Clayton and Gumbel generators, they simplify significantly to
\begin{align} \label{eq:a-function}
    a_{iq_i}(t) := (\gamma^{\theta_i}+t)^{q_i \theta_0/\theta_i-d_i} s_{d_i q_i}(\theta_0/\theta_i),
\end{align}
where $\gamma=1$ for the Clayton generator and $\gamma=0$ for the Gumbel generator. 
Note that $a_{iq_i}$ depends on $\theta_i$, and hence $b_k$ depends on $\bm{\theta}$ as well.
The derivatives of $a_{ij}(t)$ and $A_{ij}(\bm{u}) := a_{ij}\{t(\bm{u}_i)\}$ are given in Tables~\ref{tab:a-function} and \ref{tab:A-function}, respectively, while the derivatives of $B_k(\bm{u})$ are given in Table~\ref{tab:B-function}.

\begin{remark}
The general form of the function $a_{iq_i}$ in \eqref{eq:a-function} is obtained from the fact that both the Clayton and Gumbel families are subsets of the tilted outer power family; see \cite{Hofert:2010} and \cite[Section~4.1]{Hofert/Pham:2013}.
This family is characterized by generators $\psi_{\theta}$ of the form $\psi_\theta(t) = \psi_*\{(\gamma^\theta + t)^{1/\theta} - \gamma\}$ for $\gamma \in [0,\infty)$, $\theta \in [1,\infty)$ and a $\infty$-monotone generator $\psi_*$.
The Clayton generator is recovered by setting $\psi_*(t) = 1/(1 + t)$ and $\gamma=1$, and the Gumbel generator is recovered by setting $\psi_*(t)=e^{-t}$ and $\gamma=0$.
\end{remark}

\subsection{Partial derivatives of the log-likelihood} \label{sec:partial}

We now discuss how the first- and second-order partial derivatives of the log-likelihood of HACs are obtained. The expression for the second-order derivatives provides an estimator for the Fisher information matrix $\bm{J}_{\theta}$ which appeared in Section \ref{sec:sigma-estimation} via the second Bartlett identity.
In view of \eqref{eq:density-hac}, one can write the HAC log-density as $\ln c(\bm{u}) = \ln \mathfrak{B}_1(\bm{u}) + \ln \mathfrak{B}_2(\bm{u})$, where
\begin{align} \label{eq:log-density-hac}
 \mathfrak{B}_1 \equiv \mathfrak{B}_1(\bm{u}) = (-1)^d \sum_{k=K}^d B_{k} (\bm{u})\ \Psi_k(\bm{u}) \quad \text{and} \quad \mathfrak{B}_2 \equiv \mathfrak{B}_2(\bm{u}) =  \prod_{i=1}^{K} \prod_{j=1}^{d_i} (-1) \phi_i'(u_{ij}) .
\end{align}
In particular, for each $i =0,\dots,K$,
$$
    \frac{\partial \ln c(\bm{u})}{\partial \theta_s} = \sum_{k=1}^2 \frac{\partial \ln \mathfrak{B}_k}{\partial \theta_s} = \sum_{k=1}^2 \left( \frac{\partial \mathfrak{B}_k}{\partial \theta_s} \right) \mathfrak{B}_k^{-1}
$$
and
$$
    \frac{\partial^2 \ln c(\bm{u})}{\partial \theta_s \partial \theta_r} = \sum_{k=1}^2 \left\{ \left( \frac{\partial^2 \mathfrak{B}_k}{\partial \theta_s \partial \theta_r} \right) \mathfrak{B}_k^{-1} - \left( \frac{\partial \mathfrak{B}_k}{\partial \theta_s} \right) \left( \frac{\partial \mathfrak{B}_k}{\partial \theta_r} \right) \mathfrak{B}_k^{-2} \right\} .
$$
It is straightforward to verify that $\partial \mathfrak{B}_2 / \partial \theta_0 = 0$, and thus also that $\partial^2 \mathfrak{B}_2 /(\partial \theta_0 \partial \theta_i) = 0$ for all $i=0,\dots,K$.
For Clayton and Gumbel generators, $\partial \ln \mathfrak{B}_2 / \partial \theta_i = d_i/\theta_i + \sum_{j=1}^{d_i} g(u_{ij})$ and $\partial^2 \mathfrak{B}_2 / \partial \theta_i^2 = -d_i/\theta_i^2$, where $g(u_{ij}) = - \ln(u_{ij})$ for the Clayton generator and $g(u_{ij}) = \ln\{- \ln(u_{ij})\}$ for the Gumbel generator.

The partial derivatives of $\mathfrak{B}_1$ are given by
\begin{align} \label{eq:derivatives-B1-general}
\begin{split}
 \frac{\partial \mathfrak{B}_1}{\partial \theta_s} &= (-1)^d\sum_{k=d_0}^{d} \left\{\Pk{k} \btsk{s}{k} + B_{k} \Pkr{k}{s}\right\}\\
\frac{\partial^2 \mathfrak{B}_1}{\partial \theta_s \partial \theta_r} &= (-1)^d\sum_{k=d_0}^{d} \left\{\btssk{s}{r}{k} \Pk{k} + 
\btsk{s}{k} \Pkr{k}{r} + 
\btsk{r}{k} \Pkr{k}{s} +
B_{k} \Pkrs{k}{s}{r}\right\} ,
\end{split}
\end{align}
where the dependence of each term on $\bm{u}$ has been suppressed for convenience.
Explicit expressions for the Clayton and Gumbel families based on these formulas can be constructed from Tables~\ref{tab:B-function} and \ref{tab:Psi-function}, which themselves builds upon previous tables and definitions.

Along with Tables~\ref{tab:generators}--\ref{tab:Psi-function}, the equations in \eqref{eq:log-density-hac} and \eqref{eq:derivatives-B1-general} contain the necessary material for implementing functions that returns the log-likelihood and its partial derivatives.
As part of the Supplementary Material accompanying this article, we provide our own implementations written in the \texttt{R} programming language for calculating these derivatives.
In the simulations of \S\ref{sec:sim-study}, we use these to assess the accuracy of numerical estimations of $\bm{\Sigma}$, and generally rely on the more efficient estimation functions available in the \texttt{R} package \texttt{HAC} \citep{Okhrin/Ristig:2014}.


\begin{table}[ph!]
\caption{Derivatives of the generator $\psi_{\theta}$ and its inverse $\phi_{\theta}$ relevant for computing Clayton and Gumbel HAC densities and their associated Fisher information. See \eqref{eq:notation-1}--\eqref{eq:notation-2} for notation.} \label{tab:generators}
\normalsize
\setstretch{1.5}
\centering
\begin{tabular}{lll}
\hline
 & Clayton & Gumbel \\
\hline
$\psi_{\theta}(t)$ & $(1+t)^{-1/\theta}$ & $\exp(-t^{1/\theta})$\\

$\psi_{\theta}'(t)$ & $-(1/\theta)(1+t)^{-1/\theta-1}$ & $-\psi_{\theta}(t)t^{1/\theta -1}(1/\theta)$\\

$\psi_{\theta}^{(k)}(t)$ & $(-1/\theta)_k(1+t)^{-1/\theta-k}$ & $\psi_{\theta}(t) \sum_{j=1}^{k}t^{j/\theta-k}(-1)^j s_{kj}(1/\theta)$\\

$\dot{\psi}_{\theta}(t)$ & $\psi_{\theta}(t) \ln(1+t) / \theta^2$ & $\psi_{\theta}(t)t^{1/\theta}\ln (t)/\theta^2$ \\

$\ddot{\psi}_{\theta}(t)$ & $\dot{\psi}_{\theta}(t) [\ln(1+t)/\theta^2 - 2/\theta ]$ & $\dot{\psi}_{\theta}(t) [(t^{1/\theta} - 1) \ln (t)/\theta^2 - 2/\theta]$\\

$\dot{\psi}_{\theta}'(t)$ & $\psi_{\theta}'(t)[\ln(1+t)/\theta^2 - 1/\theta]$ & $\psi_{\theta}'(t) [(t^{1/\theta}-1)\ln (t)/\theta^2 - 1/\theta]$\\

\hline 

$\phi_{\theta}(u)$ & $u^{-\theta}-1$ & $(-\ln u)^{\theta}$\\

$\phi_{\theta}'(u)$ & $-\theta u^{-\theta-1}$ & $-\theta u^{-1}(-\ln u)^{\theta-1}$\\  

$\phi_{\theta}^{(2)}(u)$ & $- \phi_\theta'(u) (\theta + 1)u^{-1}$ & $\phi_\theta'(u) u^{-1} [(\theta-1)/\ln(u) - 1]$\\

$\dot{\phi}_{\theta}(u)$ & $u^{-\theta}\ln (1/u)$ & $\phi_{\theta}(u) \ln (-\ln u)$\\

$\ddot{\phi}_{\theta}(u)$ & $\dot{\phi}_{\theta}(u) \ln(1/u)$ & $\dot{\phi}_{\theta}(u) \ln (-\ln u)$\\

$\dot{\phi}_{\theta}'(u)$ & $\phi_{\theta}'(u) [1/\theta - \ln(u)]$ & $\phi_{\theta}'(u) [1/\theta - \ln(-\ln u)]$\\

\hline
\end{tabular}
\end{table}

\begin{table}[t]
\caption{Functions involving Stirling numbers of the first and second kind \citep[Sections~24.1.3-4]{Goldberg/Newman/Haynsworth:1972}, denoted $s$ and $S$, respectively, and their derivatives appearing in HAC densities and their associated Fisher information. See \eqref{eq:notation-1}--\eqref{eq:notation-2} for notation.} \label{tab:s-function}
\normalsize
\setstretch{1.5}
\centering
\begin{tabular}{ll|ll}
\hline

$s_{nk}(x)$ & $\sum_{j=k}^{n}x^js(n,j)S(j,k)$ &
$(x)_n$ & $\sum_{j=1}^{n}x^js(n,j)$\\

$s_{nk}'(x)$ & $\sum_{j=k}jx^{j-1}s(n,j)S(j,k)$ &
$(x)'_n$ & $\sum_{j=1}^{n}jx^{j-1}s(n,j)$\\

$s_{nk}^{(2)}(x)$ & $\sum_{j=k}j(j-1)x^{j-2}s(n,j)S(j,k)$ &
$(x)^{(2)}_n$ & $\sum_{j=1}^{n}j(j-1)x^{j-2}s(n,j)$\\
\hline
\end{tabular}
\end{table}

\begin{table}[t]
\caption{The function $a_{sj}$ of \eqref{eq:a-function} and its derivatives appearing in Clayton ($\gamma = 1$) and Gumbel ($\gamma = 0$) HAC densities and their associated Fisher information. See \eqref{eq:notation-1}--\eqref{eq:notation-2} for notation.} \label{tab:a-function}
\normalsize
\setstretch{2}
\centering
\begin{tabular}{ll}

\multicolumn{2}{l}{
$\xi_{s j}(t) := (\gamma^{\theta_s}+t)^{j \theta_0/\theta_s-d_s}$
\qquad 
$\zeta_s(t) := \frac{j}{\theta_s}\ln(\gamma^{\theta_s}+t)$
}\\

\hline
$a_{sj}(t)$ &  $\xi_{s j}(t) s_{d_s j}(\tfrac{\theta_0}{\theta_s})$\\

$a_{sj}'(t)$ & $(j \theta_0/\theta_s-d_s)(\gamma^{\theta_s}+t)^{-1}a_{sj}(t)$\\

$a_{sj}^{(2)}(t)$ & $(j \theta_0/\theta_s-d_s-1)(\gamma^{\theta_s}+t)^{-1} a_{sj}'(t)$\\

$\gensymb{0ex}{\ccirc}{a}{_{sj}}(t)$ & $\zeta_s(t)a_{sj}(t)+\xi_{s j}(t) s'_{d_s j}(\tfrac{\theta_0}{\theta_s})\frac{1}{\theta_s}$\\

$\gensymb{0ex}{\scriptstyle{s}}{a}{_{sj}}(t)$ & $\frac{-\theta_0}{\theta_s}\zeta_s(t)a_{sj}(t)-\xi_{s j}(t)s'_{d_s j}(\tfrac{\theta_0}{\theta_s})\frac{\theta_0}{\theta^2_s}$\\

$\gensymb{0ex}{\ccirc}{a}{'_{sj}}(t)$ & $(\gamma^{\theta_s}+t)^{-1}\left[\frac{j}{\theta_s}a_{sj}(t)+(j \theta_0/\theta_s-d_s)\gensymb{0ex}{\ccirc}{a}{_{sj}}(t)\right] $\\

$\gensymb{0ex}{\scriptstyle{s}}{a}{'_{sj}}(t)$ & $(\gamma^{\theta_s}+t)^{-1}\left[\frac{-j\theta_0}{\theta_s^2}a_{sj}(t)+(j \theta_0/\theta_s-d_s)\gensymb{0ex}{\scriptstyle{s}}{a}{_{sj}}(t)\right]$\\

$\gensymb{0ex}{\ccirc \hspace{-.1ex} \ccirc}{a}{_{sj}}(t)$ & $\zeta_s(t)\left[\frac{\partial}{\partial \theta_0}a_{sj}(t)+\xi_{s j}(t) s'_{d_s j}(\tfrac{\theta_0}{\theta_s})\frac{1}{\theta_s}\right]+\xi_{s j}(t) s^{(2)}_{d_s j}(\tfrac{\theta_0}{\theta_s})\frac{1}{\theta^2_s}$\\

$\gensymb{0ex}{\ccirc \hspace{-.15ex} \scriptstyle{s}}{a}{_{sj}}(t)$ & $\frac{\theta_0 \zeta_s(t)}{\theta_s}\left[\frac{2}{\theta_s} a_{s j}(t) - \gensymb{0ex}{\scriptstyle{s}}{a}{_{sj}}(t)\right] + \frac{{\theta_0^2}\xi_{s j}(t)}{\theta_s^4}\left\{ \theta_s \left[\zeta_s(t) + \frac{2}{\theta_0} \right] s'_{d_s j}(\tfrac{\theta_0}{\theta_s})+s^{(2)}_{d_s j}(\tfrac{\theta_0}{\theta_s})\right\}$\\
\hline
\end{tabular}
\end{table}

\begin{table}
\caption{The function $A_{sj}$ defined below \eqref{eq:a-function} and its derivatives appearing in HAC densities and their associated Fisher information. See \eqref{eq:notation-1}--\eqref{eq:notation-2} for notation.} \label{tab:A-function}
\normalsize
\setstretch{2}
\centering
\begin{tabular}{lll}
\hline
$A_{sj}(\bm{u})$ & 
$a_{sj}(t_s)$ &
$t_s \equiv t(\bm{u_s}) := \sum_{j=1}^{d_s}\psi_{{\theta_s}}(u_{sj})$\\

$\gensymb{0ex}{\ccirc}{A}{_{sj}}(\bm{u})$ &
$\gensymb{0ex}{\ccirc}{a}{_{sj}}(t_s)$ &
$\dot{t}_s \equiv \dot{t}_s(\bm{u_s}) := \sum_{j=1}^{d_s}\dot{\psi}_{{\theta_s}}(u_{sj})$\\

$\gensymb{0ex}{\scriptstyle{s}}{A}{_{sj}}(\bm{u})$ &
$\gensymb{0ex}{\scriptstyle{s}}{a}{_{sj}}(t_s)+ \dot{t}_s a_{sj}'(t_s)$ &
$\ddot{t}_s \equiv \ddot{t}_s(\bm{u_s}) := \sum_{j=1}^{d_s}\ddot{\psi}_{{\theta_s}}(u_{sj})$\\

$\gensymb{0ex}{\ccirc \ccirc}{A}{_{sj}}(\bm{u})$ &
$\gensymb{0ex}{\ccirc \ccirc}{a}{_{sj}}(t_s)$ &
\\

$\gensymb{0ex}{\ccirc \scriptstyle{s}}{A}{_{sj}}(\bm{u})$ &
$\gensymb{0ex}{\ccirc \scriptstyle{s}}{a}{_{sj}}(t_s) + \dot{t}_s\gensymb{0ex}{\ccirc \hspace{-.15ex}}{a}{_{sj}'}(t_s)$\\

$\gensymb{0ex}{\scriptstyle{s} \scriptstyle{s}}{A}{_{sj}}(\bm{u})$ &
$\gensymb{0ex}{\scriptstyle{s} \scriptstyle{s}}{a}{_{sj}}(t_s) + 2\gensymb{0ex}{\scriptstyle{s}}{a}{'_{sj}}(t_s)\ \dot{t}_s + \ddot{t}_s a_{sj}''(t_s)\ (\dot{t}_s)^2+a_{sj}'(t_s)$\\
\hline
\end{tabular}
\end{table}

\begin{table}
\caption{The function $B_k := b_k \circ \bm{t}$ defined in \eqref{eq:density-hac}--\eqref{eq:b-function} and its derivatives appearing in HAC densities and their associated Fisher information. See \eqref{eq:notation-1}--\eqref{eq:notation-2} for notation.} \label{tab:B-function}
\small
\setstretch{2}
\centering
\begin{tabular}{ll}
\multicolumn{2}{l}{
\vspace{-.25cm}
$A_{sj_s} := a_{sj_s}(t_s)$ 
\qquad 
$t_s \equiv t(\bm{u_s}) := \sum_{j=1}^{d_s}\psi_{{\theta_s}}(u_{sj})$}\\
\multicolumn{2}{l}{
$\mathcal{S} := \{1,\dots,d_0\}$ 
\qquad
$\mathcal{S}_{s} := \mathcal{S} \setminus \{s\}$
\qquad
$\mathcal{S}_{sr} := \mathcal{S} \setminus \{s,r\}$}\\

\hline
$B_k(\bm{u})$ 
& $\sum_{\bm{j} \in \mathcal{Q}_{\bm{d},k}}
\prod_{s \in \mathcal{S}} a_{sj_s}$\\

$\gensymb{0ex}{\ccirc}{B_{k}\hspace{-.6ex}}{(\bm{u})}$
& $\sum_{\bm {j}\in \mathcal{Q}_{\bm {d},k}}\sum_{s \in \mathcal{S}}\left\{\left[\prod_{m \in \mathcal{S}_s}a_{mj_m}\left(t_m\right)\right]\gensymb{0ex}{\ccirc}{A}{_{sj_s}}\right\}$\\

$\gensymb{0ex}{\scriptstyle{s}}{B_{k}\hspace{-.6ex}}{(\bm{u})}{}$
& $\sum_{\bm {j}\in \mathcal{Q}_{\bm {d},k}}\left[\prod_{m \in\mathcal{S}_s}a_{mj_m}\left(t_m\right)\right]\gensymb{0ex}{\scriptstyle{s}}{a}{_{sj_s}}$\\

$\gensymb{0ex}{\ccirc \ccirc}{B_{k}\hspace{-.6ex}}{(\bm{u})}$  
& $\sum_{\bm {j}\in \mathcal{Q}_{\bm {d},k}}\sum_{s\in\mathcal{S}}\left( \left\{\sum_{l\in\mathcal{S}_s} \gensymb{0ex}{\ccirc}{A}{_{lj_l}} \prod_{m\in\mathcal{S}_{sl}}a_{mj_m}\left(t_m\right) \right\}\gensymb{0ex}{\ccirc}{a}{_{sj_s}} +\left[\prod_{m\in\mathcal{S}_s}a_{mj_m}\left(t_m\right)\right]\gensymb{0ex}{\ccirc \ccirc}{a}{_{sj_s}}\right)$\\

$\gensymb{0ex}{\ccirc \scriptstyle{s}}{B_{k}\hspace{-.6ex}}{(\bm{u})}$ 
& $\sum_{\bm {j}\in \mathcal{Q}_{\bm {d},k}}\left( \left\{\sum_{l \in\mathcal{S}_s}\left[\prod_{ m\in\mathcal{S}_{sl}}a_{mj_m}\left(t_m\right)\right]\gensymb{0ex}{\ccirc}{a}{_{lj_l}} \right\}\gensymb{0ex}{\scriptstyle{s}}{A}{_{sj_s}} +\left[\prod_{ m\in\mathcal{S}_s}a_{mj_m}\left(t_m\right)\right]\gensymb{0ex}{\ccirc \scriptstyle{s}}{A}{_{sj_s}}\right)$\\

$\gensymb{0ex}{\scriptstyle{s} \scriptstyle{s}}{B_{k}\hspace{-.6ex}}{(\bm{u})}$ 
& $\sum_{\bm {j}\in \mathcal{Q}_{\bm {d},k}}\left[\prod_{ m\in\mathcal{S}_s}a_{mj_m}\left(t_m\right)\right]\gensymb{0ex}{\scriptstyle{s} \scriptstyle{s}}{A}{_{sj_s}}$\\

$\gensymb{0ex}{\scriptstyle{s} \scriptstyle{m}}{B_{k}\hspace{-.6ex}}{(\bm{u})}$ 
& $\sum_{\bm {j}\in \mathcal{Q}_{\bm {d},k}}\left[\prod_{ l\in\mathcal{S}_{sm}}a_{lj_l}\left(t_l\right)\right]\gensymb{0ex}{\scriptstyle{s}}{A}{_{sj_s}}\gensymb{0ex}{\scriptstyle{m}}{A}{_{mj_m}}$\\

\hline
\end{tabular}
\end{table}

\tiny
\begin{table}
\caption{The function $\Psi_{k} := \psi_0^{(k)} \circ t$ defined in \eqref{eq:density-hac} and its derivatives appearing in Clayton and Gumbel HAC densities and their associated Fisher information. See \eqref{eq:notation-1}--\eqref{eq:notation-2} for notation.} \label{tab:Psi-function}
\scriptsize
\setstretch{2}
\centering
\begin{tabular}{lll}
\hline
\multicolumn{3}{l}{Clayton:\ $f_{\theta k}(x)=(-\frac{1}{\theta})_k\left(1+x\right)^{-(k+1/\theta)}$ \qquad Gumbel:\ $f_{\theta k} (x)=\psi_{\theta}(x)\sum_{j=1}^{k}x^{j/\theta-k}(-1)^js_{kj}(1/\theta)$}\\
\hline
$\Pk{k}(\bm{u})$ &  $f_{\theta_0 k}(t)$ & $t \equiv t(\bm{u}) = \sum_{i=1}^K \phi_0(C_i(\bm{u}_i))$\\

$\Pkc{k}(\bm{u})$ & $\gensymb{-.5ex}{\ccirc}{f_{\theta_0 k}}{}(t) + \gensymb{0ex}{\ccirc}{t}{} f'_{\theta_0 k}(t)$ & $\gensymb{0ex}{\ccirc}{t}{} \equiv \gensymb{0ex}{\ccirc}{t}{}(\bm{u}) = \sum_{i=1}^K \dot{\phi}_{0}(C_i(\bm{u}_i))$\\

$\Pkr{k}{s}(\bm{u})$ & $\gensymb{0ex}{\scriptstyle{s}}{t}{} f'_{\theta_0 k}(t)$  & $\gensymb{0ex}{\scriptstyle{s}}{t}{} \equiv \gensymb{0ex}{\scriptstyle{s}}{t}{}(\bm{u}) = \phi_{0}'(C_s(\bm{u}_s)) \dot{C}_s(\bm{u}_s)$\\

$\Pkcc{k}(\bm{u})$ & $\gensymb{-.5ex}{\ccirc\hspace{-.1ex}\ccirc}{f_{\theta_0 k}}(t) +
\gensymb{0ex}{\ccirc}{t}{} \gensymb{-.5ex}{\ccirc}{f_{\theta_0 k}'}(t) + \Big(\gensymb{0ex}{\ccirc}{t}{}\Big)^2 f''_{\theta_0 k}(t) + \gensymb{0ex}{\ccirc\hspace{-0.09ex}\ccirc}{t}{} f'_{\theta_0 k}(t)$ & $\gensymb{0ex}{\ccirc\hspace{-0.1ex}\ccirc}{t}{} \equiv \gensymb{0ex}{\ccirc\hspace{-0.1ex}\ccirc}{t}{}(\bm{u}) = \sum_{i=1}^K \ddot{\phi}_{0}(C_i(\bm{u}_i))$\\

$\Pkcr{k}{s}(\bm{u})$ & $\gensymb{0ex}{\scriptstyle{s}}{t}{} \gensymb{-.5ex}{\ccirc}{f_{\theta_0 k}'}(t) + \gensymb{0ex}{\ccirc}{t}{}\ \gensymb{0ex}{\scriptstyle{s}}{t}{} f''_{\theta_0 k}(t) + \gensymb{0ex}{\ccirc\hspace{-0.15ex}\scriptstyle{s}}{t}{} f'_{\theta_0 k}(t)$ & $\gensymb{0ex}{\scriptstyle{\ccirc}\hspace{-0.1ex}\scriptstyle{s}}{t}{} \equiv \gensymb{0ex}{\scriptstyle{\ccirc}\hspace{-0.1ex}\scriptstyle{s}}{t}{}(\bm{u}) = \dot{\phi}_{0}'(C_s(\bm{u}_s)) \dot{C}_s(\bm{u}_s)$\\

$\Pkrs{k}{s}{s}(\bm{u})$ & $\Big(\gensymb{0ex}{\scriptstyle{s}}{t}{}\Big)^2 f''_{\theta_0 k}(t) + \gensymb{0ex}{\scriptstyle{s}\hspace{-0.1ex}\scriptstyle{s}}{t}{} f'_{\theta_0 k}(t)$ & $\gensymb{0ex}{\scriptstyle{s}\hspace{-0.1ex}\scriptstyle{s}}{t}{} \equiv \gensymb{0ex}{\scriptstyle{s}\hspace{-0.1ex}\scriptstyle{s}}{t}{}(\bm{u}) = \phi_{0}''(C_s(\bm{u}_s)) \{\dot{C}_s(\bm{u}_s)\}^2 + \phi_{0}'(C_s(\bm{u}_s)) \ddot{C}_s(\bm{u}_s)$\\

$\Pkrs{k}{s}{m}(\bm{u})$ & $\gensymb{0ex}{\scriptstyle{s}}{t}{}\ \gensymb{0ex}{\scriptstyle{m}}{t}{} f''_{\theta_0 k}(t)$\\
\hline
\end{tabular}
\end{table}

\clearpage
\newpage
\normalsize
\section{Example: trivariate Clayton HAC} \label{app:Clayton}

To demonstrate the relatively weak nature of the non-singularity of the Fisher information matrix and the Lipschitz condition in Assumptions~\ref{ass:stochastic}, consider the trivariate Clayton HAC model given by 
$$
    C_{\bm\theta}(\bm{u}) = C_{\theta_0}(C_{\theta_1}(u_{1}, u_{2}), u_{3}) = \Big\{ (u_{11}^{-\theta_1} + u_{12}^{-\theta_1} - 1 )^{\theta_0/\theta_1} + u_{21}^{-\theta_0} - 1\Big\}^{-1/\theta_0} , \quad \theta_0,\theta_1 \in \Theta = (0,\infty) ,
$$
which satisfies Assumptions~\ref{ass:basic-framework}.
The derivations which follows can be extended to arbitrary Clayton HACs, although the notation quickly becomes cumbersome.
To check the non-singularity of the Fisher information, we computed the latter's determinant, using the numerical method described in \S\ref{sec:sigma-estimation}, for each value of $\bm{\theta}$ on a fine grid covering $\{ (x_0,x_1) \in (0,2] : x_0 \leqslant x_1 \}$.
The results are depicted in Figure~\ref{fig:determinant}, which shows that the determinant is positive for all values of $\bm{\theta}$ on the grid, suggesting that this is likely to hold for the entire parameter space.
Furthermore, we note that as the copula density is infinitely differentiable, it is trivially differentiable in quadratic mean.

It only remains to verify the Lipschitz condition in Assumption B. 
We decompose the log-likelihood function into a sum of 5 functions, which individually satisfy the Lipschitz assumption. 
Specifically, we upper bound the absolute value of the difference of each function within a small neighborhood by a term of the kind $\lVert \bm\theta' - \bm\vartheta \Vert f(u)$, we then combine these upper bounds together through the triangle inequality and further demonstrate that these functions satisfy the square integrability condition.
The following holds for all $\bm\theta' $ such that $\lVert\bm\theta' - \bm\vartheta\lVert_2 < \epsilon(\vartheta)$ and for $\epsilon(\vartheta) = \min(\vartheta_1/2, \vartheta_0/2)$; these quantities only depend on the true data generating parameter $\bm\vartheta$. 
The copula density can be written as:
{\scriptsize
\begin{align*}
    c_{\bm{\theta}}( u_{11}, u_{12}, u_{21}) &= -\theta_1^2(u_{11}u_{12})^{-(1 + \theta_1)} \theta_0 u_{21}^{-(1 + \theta_0)} \left\{ \Big(1 + t_1(u_{11},u_{12})\Big)^{\frac{\theta_0}{\theta_1} - 2} s_{21}\left( \frac{\theta_0}{\theta_1} \right) \left( \frac{1}{\theta_0} + \frac{1}{\theta_0^2}\right) C_{\bm{\theta}}(u_{11}, u_{12}, u_{21})^{2\theta_0 + 1} \right.\\
    &- \left. \Big(1 + t_1(u_{11},u_{12})\Big)^{\frac{2\theta_0}{\theta_1} - 2} s_{22}\left( \frac{\theta_0}{\theta_1} \right) \left( \frac{1}{\theta_0} + \frac{1}{\theta_0^2}\right)\left( 2 + \frac{1}{\theta_0}\right) C_{\bm{\theta}}(u_{11}, u_{12}, u_{21})^{3\theta_0 + 1} \right\} \\
    &= \underbrace{ \theta_1^2(u_{11}u_{12})^{-(1 + \theta_1)} \theta_0 u_{21}^{-(1 + \theta_0)}}_{A(\bm\theta, \bm{u})} \underbrace{\left\{C_{\bm{\theta}}(u_{11}, u_{12}, u_{21})^{2\theta_0 + 1} \right\} }_{B(\bm\theta, \bm{u})} \underbrace{\left\{ \Big(1 + t_1(u_{11},u_{12})\Big)^{\frac{\theta_0}{\theta_1} - 2} \right\}}_{C(\bm\theta, \bm{u})} \\
    &\times\underbrace{  \left( \frac{1}{\theta_0} + \frac{1}{\theta_0^2}\right) }_{D(\bm\theta, \bm{u}
    )}  
    \times \underbrace{\left\{  s_{21}\left( \frac{\theta_0}{\theta_1} \right)  -  \Big(1 + t_1(u_{11},u_{12})\Big)^{\frac{\theta_0}{\theta_1}} s_{22}\left( \frac{\theta_0}{\theta_1} \right) \left( 2 + \frac{1}{\theta_0}\right) C_{\bm{\theta}}(u_{11}, u_{12}, u_{21})^{\theta_0} \right\}}_{E(\bm\theta, \bm{u})}
\end{align*}}
where, 
\begin{align*}
    C_{\bm{\theta}}(u_{11}, u_{12}, u_{21}) = \left\{ \left(u_{11}^{-\theta_1} + u_{12}^{-\theta_1} - 1 \right)^{\frac{\theta_0}{\theta_1}} + u_{21}^{-\theta_0} - 1\right\}^{\frac{-1}{\theta_0}},
\end{align*}
and $s_{11}(\cdot)$ and $s_{12}(\cdot)$, are polynomials whose form are given in Appendix A. 

\textbf{Controlling $\ln(A(\bm\theta, \bm{u}))$}. Consider:
\begin{align*}
    \left|\ln(A(\bm\theta', \bm{u})) - \ln(A(\bm\vartheta, \bm{u})) \right|&= \left| 2\{\ln(\theta_1') - \ln(\vartheta_1)\} + \{\ln(\theta_0') - \ln(\vartheta_0)\} \right.\\
    &\left.+(\vartheta_1 - \theta_1')\ln(u_{11}u_{12}) +(\vartheta_0 - \theta_0')\ln(u_{21}) \right|, 
\end{align*}
applying the triangle inequality, we see that the differences in logarithms:
\[ 2\left|\ln(\vartheta_1) - \ln(\theta_1') \right| \leq \frac{|\vartheta_1 - \theta'_1|}{\vartheta_1 - \epsilon_1}, \]
by a first-order Taylor expansion and any this is valid for any $0< \epsilon_1 \leq \theta_1/2$. As for
\[\left| (\vartheta_1 - \theta_1')\ln(u_{11}u_{12}) \right|\leq \left| (\vartheta_1 - \theta_1') \right|\ln(u_{11}u_{12}) \leq 2^{1/2} \lVert \bm\theta' - \bm\vartheta \rVert_2 \ln(u_{11}u_{12}) \]
The other differences are bounded by similar terms which only involves constants and logarithms of $u_{11}, u_{12}, u_{21}$.

\textbf{Controlling $\ln(B(\bm\theta, \bm{u}))$}. Consider the derivative: 

\begin{align*}
    \left|\frac{\partial}{\partial\theta}\ln(B(\bm\theta, \bm{u}))\right| &= \left|\frac{2}{\theta_0^2}\ln\{( u_{11}^{-\theta_1} + u_{12}^{-\theta_1} - 1)^{\theta_0/\theta_1} + u_{21}^{-\theta_0} - 1\}\right. \\
    &+ \left. \frac{1}{\theta_0} \left( 1 - \frac{2}{\theta_0} \right)\frac{ (u_{11}^{-\theta_1} + u_{12}^{-\theta_1} - 1)^{\theta_0/\theta_1}\ln(u_{11}^{-\theta_1} + u_{12}^{-\theta_1} - 1) + u_{21}^{-\theta_0}\ln(u_{21}^{-\theta_0}) }{(u_{11}^{-\theta_1} + u_{12}^{-\theta_1} - 1) + u_{21}^{-\theta_0} - 1}\right|\\
    &\leq \left|\frac{2}{\theta_0^2}\ln\{ u_{11}^{-\theta_1} + u_{12}^{-\theta_1}  + u_{21}^{-\theta_0} \}+ \frac{1}{\theta_0} \left( 1 + \frac{2}{\theta_0} \right) \ln(u_{11}^{-\theta_1} + u_{12}^{-\theta_1} ) +\ln(u_{21}^{-\theta_0}) \right|,
\end{align*}
as $\theta_0/\theta_1 \leq 1$, $\ln(\cdot)$ is a monotonic function and $0<x/(x+y) \leq 1$ for positive $x$ and $y$. While the partial derivative with respect to $\theta_1$:
\begin{align*}
    \left|\frac{\partial}{\partial\theta_1}\ln(B(\bm\theta, \bm{u}))\right| &\leq (1 + 2/\theta_0)\left\{ \frac{\theta_0}{\theta_1} (\ln(u_{11}^{-1}) + \ln(u_{12}^{-1}) + \frac{\theta_0}{\theta_1^2}\ln(u_{11}^{-\theta_1} + u_{12}^{-\theta_1})\right\},
\end{align*}
through similar calculations. These derivatives are bounded within the chosen radius around $\vartheta$, therefore these upper bounds may be used as the Lipchitz constant.

\textbf{Controlling $\ln(C(\bm\theta, \bm{u}))$}. Consider:
\begin{align*}
     &\left|\ln(C(\bm\theta', \bm{u})) - \ln(C(\bm\vartheta, \bm{u})) \right| \\
     &\leq \left| \frac{\theta_0'}{\theta_1'} - \frac{\vartheta_0}{\vartheta_1} \right| \ln\Big(1 + t^{\bm\vartheta}_1(u_{11},u_{12})\Big) + \left|\frac{\vartheta_0}{\vartheta_1} - 2\right|\left| \ln\Big(1 + t^{\vartheta}_1(u_{11},u_{12})\Big) - \ln\Big(1 + t^{\theta_0'}_1(u_{11},u_{12})\Big) \right| \\
     &\leq \frac{\vartheta_1|\vartheta_0 - \theta_0' |+ \vartheta_0|\vartheta_1 - \theta_1' |}{\vartheta_1(\vartheta_1 - \epsilon_1)} \ln(u_{11}^{-\vartheta_1} + u_{11}^{-\vartheta_1}) + \left|\theta_1' - \vartheta_1 \right|
 \left|\frac{\vartheta_0}{\vartheta_1} - 2\right|\left| \ln(u_{11}^{-1}) +\ln(u_{12}^{-1}) \right|\\
     &\leq \lVert \theta' - \vartheta \rVert_2 \left[\frac{\sqrt{2}\max(\vartheta_0, \vartheta_1) }{\vartheta_1(\vartheta_1 - \epsilon_1)} \ln\left(u_{11}^{-\vartheta_1} + u_{11}^{-\vartheta_1}\right)+ \sqrt{2}  \left|\frac{\vartheta_0}{\vartheta_1} - 2\right| \ln(u_{11}^{-1}) +\ln(u_{12}^{-1})  \right],
\end{align*}
where we have used the following first-order Taylor expansion, for every $u_{11}$ and $u_{12}$ fixed
\begin{align*}
     &\left|\ln\Big(1 + t^{\bm\vartheta}_1(u_{11},u_{12})\Big) - \ln\Big(1 + t^{\bm\vartheta}_1(u_{11},u_{12})\Big) \right|\\
     &= |\theta_1' - \vartheta_1 | \left\{ \frac{u_{12}^{\tilde\theta}\ln(u_{11}^{-1}) +u_{11}^{\tilde\theta}\ln(u_{12}^{-1})}{u_{11}^{\tilde\theta} + u_{12}^{\tilde\theta} - 1} \right\} \leq \ln(u_{11}^{-1}) + \ln(u_{12}^{-1}),
\end{align*}
where $\tilde\theta$ lies between $\theta_1'$ and $\vartheta_1$. Similar to the previous case, these derivatives are bounded within the chosen radius around $\vartheta$ and can therefore be used as the Lipchitz constant.

\textbf{Controlling $\ln(D(\bm\theta, \bm{u}))$}. Consider: 

\[\
\ln(D(\bm\theta, \bm{u})) = \ln\left(\frac{1}{\theta_1} + \frac{1}{\theta^2_1} \right).
\]
Using a first-order Taylor expansion we have for some $\tilde\theta$ between $\theta_1^\prime$ and $\vartheta_1$:

\begin{align*}
    \left| \ln\left(\frac{1}{\theta_1'} + \frac{1}{(\theta'_1)^2} \right) - \ln\left(\frac{1}{\vartheta_1} + \frac{1}{\vartheta_1^2} \right) \right| \leq \frac{1}{\tilde\theta} \frac{1 + 2\tilde\theta^{-2}}{ 1 + \tilde\theta^{-1} } |\theta_1 - \theta'_1| \leq \frac{|\vartheta_1 - \theta_1'|}{\vartheta_0 - \epsilon}.
\end{align*}

\textbf{Controlling $\ln(E(\bm\theta, \bm{u}))$}. 
\begin{align*}
    \ln(E(\bm\theta, \bm{u}))  &= \ln \left[-s_{11}\left(\frac{\theta_0}{\theta_1}\right)  +  \Big(1 + t_1(u_{11},u_{12})\Big)^{\frac{\theta_0}{\theta_1}} s_{22}\left( \frac{\theta_0}{\theta_1} \right) \left( 2 + \frac{1}{\theta_0}\right) C_{\bm{\theta}}(u_{11}, u_{12}, u_{21})^{\theta_0} \right]\\
      &= \ln\left[ \left( \frac{\theta_1}{\theta_0} - 1 \right) + \frac{ \left(u_{11}^{-\theta_1} + u_{12}^{-\theta_1} - 1 \right)^{\frac{\theta_0}{\theta_1}}  }{\left\{ \left(u_{11}^{-\theta_1} + u_{12}^{-\theta_1} - 1 \right)^{\frac{\theta_0}{\theta_1}} + u_{21}^{-\theta_0} - 1\right\}} \left( 2 + \frac{1}{\theta_0}\right) \right] + \ln\left(\frac{\theta_0^2}{\theta_1^2}\right)
\end{align*}
We show that the derivative of this function with respect to $\theta_0$ and $\theta_1$ is bounded within the chosen neighborhood of $\vartheta$. Let: 
\begin{align*}
    W(\bm\theta, \bm{u}) &:= \frac{ \left(u_{11}^{-\theta_1} + u_{12}^{-\theta_1} - 1 \right)^{\frac{\theta_0}{\theta_1}}  }{\left\{ \left(u_{11}^{-\theta_1} + u_{12}^{-\theta_1} - 1 \right)^{\frac{\theta_0}{\theta_1}} + u_{21}^{-\theta_0} - 1\right\}}, \\
    V(\bm\theta, \bm{u}) &:=\frac{u_{21}^{-\theta_0}  }{\left\{ \left(u_{11}^{-\theta_1} + u_{12}^{-\theta_1} - 1 \right)^{\frac{\theta_0}{\theta_1}} + u_{21}^{-\theta_0} - 1\right\}},\\
    Z(\bm\theta, \bm{u}) &= \frac{\theta_0}{\theta_1}\left\{ \frac{u_{11}^{-\theta_1}\ln(u_{11}^{-1})+u_{12}^{-\theta_1}\ln(u_{12}^{-1})}{u_{11}^{-\theta_1} +u_{12}^{-\theta_1} - 1} - \frac{\ln(u_{11}^{-\theta_1} +u_{12}^{-\theta_1} - 1)}{\theta_1} \right\}.
\end{align*}
Note $ W(\bm\theta, \bm{u})$ and $ W(\bm\theta, \bm{u})$ are positive and bounded by $1$. The partial derivatives are: 
{\small
\begin{align*}
 \left|\frac{\partial}{\partial\theta_0}  \ln(E(\bm\theta, \bm{u}))\right| &=\left| \left\{-\frac{1}{ \theta_0^2} W(\bm\theta, \bm{u}) +\frac{2 + \theta_0^{-1}}{\theta_1}W(\bm\theta, \bm{u})\ln(u_{11}^{-\theta_1}+u_{12}^{-\theta_1} -1 ) - \frac{\theta_1}{\theta_0^2}\right.\right.\\
 &\left.\left. - \left( 2 +\frac{1}{ \theta_0}\right)W(\bm\theta, \bm{u})\left\{ \theta_1^{-1}W(\bm\theta, \bm{u})\ln(u_{11}^{-\theta_1} +u_{11}^{-\theta_1} - 1 ) + V(\bm\theta, \bm{u})\ln(u_{21}^{-1}) \right\}  \right\}\right|\\
 &\times \left[ \left( \frac{\theta_1}{\theta_0} - 1 \right) + \frac{ \left(u_{11}^{-\theta_1} + u_{12}^{-\theta_1} - 1 \right)^{\frac{\theta_0}{\theta_1}}  }{\left\{ \left(u_{11}^{-\theta_1} + u_{12}^{-\theta_1} - 1 \right)^{\frac{\theta_0}{\theta_1}} + u_{21}^{-\theta_0} - 1\right\}} \left( 2 + \frac{1}{\theta_0}\right) \right]^{-1}\\
 &\leq \left\{\frac{1}{ \theta_0^2}  +\frac{2 + \theta_0^{-1}}{\theta_1}\ln(u_{11}^{-\theta_1}+u_{12}^{-\theta_1} -1 )  + \left(2+\frac{1}{ \theta_0}\right)\left\{ \theta_1^{-1}\ln(u_{11}^{-\theta_1} +u_{12}^{-\theta_1} - 1 ) + \ln(u_{21}^{-1}) \right\}  \right\},
\end{align*}}
and
 \begin{align*}
    \left| \frac{\partial}{\partial\theta_1}  \ln(E(\bm\theta, \bm{u})) \right|&= \left| \frac{2\theta_0 +1}{\theta_1 - \theta_0}  W(\bm\theta, \bm{u}) \left\{ \frac{ Z(\bm\theta, \bm{u}) -  W(\bm\theta, \bm{u}) Z(\bm\theta, \bm{u}) - 1}{1 + \frac{2\theta_0 +1}{\theta_1 - \theta_0} W(\bm\theta, \bm{u}) } \right\}\right|\\ 
     &\leq 2 \frac{2\theta_0 +1}{\theta_1 - \theta_0} |Z(\bm\theta, \bm{u})|\\
     &\leq \frac{4\theta_0 +2}{\theta_1 - \theta_0} \left\{ \ln(u^{-1}_{11}) + \ln(u^{-1}_{12}) + \theta_1^{-1}\ln(u^{-1}_{12} +u^{-1}_{12} )\right\}.
\end{align*}
Therefore taking $\epsilon < (\theta_1 - \theta_0)/2$ we get that the derivatives are bounded in a neighborhood of the true parameter, which we can use as the Lipszhit constant.

\textbf{Integrability of the envelope function.} It is sufficient to check that all of the following functions are integrable with respect to the true data generating density: $\ln(u_{11}^{-\theta_1} + u_{12}^{-\theta_1} )^2$, $\ln(u_{11}^{-\theta_1} + u_{12}^{-\theta_1} + u_{21}^{-\theta_0} )^2$, $ \ln(u_{11})^2, \ln(u_{12})^2, \ln(u_{21})^2$ (as $E[\ln(u_{11})]^2 < E\ln(u_{11})^2$ by Jensen's inequality) under the null of $\theta_0 = \theta_1$. Note that the marginal distribution of each individual $u_{11}, u_{12}, u_{21}$ are uniform $[0,1]$. Therefore: 
\[ \int_{0}^1 \ln(u_{21})^2 du_{21} < \infty.\]
 Since $\ln(u_{11}^{-\theta_1} + u_{12}^{-\theta_1}) \leq \ln(2u_{11}^{-\theta_1}) +\ln( 2u_{12}^{-\theta_1})$ on the unit square, and the fact that $x^2 + y^2 > 2xy$ for real numbers $x, y$ therefore:
\[\int_{0}^1\int_{0}^1 \ln(u_{11}^{-\theta_1}+u_{12}^{-\theta_1} )^2 dC_{\bm\theta} \leq 2\int_{0}^1\ln(2u_{11}^{-\theta_1})^2 du_{11} +2\int_{0}^1 \ln(2u_{12}^{-\theta_1})^2 du_{12} < \infty \]
for all $\theta_1 > 0$, and for $\theta_1 = \theta_0$. A similar result can be shown for the expectation of $\ln(u_{11}^{-\theta_1} + u_{12}^{-\theta_1} + u_{21}^{-\theta_0} )^2$.

\clearpage
\newpage
\section{Additional material for \S~\ref{sec:sigma-estimation}} \label{app:sigma-estimation}

\begin{figure}[h!]
    \centering
    \includegraphics[scale=.8]{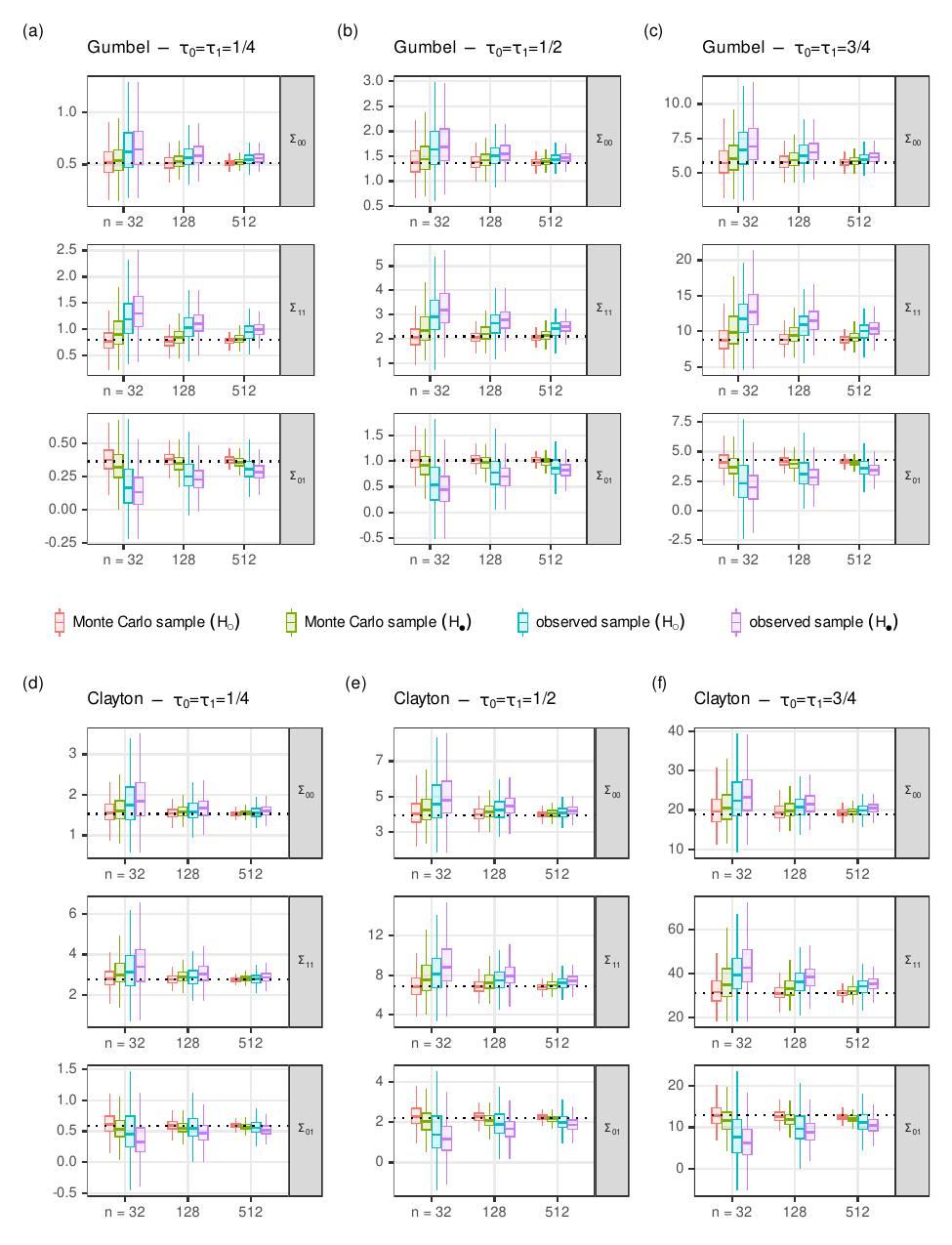}
    \caption{Boxplots comparing four estimators of $\Sigma_{00}$ based on the methodology of \S\ref{sec:sigma-estimation} for a Gumbel (a--c) and Clayton (d--f) HACs with structure $\{\{1,2\},3\}$ and three pairs of identical parameters: $(\vartheta_0, \vartheta_1) = (\vartheta, \vartheta)$ with $\vartheta \in \{\theta : \tau_{\psi}(\theta) = 1/4,1/2,3/4\}$; the null hypothesis is $H_\circ : \vartheta_0 = \vartheta_1$. The four estimators are obtained by considering either $\bm{\hat\theta}_{\circ}$ ($H_\circ$) or $\bm{\hat\theta}_{\bullet}$ ($H_\bullet$), and either the observed sample (of size $n$) or a Monte Carlo sample (of size $N = 10^5$) generated from $C_{\bm{\hat\theta}}$. The dashed lines indicate the true value of $\Sigma_{00}$, approximated using Monte Carlo sampling and the explicit formulas given in Appendix~\ref{app:two-level}}
    \label{fig:sigma-full}
\end{figure}

\clearpage
\newpage
\section{Results of the simulation study} \label{app:sim-study}

\begin{table}[h!]
\centering
\begin{tabular}{lrrrrrrrrr}
\multicolumn{10}{l}{\textbf{Simple hypothesis -- likelihood ratio test}}\\[1.25ex]
    & \multicolumn{3}{c}{Gumbel data} & \multicolumn{3}{c}{Clayton data} & \multicolumn{3}{c}{Frank data}\\[.5ex]
Model & n = 32 & 128 & 512 & n = 32 & 128 & 512 & n = 32 & 128 & 512 \\[.5ex]
  \hline\\[-1.5ex]
  Gumbel & 4 & 5 & 4 & 6 & 10 & 21 & 7 & 7 & 11 \\ 
  Clayton & 6 & 5 & 7 & 4 & 5 & 6 & 6 & 7 & 8 \\ 
  Frank & 6 & 5 & 6 & 8 & 10 & 13 & 6 & 4 & 5 \\[.5ex] 
\multicolumn{10}{l}{\textbf{Case a: $\tau_0 = \tau_1 = 1/4$}}\\[3ex] 
  Gumbel & 5 & 5 & 5 & 11 & 16 & 37 & 8 & 9 & 15 \\ 
  Clayton & 8 & 15 & 31 & 4 & 4 & 4 & 10 & 19 & 45 \\ 
  Frank & 5 & 5 & 7 & 7 & 10 & 20 & 5 & 6 & 5 \\[.5ex] 
\multicolumn{10}{l}{\textbf{Case b: $\tau_0 = \tau_1 = 1/2$}}\\[3ex] 
  Gumbel & 4 & 6 & 6 & 17 & 32 & 68 & 10 & 20 & 48 \\ 
  Clayton & 16 & 32 & 70 & 5 & 4 & 6 & 24 & 54 & 94 \\ 
  Frank & 6 & 8 & 13 & 10 & 16 & 32 & 4 & 4 & 4 \\[.5ex] 
\multicolumn{10}{l}{\textbf{Case c: $\tau_0 = \tau_1 = 3/4$}}\\[3ex] 
  Gumbel & 27 & 73 & 99 & 23 & 50 & 94 & 23 & 51 & 96 \\ 
  Clayton & 18 & 42 & 89 & 32 & 73 & 100 & 18 & 42 & 88 \\ 
  Frank & 24 & 61 & 98 & 27 & 65 & 99 & 26 & 62 & 99 \\[.5ex] 
\multicolumn{10}{l}{\textbf{Case d: $\tau_0 = 1/4\ <\ \tau_1 = 1/4 + 1/10$}}\\[3ex] 
  Gumbel & 47 & 93 & 100 & 33 & 74 & 99 & 34 & 77 & 100 \\ 
  Clayton & 36 & 72 & 99 & 53 & 95 & 100 & 33 & 71 & 99 \\ 
  Frank & 41 & 89 & 100 & 43 & 91 & 100 & 44 & 89 & 100 \\[.5ex] 
\multicolumn{10}{l}{\textbf{Case e: $\tau_0 = 1/2\ <\ \tau_1 = 1/2 + 1/10$}}\\[3ex] 
  Gumbel & 93 & 100 & 100 & 72 & 100 & 100 & 78 & 100 & 100 \\ 
  Clayton & 78 & 100 & 100 & 94 & 100 & 100 & 67 & 98 & 100 \\ 
  Frank & 90 & 100 & 100 & 90 & 100 & 100 & 92 & 100 & 100 \\[.5ex] 
\multicolumn{10}{l}{\textbf{Case f: $\tau_0 = 3/4\ <\ \tau_1 = 3/4 + 1/10$}} 
\end{tabular}
\caption{Size (a--c) and power (d--f) in finite samples of the (unconditional) test of $H_{\circ}:\vartheta_0=\vartheta_1$ based on Corollary~\ref{cor:simple}.
The tree structure is set to $\mathcal{G} = \{ \{ 1,2\}, 3\}$ (two parameters) and three generator families (Gumbel, Clayton and Frank) and various values of $\vartheta_0 = \tau_{\psi}^{-1}(\tau_0)$ and $\vartheta_1 = \tau_{\psi}^{-1}(\tau_1)$ (as described in Section~\ref{sec:power-local}) are considered.
The same three generator families are considered for modeling the data.
Rejection rates are reported in \%.} \label{tab:simple-uncond}
\end{table}
\pagebreak

\begin{table}[h!]
\centering
\begin{tabular}{lrrrrrrrrr}
\multicolumn{10}{l}{\textbf{Simple hypothesis -- conditional likelihood ratio test}}\\[1.25ex]
    & \multicolumn{3}{c}{Gumbel data} & \multicolumn{3}{c}{Clayton data} & \multicolumn{3}{c}{Frank data}\\[.5ex]
Model & n = 32 & 128 & 512 & n = 32 & 128 & 512 & n = 32 & 128 & 512 \\[.5ex]
  \hline\\[-1.5ex]
Gumbel & 2 & 3 & 2 & 4 & 6 & 14 & 3 & 4 & 6 \\ 
  Clayton & 3 & 2 & 4 & 2 & 3 & 3 & 4 & 4 & 5 \\ 
  Frank & 3 & 3 & 3 & 4 & 6 & 8 & 3 & 1 & 3 \\[.5ex] 
\multicolumn{10}{l}{\textbf{Case a: $\tau_0 = \tau_1 = 1/4$}}\\[3ex] 
  Gumbel & 2 & 2 & 3 & 7 & 11 & 28 & 5 & 5 & 11 \\ 
  Clayton & 4 & 9 & 23 & 2 & 2 & 2 & 6 & 13 & 35 \\ 
  Frank & 3 & 3 & 4 & 4 & 6 & 13 & 3 & 2 & 3 \\[.5ex] 
\multicolumn{10}{l}{\textbf{Case b: $\tau_0 = \tau_1 = 1/2$}}\\[3ex] 
  Gumbel & 2 & 2 & 3 & 11 & 26 & 60 & 7 & 15 & 40 \\ 
  Clayton & 11 & 26 & 62 & 2 & 2 & 3 & 18 & 47 & 91 \\ 
  Frank & 4 & 4 & 8 & 6 & 11 & 23 & 2 & 2 & 2 \\[.5ex] 
\multicolumn{10}{l}{\textbf{Case c: $\tau_0 = \tau_1 = 3/4$}}\\[3ex] 
  Gumbel & 20 & 62 & 99 & 16 & 41 & 91 & 15 & 38 & 92 \\ 
  Clayton & 12 & 32 & 81 & 21 & 62 & 99 & 12 & 33 & 82 \\ 
  Frank & 16 & 50 & 97 & 18 & 53 & 99 & 18 & 49 & 97 \\[.5ex] 
\multicolumn{10}{l}{\textbf{Case d: $\tau_0 = 1/4\ <\ \tau_1 = 1/4 + 1/10$}}\\[3ex] 
  Gumbel & 38 & 89 & 100 & 26 & 66 & 98 & 25 & 70 & 99 \\ 
  Clayton & 26 & 65 & 98 & 40 & 92 & 100 & 25 & 64 & 99 \\ 
  Frank & 32 & 82 & 100 & 32 & 85 & 100 & 32 & 83 & 100 \\[.5ex] 
\multicolumn{10}{l}{\textbf{Case e: $\tau_0 = 1/2\ <\ \tau_1 = 1/2 + 1/10$}}\\[3ex] 
  Gumbel & 88 & 100 & 100 & 65 & 99 & 100 & 72 & 100 & 100 \\ 
  Clayton & 70 & 99 & 100 & 90 & 100 & 100 & 60 & 97 & 100 \\ 
  Frank & 85 & 100 & 100 & 85 & 100 & 100 & 86 & 100 & 100 \\[.5ex] 
\multicolumn{10}{l}{\textbf{Case f: $\tau_0 = 3/4\ <\ \tau_1 = 3/4 + 1/10$}} 
\end{tabular}
\caption{Size (a--c) and power (d--f) in finite samples of the conditional test of $H_{\circ}:\vartheta_0=\vartheta_1$ based on Remark~\ref{rem:cond}.
The tree structure is set to $\mathcal{G} = \{ \{ 1,2\}, 3\}$ (two parameters) and three generator families (Gumbel, Clayton and Frank) and various values of $\vartheta_0 = \tau_{\psi}^{-1}(\tau_0)$ and $\vartheta_1 = \tau_{\psi}^{-1}(\tau_1)$ (as described in Section~\ref{sec:power-local}) are considered.
The same three generator families are considered for modeling the data.
Rejection rates are reported in \%.} \label{tab:simple-cond}
\end{table}
\pagebreak

\begin{table}[h!]
\small\centering
\begin{tabular}{lrrrrrrrrr}
\multicolumn{10}{l}{\textbf{Intersection hypothesis -- likelihood ratio test}}\\[.25ex]
\multicolumn{10}{l}{$\bm{\hat\Sigma}$ computed at $\bm{\hat\theta_\circ}$ using Monte Carlo sampling ($10^5$ replicates)}\\[1.25ex]
    & \multicolumn{3}{c}{Gumbel data} & \multicolumn{3}{c}{Clayton data} & \multicolumn{3}{c}{Frank data}\\[.5ex]
Model & n = 32 & 128 & 512 & n = 32 & 128 & 512 & n = 32 & 128 & 512 \\[.5ex]
  \hline\\[-1.5ex] 
Gumbel & 6 & 5 & 4 & 9 & 13 & 26 & 5 & 6 & 10 \\ 
  Clayton & 7 & 6 & 7 & 6 & 5 & 4 & 7 & 7 & 8 \\ 
  Frank & 6 & 5 & 4 & 7 & 10 & 22 & 4 & 5 & 4 \\[.2ex] 
\multicolumn{10}{l}{\textbf{Case a: $\tau_0 = \tau_1 = \tau_2 = 1/4$}}\\[1.2ex]  
  Gumbel & 5 & 4 & 5 & 13 & 22 & 52 & 8 & 10 & 22 \\ 
  Clayton & 10 & 21 & 53 & 5 & 5 & 5 & 15 & 37 & 82 \\ 
  Frank & 7 & 6 & 9 & 9 & 11 & 28 & 4 & 4 & 5 \\[.2ex] 
\multicolumn{10}{l}{\textbf{Case b: $\tau_0 = \tau_1 = \tau_2 = 1/2$}}\\[1.2ex]  
  Gumbel & 4 & 5 & 4 & 27 & 55 & 95 & 16 & 36 & 76 \\ 
  Clayton & 27 & 63 & 98 & 4 & 5 & 6 & 51 & 94 & 100 \\ 
  Frank & 6 & 9 & 16 & 12 & 25 & 52 & 5 & 4 & 4 \\[.2ex] 
\multicolumn{10}{l}{\textbf{Case c: $\tau_0 = \tau_1 = \tau_2 = 3/4$}}\\[1.2ex]  
  Gumbel & 28 & 68 & 99 & 22 & 47 & 93 & 19 & 48 & 95 \\ 
  Clayton & 17 & 37 & 87 & 27 & 72 & 100 & 18 & 39 & 91 \\ 
  Frank & 21 & 60 & 99 & 25 & 65 & 100 & 24 & 56 & 98 \\[.2ex] 
\multicolumn{10}{l}{\textbf{Case d: $\tau_0 = \tau_1 = 1/4\ <\ \tau_2 = 1/4 + 1/10$}}\\[1.2ex]  
  Gumbel & 40 & 90 & 100 & 33 & 77 & 100 & 32 & 75 & 100 \\ 
  Clayton & 37 & 79 & 100 & 44 & 95 & 100 & 40 & 84 & 100 \\ 
  Frank & 40 & 87 & 100 & 42 & 93 & 100 & 37 & 90 & 100 \\[.2ex] 
\multicolumn{10}{l}{\textbf{Case e: $\tau_0 = \tau_1 = 1/2\ <\ \tau_2 = 1/2 + 1/10$}}\\[1.2ex]  
  Gumbel & 94 & 100 & 100 & 75 & 100 & 100 & 79 & 100 & 100 \\ 
  Clayton & 86 & 100 & 100 & 94 & 100 & 100 & 83 & 100 & 100 \\ 
  Frank & 90 & 100 & 100 & 91 & 100 & 100 & 91 & 100 & 100 \\[.2ex] 
\multicolumn{10}{l}{\textbf{Case f: $\tau_0 = \tau_1 = 3/4\ <\ \tau_2 = 3/4 + 1/10$}}\\[1.2ex]  
  Gumbel & 45 & 91 & 100 & 29 & 70 & 99 & 32 & 73 & 100 \\ 
  Clayton & 26 & 62 & 99 & 47 & 94 & 100 & 29 & 67 & 100 \\ 
  Frank & 39 & 86 & 100 & 40 & 87 & 100 & 37 & 84 & 100 \\[.2ex] 
\multicolumn{10}{l}{\textbf{Case g: $\tau_0 = 1/4 \ <\ \tau_1 = \tau_2 = 1/4 + 1/10$}}\\[1.2ex]  
  Gumbel & 71 & 100 & 100 & 52 & 93 & 100 & 55 & 95 & 100 \\ 
  Clayton & 56 & 98 & 100 & 75 & 100 & 100 & 59 & 98 & 100 \\ 
  Frank & 65 & 99 & 100 & 63 & 100 & 100 & 66 & 99 & 100 \\[.2ex] 
\multicolumn{10}{l}{\textbf{Case h: $\tau_0 = 1/2 \ <\ \tau_1 = \tau_2 = 1/2 + 1/10$}}\\[1.2ex]  
  Gumbel & 100 & 100 & 100 & 94 & 100 & 100 & 94 & 100 & 100 \\ 
  Clayton & 97 & 100 & 100 & 100 & 100 & 100 & 97 & 100 & 100 \\ 
  Frank & 99 & 100 & 100 & 99 & 100 & 100 & 99 & 100 & 100 \\[.2ex] 
\multicolumn{10}{l}{\textbf{Case i: $\tau_0 = 3/4 \ <\ \tau_1 = \tau_2 = 3/4 + 1/10$}} 
\end{tabular}
\caption{Size (a--c) and power (d--i) in finite samples of the (unconditional) test of $H_{\circ}:\vartheta_0=\vartheta_1=\vartheta_2$ based on Corollary~\ref{cor:simple-2}, using the estimator of $\bm{\Sigma}$ computed at the null parameter ($\bm{\hat\theta}_{\circ}$) based on a Monte Carlo sample ($10^5$ replicates) from the null model.
The tree structure is set to $\mathcal{G} = \{ \{ 1,2\}, \{3,4\}\}$ (three parameters) and three generator families (Gumbel, Clayton and Frank) and various values of $\vartheta_i = \tau_{\psi}^{-1}(\tau_i)$ ($i=0,1,2$; as described in Section~\ref{sec:power-local}) are considered.
The same three generator families are considered for modeling the data.
Rejection rates are reported in \%.} \label{tab:int-0-mc}
\end{table}
\pagebreak

\begin{table}[h!]
\small\centering
\begin{tabular}{lrrrrrrrrr}
\multicolumn{10}{l}{\textbf{Intersection hypothesis -- likelihood ratio test}}\\[.25ex]
\multicolumn{10}{l}{$\bm{\hat\Sigma}$ computed at $\bm{\hat\theta_\bullet}$ using Monte Carlo sampling ($10^5$ replicates)}\\[1.25ex]
    & \multicolumn{3}{c}{Gumbel data} & \multicolumn{3}{c}{Clayton data} & \multicolumn{3}{c}{Frank data}\\[.5ex]
Model & n = 32 & 128 & 512 & n = 32 & 128 & 512 & n = 32 & 128 & 512 \\[.5ex]
  \hline\\[-1.5ex] 
Gumbel & 9 & 9 & 7 & 13 & 18 & 33 & 9 & 9 & 14 \\ 
  Clayton & 10 & 11 & 11 & 9 & 7 & 8 & 11 & 10 & 12 \\ 
  Frank & 9 & 9 & 8 & 11 & 16 & 31 & 7 & 8 & 8 \\[.2ex] 
\multicolumn{10}{l}{\textbf{Case a: $\tau_0 = \tau_1 = \tau_2 = 1/4$}}\\[1.2ex] 
  Gumbel & 8 & 7 & 9 & 20 & 30 & 62 & 12 & 15 & 30 \\ 
  Clayton & 16 & 31 & 63 & 7 & 8 & 9 & 20 & 46 & 88 \\ 
  Frank & 10 & 10 & 13 & 13 & 19 & 38 & 8 & 7 & 9 \\[.2ex] 
\multicolumn{10}{l}{\textbf{Case b: $\tau_0 = \tau_1 = \tau_2 = 1/2$}}\\[1.2ex] 
  Gumbel & 6 & 8 & 7 & 33 & 61 & 96 & 22 & 44 & 82 \\ 
  Clayton & 35 & 70 & 99 & 8 & 8 & 8 & 57 & 96 & 100 \\ 
  Frank & 11 & 15 & 24 & 19 & 33 & 63 & 8 & 7 & 6 \\[.2ex] 
\multicolumn{10}{l}{\textbf{Case c: $\tau_0 = \tau_1 = \tau_2 = 3/4$}}\\[1.2ex] 
  Gumbel & 36 & 75 & 100 & 28 & 56 & 95 & 26 & 57 & 97 \\ 
  Clayton & 24 & 47 & 92 & 34 & 80 & 100 & 24 & 49 & 95 \\ 
  Frank & 26 & 68 & 100 & 33 & 74 & 100 & 30 & 66 & 99 \\[.2ex] 
\multicolumn{10}{l}{\textbf{Case d: $\tau_0 = \tau_1 = 1/4\ <\ \tau_2 = 1/4 + 1/10$}}\\[1.2ex]  
  Gumbel & 48 & 94 & 100 & 42 & 84 & 100 & 42 & 82 & 100 \\ 
  Clayton & 44 & 85 & 100 & 52 & 97 & 100 & 48 & 90 & 100 \\ 
  Frank & 50 & 92 & 100 & 50 & 95 & 100 & 46 & 93 & 100 \\[.2ex] 
\multicolumn{10}{l}{\textbf{Case e: $\tau_0 = \tau_1 = 1/2\ <\ \tau_2 = 1/2 + 1/10$}}\\[1.2ex]  
  Gumbel & 96 & 100 & 100 & 83 & 100 & 100 & 84 & 100 & 100 \\ 
  Clayton & 88 & 100 & 100 & 97 & 100 & 100 & 88 & 100 & 100 \\[.2ex] 
  Frank & 93 & 100 & 100 & 94 & 100 & 100 & 94 & 100 & 100 \\ 
\multicolumn{10}{l}{\textbf{Case f: $\tau_0 = \tau_1 = 3/4\ <\ \tau_2 = 3/4 + 1/10$}}\\[1.2ex]  
  Gumbel & 53 & 93 & 100 & 39 & 76 & 100 & 41 & 80 & 100 \\ 
  Clayton & 35 & 72 & 100 & 56 & 96 & 100 & 36 & 75 & 100 \\ 
  Frank & 48 & 91 & 100 & 50 & 91 & 100 & 46 & 90 & 100 \\[.2ex] 
\multicolumn{10}{l}{\textbf{Case g: $\tau_0 = 1/4 \ <\ \tau_1 = \tau_2 = 1/4 + 1/10$}}\\[1.2ex]  
  Gumbel & 78 & 100 & 100 & 61 & 96 & 100 & 63 & 97 & 100 \\ 
  Clayton & 64 & 98 & 100 & 80 & 100 & 100 & 68 & 99 & 100 \\ 
  Frank & 72 & 99 & 100 & 72 & 100 & 100 & 74 & 100 & 100 \\[.2ex] 
\multicolumn{10}{l}{\textbf{Case h: $\tau_0 = 1/2 \ <\ \tau_1 = \tau_2 = 1/2 + 1/10$}}\\[1.2ex]  
  Gumbel & 100 & 100 & 100 & 96 & 100 & 100 & 96 & 100 & 100 \\ 
  Clayton & 98 & 100 & 100 & 100 & 100 & 100 & 98 & 100 & 100 \\ 
  Frank & 99 & 100 & 100 & 99 & 100 & 100 & 99 & 100 & 100 \\[.2ex] 
\multicolumn{10}{l}{\textbf{Case i: $\tau_0 = 3/4 \ <\ \tau_1 = \tau_2 = 3/4 + 1/10$}}
\end{tabular}
\caption{Size (a--c) and power (d--i) in finite samples of the (unconditional) test of $H_{\circ}:\vartheta_0=\vartheta_1=\vartheta_2$ based on Corollary~\ref{cor:simple-2}, using the estimator of $\bm{\Sigma}$ computed at the alternative parameter ($\bm{\hat\theta}_{\bullet}$) based on a Monte Carlo sample ($10^5$ replicates) from the alternative model.
The tree structure is set to $\mathcal{G} = \{ \{ 1,2\}, \{3,4\}\}$ (three parameters) and three generator families (Gumbel, Clayton and Frank) and various values of $\vartheta_i = \tau_{\psi}^{-1}(\tau_i)$ ($i=0,1,2$; as described in Section~\ref{sec:power-local}) are considered.
The same three generator families are considered for modeling the data.
Rejection rates are reported in \%.} \label{tab:int-1-mc}
\end{table}
\pagebreak

\begin{table}[h!]
\small\centering
\begin{tabular}{lrrrrrrrrr}
\multicolumn{10}{l}{\textbf{Intersection hypothesis -- likelihood ratio test}}\\[.25ex]
\multicolumn{10}{l}{$\bm{\hat\Sigma}$ computed at $\bm{\hat\theta_\circ}$ based on the observed data}\\[1.25ex]
    & \multicolumn{3}{c}{Gumbel data} & \multicolumn{3}{c}{Clayton data} & \multicolumn{3}{c}{Frank data}\\[.5ex]
Model & n = 32 & 128 & 512 & n = 32 & 128 & 512 & n = 32 & 128 & 512 \\[.5ex]
  \hline\\[-1.5ex]
Gumbel & 6 & 5 & 4 & 9 & 13 & 26 & 5 & 6 & 10 \\ 
  Clayton & 7 & 7 & 7 & 6 & 5 & 4 & 8 & 7 & 8 \\ 
  Frank & 6 & 5 & 5 & 7 & 11 & 22 & 4 & 5 & 4 \\[.2ex] 
\multicolumn{10}{l}{\textbf{Case a: $\tau_0 = \tau_1 = \tau_2 = 1/4$}}\\[1.2ex] 
  Gumbel & 5 & 4 & 5 & 13 & 23 & 52 & 8 & 10 & 22 \\ 
  Clayton & 10 & 21 & 53 & 5 & 5 & 5 & 15 & 37 & 82 \\ 
  Frank & 7 & 6 & 9 & 9 & 11 & 28 & 5 & 4 & 5 \\[.2ex] 
\multicolumn{10}{l}{\textbf{Case b: $\tau_0 = \tau_1 = \tau_2 = 1/2$}}\\[1.2ex] 
  Gumbel & 4 & 5 & 4 & 27 & 55 & 95 & 16 & 36 & 76 \\ 
  Clayton & 27 & 63 & 98 & 4 & 5 & 6 & 50 & 94 & 100 \\ 
  Frank & 6 & 9 & 16 & 12 & 25 & 52 & 5 & 4 & 4 \\[.2ex] 
\multicolumn{10}{l}{\textbf{Case c: $\tau_0 = \tau_1 = \tau_2 = 3/4$}}\\[1.2ex] 
  Gumbel & 28 & 68 & 99 & 22 & 48 & 93 & 19 & 48 & 95 \\ 
  Clayton & 18 & 38 & 87 & 27 & 72 & 100 & 18 & 40 & 91 \\ 
  Frank & 21 & 60 & 99 & 25 & 65 & 100 & 24 & 57 & 99 \\[.2ex] 
\multicolumn{10}{l}{\textbf{Case d: $\tau_0 = \tau_1 = 1/4\ <\ \tau_2 = 1/4 + 1/10$}}\\[1.2ex]  
  Gumbel & 41 & 90 & 100 & 33 & 77 & 100 & 32 & 75 & 100 \\ 
  Clayton & 37 & 79 & 100 & 45 & 95 & 100 & 40 & 84 & 100 \\ 
  Frank & 40 & 88 & 100 & 42 & 93 & 100 & 37 & 90 & 100 \\[.2ex] 
\multicolumn{10}{l}{\textbf{Case e: $\tau_0 = \tau_1 = 1/2\ <\ \tau_2 = 1/2 + 1/10$}}\\[1.2ex]  
  Gumbel & 94 & 100 & 100 & 75 & 100 & 100 & 79 & 100 & 100 \\ 
  Clayton & 86 & 100 & 100 & 94 & 100 & 100 & 83 & 100 & 100 \\ 
  Frank & 90 & 100 & 100 & 91 & 100 & 100 & 92 & 100 & 100 \\[.2ex] 
\multicolumn{10}{l}{\textbf{Case f: $\tau_0 = \tau_1 = 3/4\ <\ \tau_2 = 3/4 + 1/10$}}\\[1.2ex]  
  Gumbel & 45 & 91 & 100 & 29 & 70 & 99 & 32 & 74 & 100 \\ 
  Clayton & 27 & 63 & 99 & 48 & 94 & 100 & 29 & 68 & 100 \\ 
  Frank & 40 & 86 & 100 & 40 & 87 & 100 & 37 & 84 & 100 \\[.2ex] 
\multicolumn{10}{l}{\textbf{Case g: $\tau_0 = 1/4 \ <\ \tau_1 = \tau_2 = 1/4 + 1/10$}}\\[1.2ex]  
  Gumbel & 72 & 100 & 100 & 52 & 93 & 100 & 56 & 95 & 100 \\ 
  Clayton & 57 & 98 & 100 & 75 & 100 & 100 & 60 & 98 & 100 \\ 
  Frank & 65 & 99 & 100 & 64 & 100 & 100 & 66 & 99 & 100 \\[.2ex] 
\multicolumn{10}{l}{\textbf{Case h: $\tau_0 = 1/2 \ <\ \tau_1 = \tau_2 = 1/2 + 1/10$}}\\[1.2ex]  
  Gumbel & 100 & 100 & 100 & 94 & 100 & 100 & 94 & 100 & 100 \\ 
  Clayton & 97 & 100 & 100 & 100 & 100 & 100 & 97 & 100 & 100 \\ 
  Frank & 99 & 100 & 100 & 99 & 100 & 100 & 99 & 100 & 100 \\[.2ex] 
\multicolumn{10}{l}{\textbf{Case i: $\tau_0 = 3/4 \ <\ \tau_1 = \tau_2 = 3/4 + 1/10$}}
\end{tabular}
\caption{Size (a--c) and power (d--i) in finite samples of the (unconditional) test of $H_{\circ}:\vartheta_0=\vartheta_1=\vartheta_2$ based on Corollary~\ref{cor:simple-2}, using the estimator of $\bm{\Sigma}$ computed at the null parameter ($\bm{\hat\theta}_{\circ}$) based on the observed sample.
The tree structure is set to $\mathcal{G} = \{ \{ 1,2\}, \{3,4\}\}$ (three parameters) and three generator families (Gumbel, Clayton and Frank) and various values of $\vartheta_i = \tau_{\psi}^{-1}(\tau_i)$ ($i=0,1,2$; as described in Section~\ref{sec:power-local}) are considered.
The same three generator families are considered for modeling the data.
Rejection rates are reported in \%.} \label{tab:int-0-obs}
\end{table}
\pagebreak

\begin{table}[h!]
\small\centering
\begin{tabular}{lrrrrrrrrr}
\multicolumn{10}{l}{\textbf{Intersection hypothesis -- likelihood ratio test}}\\[.25ex]
\multicolumn{10}{l}{$\bm{\hat\Sigma}$ computed at $\bm{\hat\theta_\bullet}$ based on the observed data}\\[1.25ex]
    & \multicolumn{3}{c}{Gumbel data} & \multicolumn{3}{c}{Clayton data} & \multicolumn{3}{c}{Frank data}\\[.5ex]
Model & n = 32 & 128 & 512 & n = 32 & 128 & 512 & n = 32 & 128 & 512 \\[.5ex]
  \hline\\[-1.5ex] 
  Gumbel & 9 & 9 & 7 & 12 & 19 & 34 & 8 & 9 & 14 \\ 
  Clayton & 10 & 11 & 11 & 10 & 7 & 8 & 11 & 10 & 12 \\ 
  Frank & 9 & 9 & 8 & 10 & 16 & 30 & 7 & 8 & 8 \\[.2ex] 
\multicolumn{10}{l}{\textbf{Case a: $\tau_0 = \tau_1 = \tau_2 = 1/4$}}\\[1.2ex] 
  Gumbel & 8 & 7 & 8 & 20 & 29 & 62 & 12 & 15 & 29 \\ 
  Clayton & 16 & 31 & 64 & 7 & 8 & 9 & 21 & 46 & 88 \\ 
  Frank & 10 & 10 & 13 & 12 & 18 & 38 & 8 & 7 & 8 \\[.2ex] 
\multicolumn{10}{l}{\textbf{Case b: $\tau_0 = \tau_1 = \tau_2 = 1/2$}}\\[1.2ex] 
  Gumbel & 6 & 9 & 7 & 33 & 61 & 96 & 22 & 43 & 82 \\ 
  Clayton & 35 & 70 & 99 & 7 & 8 & 8 & 58 & 96 & 100 \\ 
  Frank & 11 & 15 & 24 & 18 & 33 & 63 & 9 & 7 & 6 \\[.2ex] 
\multicolumn{10}{l}{\textbf{Case c: $\tau_0 = \tau_1 = \tau_2 = 3/4$}}\\[1.2ex] 
  Gumbel & 36 & 74 & 100 & 28 & 56 & 95 & 25 & 56 & 97 \\ 
  Clayton & 24 & 47 & 92 & 34 & 79 & 100 & 23 & 50 & 95 \\ 
  Frank & 26 & 68 & 100 & 33 & 74 & 100 & 30 & 66 & 99 \\[.2ex] 
\multicolumn{10}{l}{\textbf{Case d: $\tau_0 = \tau_1 = 1/4\ <\ \tau_2 = 1/4 + 1/10$}}\\[1.2ex]  
  Gumbel & 48 & 94 & 100 & 41 & 83 & 100 & 41 & 82 & 100 \\ 
  Clayton & 44 & 85 & 100 & 52 & 97 & 100 & 49 & 90 & 100 \\ 
  Frank & 49 & 92 & 100 & 50 & 95 & 100 & 46 & 94 & 100 \\[.2ex] 
\multicolumn{10}{l}{\textbf{Case e: $\tau_0 = \tau_1 = 1/2\ <\ \tau_2 = 1/2 + 1/10$}}\\[1.2ex]  
  Gumbel & 96 & 100 & 100 & 82 & 100 & 100 & 83 & 100 & 100 \\ 
  Clayton & 88 & 100 & 100 & 97 & 100 & 100 & 88 & 100 & 100 \\ 
  Frank & 93 & 100 & 100 & 94 & 100 & 100 & 94 & 100 & 100 \\[.2ex] 
\multicolumn{10}{l}{\textbf{Case f: $\tau_0 = \tau_1 = 3/4\ <\ \tau_2 = 3/4 + 1/10$}}\\[1.2ex]  
  Gumbel & 53 & 93 & 100 & 38 & 76 & 100 & 40 & 80 & 100 \\ 
  Clayton & 36 & 72 & 100 & 56 & 96 & 100 & 37 & 75 & 100 \\ 
  Frank & 48 & 91 & 100 & 49 & 91 & 100 & 47 & 90 & 100 \\[.2ex] 
\multicolumn{10}{l}{\textbf{Case g: $\tau_0 = 1/4 \ <\ \tau_1 = \tau_2 = 1/4 + 1/10$}}\\[1.2ex]  
  Gumbel & 78 & 100 & 100 & 60 & 96 & 100 & 62 & 97 & 100 \\ 
  Clayton & 64 & 98 & 100 & 81 & 100 & 100 & 68 & 99 & 100 \\ 
  Frank & 73 & 99 & 100 & 71 & 100 & 100 & 74 & 100 & 100 \\[.2ex] 
\multicolumn{10}{l}{\textbf{Case h: $\tau_0 = 1/2 \ <\ \tau_1 = \tau_2 = 1/2 + 1/10$}}\\[1.2ex]  
  Gumbel & 100 & 100 & 100 & 96 & 100 & 100 & 96 & 100 & 100 \\ 
  Clayton & 98 & 100 & 100 & 100 & 100 & 100 & 98 & 100 & 100 \\ 
  Frank & 99 & 100 & 100 & 99 & 100 & 100 & 99 & 100 & 100 \\[.2ex] 
\multicolumn{10}{l}{\textbf{Case i: $\tau_0 = 3/4 \ <\ \tau_1 = \tau_2 = 3/4 + 1/10$}}
\end{tabular}
\caption{Size (a--c) and power (d--i) in finite samples of the (unconditional) test of $H_{\circ}:\vartheta_0=\vartheta_1=\vartheta_2$ based on Corollary~\ref{cor:simple-2}, using the estimator of $\bm{\Sigma}$ computed at the alternative parameter ($\bm{\hat\theta}_{\bullet}$) based on the observed sample.
The tree structure is set to $\mathcal{G} = \{ \{ 1,2\}, \{3,4\}\}$ (three parameters) and three generator families (Gumbel, Clayton and Frank) and various values of $\vartheta_i = \tau_{\psi}^{-1}(\tau_i)$ ($i=0,1,2$; as described in Section~\ref{sec:power-local}) are considered.
The same three generator families are considered for modeling the data.
Rejection rates are reported in \%.} \label{tab:int-1-obs}
\end{table}
\pagebreak

\begin{table}[h!]
\small\centering
\begin{tabular}{lrrrrrrrrr}
\multicolumn{10}{l}{\textbf{Union hypothesis -- likelihood ratio test}}\\[1.25ex]
    & \multicolumn{3}{c}{Gumbel data} & \multicolumn{3}{c}{Clayton data} & \multicolumn{3}{c}{Frank data}\\[.5ex]
Model & n = 32 & 128 & 512 & n = 32 & 128 & 512 & n = 32 & 128 & 512 \\[.5ex]
  \hline\\[-1.5ex] 
Gumbel & 1 & 0 & 0 & 0 & 1 & 3 & 0 & 0 & 0 \\ 
  Clayton & 0 & 0 & 0 & 0 & 0 & 0 & 0 & 0 & 1 \\ 
  Frank & 0 & 0 & 0 & 1 & 1 & 4 & 0 & 0 & 0 \\[.2ex] 
\multicolumn{10}{l}{\textbf{Case a: $\tau_0 = \tau_1 = \tau_2 = 1/4$}}\\[1.2ex] 
  Gumbel & 0 & 0 & 0 & 1 & 3 & 14 & 0 & 1 & 3 \\ 
  Clayton & 0 & 0 & 8 & 0 & 0 & 0 & 1 & 2 & 24 \\ 
  Frank & 0 & 0 & 1 & 0 & 1 & 3 & 0 & 0 & 1 \\[.2ex] 
\multicolumn{10}{l}{\textbf{Case b: $\tau_0 = \tau_1 = \tau_2 = 1/2$}}\\[1.2ex] 
  Gumbel & 0 & 0 & 0 & 2 & 15 & 63 & 1 & 6 & 33 \\ 
  Clayton & 2 & 12 & 70 & 0 & 0 & 0 & 3 & 45 & 99 \\ 
  Frank & 0 & 0 & 2 & 1 & 3 & 14 & 0 & 0 & 0 \\[.2ex] 
\multicolumn{10}{l}{\textbf{Case c: $\tau_0 = \tau_1 = \tau_2 = 3/4$}}\\[1.2ex] 
  Gumbel & 1 & 4 & 5 & 2 & 5 & 17 & 1 & 3 & 8 \\ 
  Clayton & 1 & 3 & 6 & 2 & 6 & 4 & 1 & 4 & 6 \\ 
  Frank & 1 & 4 & 6 & 2 & 8 & 14 & 2 & 4 & 6 \\[.2ex] 
\multicolumn{10}{l}{\textbf{Case d: $\tau_0 = \tau_1 = 1/4\ <\ \tau_2 = 1/4 + 1/10$}}\\[1.2ex]  
  Gumbel & 2 & 3 & 6 & 4 & 14 & 36 & 2 & 7 & 18 \\ 
  Clayton & 3 & 11 & 36 & 3 & 4 & 5 & 3 & 15 & 59 \\ 
  Frank & 2 & 6 & 6 & 4 & 13 & 20 & 3 & 4 & 4 \\[.2ex] 
\multicolumn{10}{l}{\textbf{Case e: $\tau_0 = \tau_1 = 1/2\ <\ \tau_2 = 1/2 + 1/10$}}\\[1.2ex]  
  Gumbel & 5 & 4 & 5 & 13 & 38 & 80 & 9 & 25 & 59 \\ 
  Clayton & 14 & 40 & 82 & 4 & 4 & 4 & 17 & 68 & 98 \\ 
  Frank & 6 & 9 & 13 & 10 & 18 & 37 & 4 & 4 & 5 \\[.2ex] 
\multicolumn{10}{l}{\textbf{Case f: $\tau_0 = \tau_1 = 3/4\ <\ \tau_2 = 3/4 + 1/10$}}\\[1.2ex]  
  Gumbel & 13 & 56 & 100 & 5 & 26 & 90 & 6 & 30 & 93 \\ 
  Clayton & 4 & 19 & 83 & 14 & 61 & 100 & 4 & 20 & 88 \\ 
  Frank & 8 & 45 & 99 & 10 & 49 & 100 & 8 & 45 & 99 \\[.2ex] 
\multicolumn{10}{l}{\textbf{Case g: $\tau_0 = 1/4 \ <\ \tau_1 = \tau_2 = 1/4 + 1/10$}}\\[1.2ex]  
  Gumbel & 28 & 90 & 100 & 16 & 61 & 100 & 15 & 68 & 100 \\ 
  Clayton & 13 & 61 & 100 & 32 & 94 & 100 & 10 & 55 & 100 \\ 
  Frank & 22 & 87 & 100 & 22 & 87 & 100 & 25 & 88 & 100 \\[.2ex] 
\multicolumn{10}{l}{\textbf{Case h: $\tau_0 = 1/2 \ <\ \tau_1 = \tau_2 = 1/2 + 1/10$}}\\[1.2ex]  
  Gumbel & 89 & 100 & 100 & 58 & 98 & 100 & 69 & 99 & 100 \\ 
  Clayton & 65 & 100 & 100 & 92 & 100 & 100 & 48 & 97 & 100 \\ 
  Frank & 84 & 100 & 100 & 84 & 100 & 100 & 89 & 100 & 100 \\[.2ex] 
\multicolumn{10}{l}{\textbf{Case i: $\tau_0 = 3/4 \ <\ \tau_1 = \tau_2 = 3/4 + 1/10$}}
\end{tabular}
\caption{Size (a--e) and power (f--i) in finite samples of the (unconditional) test of $H_{\circ}:\vartheta_0 \in \{\vartheta_1,\vartheta_2\}$ based on Corollary~\ref{cor:simple-3} (reference distribution: $W_1 \chi_1^2$ with $W_1 \sim  \mathcal{B}(1/2)$).
The tree structure is set to $\mathcal{G} = \{ \{ 1,2\}, \{3,4\}\}$ (three parameters) and three generator families (Gumbel, Clayton and Frank) and various values of $\vartheta_i = \tau_{\psi}^{-1}(\tau_i)$ ($i=0,1,2$; as described in Section~\ref{sec:power-local}) are considered.
The same three generator families are considered for modeling the data.
Rejection rates are reported in \%.} \label{tab:union}
\end{table}
\pagebreak

\begin{table}[h!]
\centering
\begin{tabular}{lrrrrrrrrr}
\multicolumn{10}{l}{\textbf{Simple hypothesis w/ nuisance -- likelihood ratio test} (assuming $\vartheta_0 = \vartheta_2$)}\\[.25ex]
\multicolumn{10}{l}{\textbf{Cases a--f}}\\[1.25ex]
    & \multicolumn{3}{c}{Gumbel data} & \multicolumn{3}{c}{Clayton data} & \multicolumn{3}{c}{Frank data}\\[.5ex]
Model & n = 32 & 128 & 512 & n = 32 & 128 & 512 & n = 32 & 128 & 512 \\[.5ex]
  \hline\\[-1.5ex] 
Gumbel & 6 & 5 & 3 & 8 & 10 & 22 & 7 & 7 & 10 \\ 
  Clayton & 6 & 6 & 7 & 5 & 6 & 6 & 6 & 8 & 10 \\ 
  Frank & 4 & 6 & 5 & 6 & 10 & 14 & 4 & 5 & 5 \\[.5ex] 
\multicolumn{10}{l}{\textbf{Case a: $\tau_0 = \tau_1 = \tau_2 = 1/4$}}\\[3ex] 
  Gumbel & 5 & 4 & 6 & 10 & 18 & 38 & 6 & 9 & 16 \\ 
  Clayton & 10 & 16 & 38 & 6 & 5 & 5 & 13 & 26 & 56 \\ 
  Frank & 5 & 6 & 9 & 8 & 11 & 20 & 5 & 5 & 5 \\[.5ex] 
\multicolumn{10}{l}{\textbf{Case b: $\tau_0 = \tau_1 = \tau_2 = 1/2$}}\\[3ex] 
  Gumbel & 4 & 4 & 4 & 19 & 38 & 77 & 13 & 25 & 59 \\ 
  Clayton & 22 & 44 & 83 & 6 & 4 & 6 & 33 & 70 & 99 \\ 
  Frank & 7 & 6 & 16 & 13 & 18 & 39 & 5 & 5 & 6 \\[.5ex] 
\multicolumn{10}{l}{\textbf{Case c: $\tau_0 = \tau_1 = \tau_2 = 3/4$}}\\[3ex] 
  Gumbel & 6 & 5 & 4 & 8 & 10 & 18 & 5 & 7 & 7 \\ 
  Clayton & 6 & 5 & 9 & 5 & 6 & 6 & 6 & 7 & 10 \\ 
  Frank & 5 & 4 & 6 & 6 & 8 & 14 & 6 & 6 & 5 \\[.5ex] 
\multicolumn{10}{l}{\textbf{Case d: $\tau_0 = \tau_1 = 1/4 \ <\ \tau_2 = 1/4 + 1/10$}}\\[3ex] 
  Gumbel & 5 & 6 & 6 & 12 & 17 & 40 & 9 & 11 & 17 \\ 
  Clayton & 10 & 20 & 43 & 6 & 7 & 10 & 12 & 29 & 65 \\ 
  Frank & 5 & 6 & 10 & 8 & 12 & 22 & 5 & 6 & 6 \\[.5ex] 
\multicolumn{10}{l}{\textbf{Case e: $\tau_0 = \tau_1 = 1/2 \ <\ \tau_2 = 1/2 + 1/10$}}\\[3ex] 
  Gumbel & 6 & 8 & 8 & 19 & 41 & 80 & 17 & 26 & 61 \\ 
  Clayton & 20 & 46 & 86 & 6 & 7 & 10 & 34 & 71 & 99 \\ 
  Frank & 7 & 12 & 20 & 13 & 19 & 46 & 5 & 6 & 10 \\[.5ex] 
\multicolumn{10}{l}{\textbf{Case f: $\tau_0 = \tau_1 = 3/4 \ <\ \tau_2 = 3/4 + 1/10$}}
\end{tabular}
\caption{Size (power in Figure~\ref{tab:nuisance-simplified-2}) in finite samples of the (unconditional) test of $H_{\circ}:\vartheta_0 = \vartheta_1$ based on Corollary~\ref{cor:simple}.
The tree structure is set to $\mathcal{G} = \{ \{ 1,2\}, \{3,4\}\}$ (three parameters, $\vartheta_2$ is nuisance) and three generator families (Gumbel, Clayton and Frank) and various values of $\vartheta_i = \tau_{\psi}^{-1}(\tau_i)$ ($i=0,1,2$; as described in Section~\ref{sec:power-local}) are considered.
During the testing procedure, it is assumed that $\vartheta_2 = \vartheta_0$.
The same three generator families are considered for modeling the data.
Rejection rates are reported in \%.} \label{tab:nuisance-simplified-1}
\end{table}
\pagebreak

\begin{table}[h!]
\centering
\begin{tabular}{lrrrrrrrrr}
\multicolumn{10}{l}{\textbf{Simple hypothesis w/ nuisance -- likelihood ratio test} (assuming $\vartheta_0 = \vartheta_2$)}\\[.25ex]
\multicolumn{10}{l}{\textbf{Cases k--l}}\\[1.25ex]
    & \multicolumn{3}{c}{Gumbel data} & \multicolumn{3}{c}{Clayton data} & \multicolumn{3}{c}{Frank data}\\[.5ex]
Model & n = 32 & 128 & 512 & n = 32 & 128 & 512 & n = 32 & 128 & 512 \\[.5ex]
  \hline\\[-1.5ex] 
  Gumbel & 33 & 74 & 100 & 26 & 55 & 96 & 23 & 57 & 96 \\ 
  Clayton & 21 & 45 & 92 & 36 & 80 & 100 & 20 & 51 & 94 \\ 
  Frank & 28 & 67 & 100 & 30 & 72 & 100 & 26 & 69 & 100 \\[.5ex] 
\multicolumn{10}{l}{\textbf{Case g: $\tau_0 = \tau_2 = 1/4 \ <\ \tau_1 = 1/4 + 1/10$}}\\[3ex] 
  Gumbel & 55 & 95 & 100 & 39 & 78 & 100 & 39 & 83 & 100 \\ 
  Clayton & 43 & 82 & 100 & 59 & 98 & 100 & 40 & 80 & 100 \\ 
  Frank & 48 & 94 & 100 & 51 & 96 & 100 & 51 & 94 & 100 \\[.5ex] 
\multicolumn{10}{l}{\textbf{Case h: $\tau_0 = \tau_2 = 1/2 \ <\ \tau_1 = 1/2 + 1/10$}}\\[3ex] 
  Gumbel & 95 & 100 & 100 & 80 & 100 & 100 & 82 & 100 & 100 \\ 
  Clayton & 86 & 100 & 100 & 98 & 100 & 100 & 75 & 99 & 100 \\ 
  Frank & 94 & 100 & 100 & 94 & 100 & 100 & 96 & 100 & 100 \\[.5ex] 
\multicolumn{10}{l}{\textbf{Case i: $\tau_0 = \tau_2 = 3/4 \ <\ \tau_1 = 3/4 + 1/10$}}\\[3ex] 
  Gumbel & 33 & 75 & 100 & 22 & 51 & 94 & 22 & 53 & 96 \\ 
  Clayton & 21 & 44 & 92 & 34 & 78 & 100 & 20 & 48 & 94 \\ 
  Frank & 28 & 66 & 99 & 31 & 70 & 100 & 25 & 68 & 99 \\[.5ex] 
\multicolumn{10}{l}{\textbf{Case j: $\tau_0 = 1/4 \ <\ \tau_1 = \tau_2 = 1/4 + 1/10$}}\\[3ex] 
  Gumbel & 56 & 94 & 100 & 38 & 78 & 100 & 41 & 81 & 100 \\ 
  Clayton & 39 & 82 & 100 & 59 & 97 & 100 & 39 & 80 & 100 \\ 
  Frank & 48 & 94 & 100 & 51 & 94 & 100 & 50 & 94 & 100 \\[.5ex] 
\multicolumn{10}{l}{\textbf{Case k: $\tau_0 = 1/2 \ <\ \tau_1 = \tau_2 = 1/2 + 1/10$}}\\[3ex] 
  Gumbel & 96 & 100 & 100 & 74 & 100 & 100 & 82 & 100 & 100 \\ 
  Clayton & 87 & 100 & 100 & 97 & 100 & 100 & 72 & 99 & 100 \\ 
  Frank & 93 & 100 & 100 & 94 & 100 & 100 & 94 & 100 & 100 \\[.5ex] 
\multicolumn{10}{l}{\textbf{Case l: $\tau_0 = 3/4 \ <\ \tau_1 = \tau_2 = 3/4 + 1/10$}} 
\end{tabular}
\caption{Power (size in Figure~\ref{tab:nuisance-simplified-1}) in finite samples of the (unconditional) test of $H_{\circ}:\vartheta_0 = \vartheta_1$ based on Corollary~\ref{cor:simple}.
The tree structure is set to $\mathcal{G} = \{ \{ 1,2\}, \{3,4\}\}$ (three parameters, $\vartheta_2$ is nuisance) and three generator families (Gumbel, Clayton and Frank) and various values of $\vartheta_i = \tau_{\psi}^{-1}(\tau_i)$ ($i=0,1,2$; as described in Section~\ref{sec:power-local}) are considered.
During the testing procedure, it is assumed that the $\vartheta_2 = \vartheta_0$.
The same three generator families are considered for modeling the data.} \label{tab:nuisance-simplified-2}
\end{table}
\pagebreak

\begin{table}[h!]
\centering
\begin{tabular}{lrrrrrrrrr}
\multicolumn{10}{l}{\textbf{Simple hypothesis w/ nuisance -- likelihood ratio test} (hybrid)}\\[.25ex]
\multicolumn{10}{l}{\textbf{Cases a--f}}\\[1.25ex]
    & \multicolumn{3}{c}{Gumbel data} & \multicolumn{3}{c}{Clayton data} & \multicolumn{3}{c}{Frank data}\\[.5ex]
Model & n = 32 & 128 & 512 & n = 32 & 128 & 512 & n = 32 & 128 & 512 \\[.5ex]
  \hline\\[-1.5ex]
Gumbel & 5 & 5 & 3 & 8 & 10 & 22 & 7 & 7 & 10 \\ 
  Clayton & 6 & 6 & 6 & 5 & 5 & 6 & 6 & 7 & 10 \\ 
  Frank & 4 & 6 & 5 & 6 & 10 & 14 & 4 & 5 & 5 \\[.5ex] 
\multicolumn{10}{l}{\textbf{Case a: $\tau_0 = \tau_1 = \tau_2 = 1/4$}}\\[3ex] 
  Gumbel & 5 & 3 & 5 & 10 & 18 & 38 & 6 & 9 & 16 \\ 
  Clayton & 10 & 16 & 36 & 5 & 5 & 5 & 13 & 26 & 54 \\ 
  Frank & 5 & 6 & 8 & 8 & 11 & 20 & 5 & 5 & 5 \\[.5ex] 
\multicolumn{10}{l}{\textbf{Case b: $\tau_0 = \tau_1 = \tau_2 = 1/2$}}\\[3ex] 
  Gumbel & 4 & 4 & 4 & 19 & 38 & 77 & 13 & 25 & 59 \\ 
  Clayton & 23 & 43 & 83 & 6 & 4 & 6 & 33 & 70 & 99 \\ 
  Frank & 7 & 6 & 15 & 13 & 18 & 38 & 5 & 5 & 6 \\[.5ex] 
\multicolumn{10}{l}{\textbf{Case c: $\tau_0 = \tau_1 = \tau_2 = 3/4$}}\\[3ex] 
  Gumbel & 6 & 5 & 4 & 8 & 10 & 19 & 6 & 7 & 8 \\ 
  Clayton & 6 & 5 & 8 & 5 & 6 & 6 & 6 & 6 & 9 \\ 
  Frank & 5 & 4 & 6 & 6 & 8 & 16 & 6 & 6 & 5 \\[.5ex] 
\multicolumn{10}{l}{\textbf{Case d: $\tau_0 = \tau_1 = 1/4 \ <\ \tau_2 = 1/4 + 1/10$}}\\[3ex] 
  Gumbel & 4 & 5 & 4 & 11 & 17 & 38 & 8 & 10 & 16 \\ 
  Clayton & 9 & 18 & 37 & 5 & 6 & 6 & 12 & 27 & 61 \\ 
  Frank & 5 & 4 & 7 & 7 & 11 & 18 & 5 & 5 & 5 \\[.5ex] 
\multicolumn{10}{l}{\textbf{Case e: $\tau_0 = \tau_1 = 1/2 \ <\ \tau_2 = 1/2 + 1/10$}}\\[3ex] 
  Gumbel & 5 & 5 & 4 & 18 & 39 & 78 & 15 & 24 & 57 \\ 
  Clayton & 18 & 42 & 82 & 4 & 4 & 5 & 33 & 71 & 99 \\ 
  Frank & 6 & 9 & 12 & 11 & 15 & 38 & 4 & 5 & 6 \\[.5ex] 
\multicolumn{10}{l}{\textbf{Case f: $\tau_0 = \tau_1 = 3/4 \ <\ \tau_2 = 3/4 + 1/10$}}
\end{tabular}
\caption{Size (power in Figure~\ref{tab:nuisance-hybrid-2}) in finite samples of the (unconditional) test of $H_{\circ}:\vartheta_0 = \vartheta_1$ based on Remark~\ref{rem:hybrid}.
The tree structure is set to $\mathcal{G} = \{ \{ 1,2\}, \{3,4\}\}$ (three parameters, $\vartheta_2$ is nuisance) and three generator families (Gumbel, Clayton and Frank) and various values of $\vartheta_i = \tau_{\psi}^{-1}(\tau_i)$ ($i=0,1,2$; as described in Section~\ref{sec:power-local}) are considered.
The same three generator families are considered for modeling the data.
Rejection rates are reported in \%.} \label{tab:nuisance-hybrid-1}
\end{table}
\pagebreak

\begin{table}[ph!]
\centering
\begin{tabular}{lrrrrrrrrr}
\multicolumn{10}{l}{\textbf{Simple hypothesis w/ nuisance -- likelihood ratio test} (hybrid)}\\[.25ex]
\multicolumn{10}{l}{\textbf{Cases g--l}}\\[1.25ex]
    & \multicolumn{3}{c}{Gumbel data} & \multicolumn{3}{c}{Clayton data} & \multicolumn{3}{c}{Frank data}\\[.5ex]
Model & n = 32 & 128 & 512 & n = 32 & 128 & 512 & n = 32 & 128 & 512 \\[.5ex]
  \hline\\[-1.5ex]
  Gumbel & 33 & 74 & 100 & 26 & 56 & 96 & 23 & 57 & 96 \\ 
  Clayton & 21 & 45 & 92 & 36 & 80 & 100 & 20 & 50 & 94 \\ 
  Frank & 28 & 67 & 100 & 30 & 72 & 100 & 26 & 69 & 100 \\[.5ex] 
\multicolumn{10}{l}{\textbf{Case g: $\tau_0 = \tau_2 = 1/4 \ <\ \tau_1 = 1/4 + 1/10$}}\\[3ex] 
  Gumbel & 54 & 95 & 100 & 39 & 77 & 100 & 38 & 83 & 100 \\ 
  Clayton & 42 & 81 & 100 & 58 & 98 & 100 & 40 & 78 & 99 \\ 
  Frank & 48 & 93 & 100 & 51 & 96 & 100 & 50 & 94 & 100 \\[.5ex] 
\multicolumn{10}{l}{\textbf{Case h: $\tau_0 = \tau_2 = 1/2 \ <\ \tau_1 = 1/2 + 1/10$}}\\[3ex] 
  Gumbel & 95 & 100 & 100 & 80 & 100 & 100 & 82 & 100 & 100 \\ 
  Clayton & 85 & 100 & 100 & 98 & 100 & 100 & 74 & 99 & 100 \\ 
  Frank & 93 & 100 & 100 & 93 & 100 & 100 & 96 & 100 & 100 \\[.5ex] 
\multicolumn{10}{l}{\textbf{Case i: $\tau_0 = \tau_2 = 3/4 \ <\ \tau_1 = 3/4 + 1/10$}}\\[3ex] 
  Gumbel & 33 & 75 & 100 & 22 & 52 & 95 & 23 & 54 & 96 \\ 
  Clayton & 21 & 44 & 92 & 34 & 78 & 100 & 20 & 48 & 94 \\ 
  Frank & 28 & 65 & 99 & 31 & 70 & 100 & 25 & 68 & 99 \\[.5ex] 
\multicolumn{10}{l}{\textbf{Case j: $\tau_0 = 1/4 \ <\ \tau_1 = \tau_2 = 1/4 + 1/10$}}\\[3ex] 
  Gumbel & 55 & 93 & 100 & 37 & 77 & 100 & 40 & 81 & 100 \\ 
  Clayton & 38 & 80 & 100 & 57 & 97 & 100 & 38 & 79 & 100 \\ 
  Frank & 47 & 92 & 100 & 50 & 93 & 100 & 49 & 93 & 100 \\[.5ex] 
\multicolumn{10}{l}{\textbf{Case k: $\tau_0 = 1/2 \ <\ \tau_1 = \tau_2 = 1/2 + 1/10$}}\\[3ex] 
  Gumbel & 95 & 100 & 100 & 73 & 99 & 100 & 80 & 100 & 100 \\ 
  Clayton & 85 & 100 & 100 & 96 & 100 & 100 & 72 & 99 & 100 \\ 
  Frank & 92 & 100 & 100 & 93 & 100 & 100 & 93 & 100 & 100 \\[.5ex] 
\multicolumn{10}{l}{\textbf{Case l: $\tau_0 = 3/4 \ <\ \tau_1 = \tau_2 = 3/4 + 1/10$}} 
\end{tabular}
\caption{Power (size in Figure~\ref{tab:nuisance-hybrid-1}) in finite samples of the (unconditional) test of $H_{\circ}:\vartheta_0 = \vartheta_1$ based on Remark~\ref{rem:hybrid}.
The tree structure is set to $\mathcal{G} = \{ \{ 1,2\}, \{3,4\}\}$ (three parameters, $\vartheta_2$ is nuisance) and three generator families (Gumbel, Clayton and Frank) and various values of $\vartheta_i = \tau_{\psi}^{-1}(\tau_i)$ ($i=0,1,2$; as described in Section~\ref{sec:power-local}) are considered.
The same three generator families are considered for modeling the data.
Rejection rates are reported in \%.} \label{tab:nuisance-hybrid-2}
\end{table}


\begin{thebibliography}{}

\bibitem[Abdallah et~al., 2015]{Abdallah/Boucher/Cossette:2015}
Abdallah, A., Boucher, J.-P., and Cossette, H. (2015).
\newblock Modeling dependence between loss triangles with hierarchical
  {A}rchimedean copulas.
\newblock {\em ASTIN Bull.}, 45:577–599.

\bibitem[Azadbakhsh et~al., 2021]{Azadbakhsh/al:2021}
Azadbakhsh, M., Gao, X., and Jankowski, H. (2021).
\newblock Composite likelihood ratio testing under nonstandard conditions using
  tangent cones.
\newblock {\em Stat}, 10:e375.

\bibitem[Bartholomew, 1959a]{Bartholomew:1959a}
Bartholomew, D.~J. (1959a).
\newblock A test of homogeneity for ordered alternatives.
\newblock {\em Biometrika}, 46:36--48.

\bibitem[Bartholomew, 1959b]{Bartholomew:1959b}
Bartholomew, D.~J. (1959b).
\newblock A test of homogeneity for ordered alternatives. {II}.
\newblock {\em Biometrika}, 46:328--335.

\bibitem[Bartholomew, 1961]{Bartholomew:1961}
Bartholomew, D.~J. (1961).
\newblock A test of homogeneity of means under restricted alternatives.
\newblock {\em J. R. Stat. Soc. B}, 23:239--272.

\bibitem[Chant, 1974]{Chant:1974}
Chant, D. (1974).
\newblock On asymptotic tests of composite hypotheses in nonstandard
  conditions.
\newblock {\em Biometrika}, 61:291--298.

\bibitem[Chaoubi et~al., 2021]{Chaboui/al:2021}
Chaoubi, I., Cossette, H., Marceau, {\'E}., and Robert, C.~Y. (2021).
\newblock Hierarchical copulas with {A}rchimedean blocks and asymmetric
  between-block pairs.
\newblock {\em Comput. Stat. Data Anal.}, 154:107071.

\bibitem[Chernoff, 1954]{Chernoff:1954}
Chernoff, H. (1954).
\newblock On the distribution of the likelihood ratio.
\newblock {\em Ann. Math. Stat.}, 25:573--578.

\bibitem[Cossette et~al., 2017]{Cossette/Gadoury/Marceau/Mtalai:2017}
Cossette, H., Gadoury, S.-P., Marceau, {\'E}., and Mtalai, I. (2017).
\newblock {Hierarchical Archimedean copulas through multivariate compound
  distributions}.
\newblock {\em Insur. Math. Econ.}, 76:1--13.

\bibitem[Cossette et~al., 2019a]{Cossette/Gadoury/Marceau/Robert:2019}
Cossette, H., Gadoury, S.-P., Marceau, {\'E}., and Robert, C.~Y. (2019a).
\newblock {Composite likelihood estimation method for hierarchical Archimedean
  copulas defined with multivariate compound distributions}.
\newblock {\em J. Multivar. Anal.}, 172:59--83.
\newblock Dependence Models.

\bibitem[Cossette et~al., 2019b]{Cossette/Marceau/Mtalai:2019}
Cossette, H., Marceau, {\'E}., and Mtalai, I. (2019b).
\newblock Collective risk models with dependence.
\newblock {\em Insur. Math. Econ.}, 87:153--168.

\bibitem[Genest et~al., 1995]{Genest/Ghoudi/Rivest:1995}
Genest, C., Ghoudi, K., and Rivest, L.-P. (1995).
\newblock {A semiparametric estimation procedure of dependence parameters in
  multivariate families of distributions}.
\newblock {\em Biometrika}, 82:543--552.

\bibitem[Genest and Rivest, 1993]{Genest/Rivest:1993}
Genest, C. and Rivest, L.-P. (1993).
\newblock Statistical inference procedures for bivariate {A}rchimedean copulas.
\newblock {\em J. Am. Stat. Assoc.}, 88:1034--1043.

\bibitem[Geyer, 1994]{Geyer:1994}
Geyer, C.~J. (1994).
\newblock On the asymptotics of constrained {$M$}-estimation.
\newblock {\em Ann. Stat.}, 22:1993--2010.

\bibitem[Goldberg et~al., 1972]{Goldberg/Newman/Haynsworth:1972}
Goldberg, K., Newman, M., and Haynsworth, E.~V. (1972).
\newblock Combinatorial analysis.
\newblock In Abramowitz, M. and Stegun, I.~A., editors, {\em Handbook of
  Mathematical Functions}, chapter~24. National Bureau of Standards. Applied
  Mathematics Series 55. Tenth Printing.

\bibitem[Grothe and Hofert, 2015]{Grothe/Hofert:2015}
Grothe, O. and Hofert, M. (2015).
\newblock Construction and sampling of {A}rchimedean and nested {A}rchimedean
  {L}évy copulas.
\newblock {\em J. Multivar. Anal.}, 138:182--198.

\bibitem[Górecki and Hofert, 2023]{Gorecki/Hofert:2023}
Górecki, J. and Hofert, M. (2023).
\newblock {Composite pseudo-likelihood estimation for pair-tractable copulas
  such as Archimedean, Archimax and related hierarchical extensions}.
\newblock {\em J. Stat. Comput. Simul.}, 93:2321--2355.

\bibitem[Górecki et~al., 2017a]{Gorecki/Hofert/Holena:2017-DM}
Górecki, J., Hofert, M., and Holeňa, M. (2017a).
\newblock {Kendall’s tau and agglomerative clustering for structure
  determination of hierarchical Archimedean copulas}.
\newblock {\em Dependence Modeling}, 5:75--87.

\bibitem[Górecki et~al., 2017b]{Gorecki/Hofert/Holena:2017-JSCS}
Górecki, J., Hofert, M., and Holeňa, M. (2017b).
\newblock {On structure, family and parameter estimation of hierarchical
  Archimedean copulas}.
\newblock {\em J. Stat. Comput. Simul.}, 87:3261--3324.

\bibitem[Górecki et~al., 2021]{Gorecki/Hofert/Okhrin:2021}
Górecki, J., Hofert, M., and Okhrin, O. (2021).
\newblock {Outer power transformations of hierarchical Archimedean copulas:
  Construction, sampling and estimation}.
\newblock {\em Comput. Stat. Data Anal.}, 155:107109.

\bibitem[Hering et~al., 2010]{Hering/al:2010}
Hering, C., Hofert, M., Mai, J.-F., and Scherer, M. (2010).
\newblock {Constructing hierarchical Archimedean copulas with L{\'e}vy
  subordinators}.
\newblock {\em J. Multivar. Anal.}, 101:1428--1433.

\bibitem[Hofert, 2010]{Hofert:2010}
Hofert, M. (2010).
\newblock {\em Sampling nested {A}rchimedean copulas with applications to CDO
  pricing}.
\newblock PhD thesis, Universität Ulm.

\bibitem[Hofert, 2011]{Hofert:2011}
Hofert, M. (2011).
\newblock Efficiently sampling nested {A}rchimedean copulas.
\newblock {\em Comput. Stat. Data Anal.}, 55:57--70.

\bibitem[Hofert et~al., 2023]{copula:2023}
Hofert, M., Kojadinovic, I., Maechler, M., and Yan, J. (2023).
\newblock {\em copula: Multivariate Dependence with Copulas}.
\newblock R package version 1.1-2.

\bibitem[Hofert and M{\"a}chler, 2011]{Hofert/Machler:2011}
Hofert, M. and M{\"a}chler, M. (2011).
\newblock {Nested Archimedean copulas meet R: The nacopula package}.
\newblock {\em J. Stat. Software}, 39:1--20.

\bibitem[Hofert et~al., 2012]{Hofert/Machler/McNeil:2012}
Hofert, M., Mächler, M., and McNeil, A.~J. (2012).
\newblock {Likelihood inference for Archimedean copulas in high dimensions
  under known margins}.
\newblock {\em J. Multivar. Anal.}, 110:133--150.

\bibitem[Hofert and Pham, 2013]{Hofert/Pham:2013}
Hofert, M. and Pham, D. (2013).
\newblock {Densities of nested Archimedean copulas}.
\newblock {\em J. Multivar. Anal.}, 118:37--52.

\bibitem[Hofert and Scherer, 2011]{Hofert/Scherer:2011}
Hofert, M. and Scherer, M. (2011).
\newblock {CDO} pricing with nested {A}rchimedean copulas.
\newblock {\em Quant. Finance}, 11:775--787.

\bibitem[Holeňa et~al., 2015]{Holena/Bajer/Scavnicky:2015}
Holeňa, M., Bajer, L., and Ščavnický, M. (2015).
\newblock Using copulas in data mining based on the observational calculus.
\newblock {\em IEEE Trans. Knowl. Data Eng.}, 27:2851--2864.

\bibitem[Huang et~al., 2020]{Huang/Ning/Cai/al:2020}
Huang, J., Ning, Y., Cai, Y., Liang, K.-Y., and Chen, Y. (2020).
\newblock Composite likelihood inference under boundary conditions.
\newblock {\em Stat. Sin.}, 30:1005--1025.

\bibitem[Joe, 1997]{Joe:1997}
Joe, H. (1997).
\newblock {\em Multivariate models and multivariate dependence concepts}.
\newblock CRC press.

\bibitem[Kudo, 1963]{Kudo:1963}
Kudo, A. (1963).
\newblock A multivariate analogue of the one-sided test.
\newblock {\em Biometrika}, 50:403--418.

\bibitem[Li et~al., 2021]{Li/Balasooriya/Liu:2021}
Li, J., Balasooriya, U., and Liu, J. (2021).
\newblock Using hierarchical {A}rchimedean copulas for modelling mortality
  dependence and pricing mortality-linked securities.
\newblock {\em Ann. Actuar. Sci.}, 15:505–518.

\bibitem[Lindsay, 1988]{Lindsay:1988}
Lindsay, B.~G. (1988).
\newblock Composite likelihood methods.
\newblock {\em Comtemp. Math.}, 80:221--239.

\bibitem[Matsypura et~al., 2016]{Matsypura/al:2016}
Matsypura, D., Neo, E., and Prokhorov, A. (2016).
\newblock Estimation of hierarchical {A}rchimedean copulas as a shortest path
  problem.
\newblock {\em Econ. Lett.}, 149:131--134.

\bibitem[McNeil, 2008]{McNeil:2008}
McNeil, A.~J. (2008).
\newblock Sampling nested {A}rchimedean copulas.
\newblock {\em J. Stat. Comput. Simul.}, 78:567--581.

\bibitem[McNeil and Nešlehová, 2009]{McNeil/Neslehova:2009}
McNeil, A.~J. and Nešlehová, J. (2009).
\newblock {Multivariate Archimedean copulas, d-monotone functions and
  $\ell_1$-norm symmetric distributions}.
\newblock {\em Ann. Stat.}, 37:3059--3097.

\bibitem[Moran, 1971]{Moran:1971b}
Moran, P. A.~P. (1971).
\newblock Maximum-likelihood estimation in non-standard conditions.
\newblock {\em Math. Proc. Cambridge Philos. Soc.}, 70:441--450.

\bibitem[Nelsen et~al., 2003]{Nelsen/al:2003}
Nelsen, R.~B., Quesada-Molina, J.~J., Rodr\'{i}guez-Lallena, J.~A., and
  \'{U}beda Flores, M. (2003).
\newblock Kendall distribution functions.
\newblock {\em Stat. Probab. Lett}, 65:263--268.

\bibitem[Okhrin et~al., 2013]{Okhrin/Okhrin/Schmid:2013}
Okhrin, O., Okhrin, Y., and Schmid, W. (2013).
\newblock On the structure and estimation of hierarchical {A}rchimedean
  copulas.
\newblock {\em J. Econom.}, 173:189--204.

\bibitem[Okhrin and Ristig, 2014]{Okhrin/Ristig:2014}
Okhrin, O. and Ristig, A. (2014).
\newblock {Hierarchical Archimedean Copulae: The {HAC} Package}.
\newblock {\em J. Stat. Software}, 58:1--20.

\bibitem[Okhrin and Ristig, 2024]{Okhrin/Ristig:2024}
Okhrin, O. and Ristig, A. (2024).
\newblock Penalized estimation of hierarchical {A}rchimedean copula.
\newblock {\em J. Multivar. Anal.}, 201:105274.

\bibitem[Perreault, 2020]{Perreault:2020}
Perreault, S. (2020).
\newblock {\em Structures de corr{\'e}lation partiellement {\'e}changeables}.
\newblock PhD thesis, Universit{\'e} Laval.

\bibitem[Perreault et~al., 2019]{Perreault/Duchesne/Neslehova:2019}
Perreault, S., Duchesne, T., and Ne{\v{s}}lehov{\'a}, J.~G. (2019).
\newblock Detection of block-exchangeable structure in large-scale correlation
  matrices.
\newblock {\em J. Multivar. Anal.}, 169:400--422.

\bibitem[Perreault et~al., 2023]{Perreault/Neslehova/Duchesne:2023}
Perreault, S., Nešlehová, J.~G., and Duchesne, T. (2023).
\newblock Hypothesis tests for structured rank correlation matrices.
\newblock {\em J. Am. Stat. Assoc.}, 118:2889--2900.

\bibitem[Rezapour, 2015]{Rezapour:2015}
Rezapour, M. (2015).
\newblock {On the construction of nested Archimedean copulas for d-monotone
  generators}.
\newblock {\em Stat. Probab. Lett}, 101:21--32.

\bibitem[Savu and Trede, 2010]{Savu/Trede:2010}
Savu, C. and Trede, M. (2010).
\newblock Hierarchies of {A}rchimedean copulas.
\newblock {\em Quant. Finance}, 10:295--304.

\bibitem[Segers and Uyttendaele, 2014]{Segers/Uyttendaele:2014}
Segers, J. and Uyttendaele, N. (2014).
\newblock {Nonparametric estimation of the tree structure of a nested
  Archimedean copula}.
\newblock {\em Comput. Stat. Data Anal.}, 72:190--204.

\bibitem[Self and Liang, 1987]{Self/Liang:1987}
Self, S.~G. and Liang, K.-Y. (1987).
\newblock Asymptotic properties of maximum likelihood estimators and likelihood
  ratio tests under nonstandard conditions.
\newblock {\em J. Am. Stat. Assoc.}, 82:605--610.

\bibitem[Shapiro, 1985]{Shapiro:1985}
Shapiro, A. (1985).
\newblock Asymptotic distribution of test statistics in the analysis of moment
  structures under inequality constraints.
\newblock {\em Biometrika}, 72:133--144.

\bibitem[Shapiro, 2000]{Shapiro:2000}
Shapiro, A. (2000).
\newblock On the asymptotics of constrained local {$M$}-estimators.
\newblock {\em Ann. Stat.}, 28:948--960.

\bibitem[Susko, 2013]{Susko:2013}
Susko, E. (2013).
\newblock {Likelihood ratio tests with boundary constraints using
  data-dependent degrees of freedom}.
\newblock {\em Biometrika}, 100:1019--1023.

\bibitem[Uyttendaele, 2018]{Uyttendaele:2018}
Uyttendaele, N. (2018).
\newblock {On the estimation of nested Archimedean copulas: a theoretical and
  an experimental comparison}.
\newblock {\em Comput. Stat.}, 33:1047--1070.

\bibitem[van~der Vaart, 1998]{vanderVaart:1998}
van~der Vaart, A.~W. (1998).
\newblock {\em Asymptotic Statistics}.
\newblock Cambridge University Press.

\bibitem[Varin, 2008]{Varin:2008}
Varin, C. (2008).
\newblock On composite marginal likelihoods.
\newblock {\em AStA Adv. Stat. Anal.}, 92:1--28.

\bibitem[Varin et~al., 2011]{Varin/Reid/Firth:2011}
Varin, C., Reid, N., and Firth, D. (2011).
\newblock An overview of composite likelihood methods.
\newblock {\em Statistica Sinica}, 21:5--42.

\bibitem[Whelan, 2004]{Whelan:2004}
Whelan, N. (2004).
\newblock Sampling from {A}rchimedean copulas.
\newblock {\em Quant. Finance}, 4:339--352.

\bibitem[Zhang et~al., 2021]{Zhang/Jin/Bai:2021}
Zhang, W., Jin, B., and Bai, Z. (2021).
\newblock {Learning block structures in U-statistic-based matrices}.
\newblock {\em Biometrika}, 108(4):933--946.

\bibitem[Zhu et~al., 2016]{Zhu/Wang/Tan:2016}
Zhu, W., Wang, C.-W., and Tan, K.~S. (2016).
\newblock {Structure and estimation of L{\'e}vy subordinated hierarchical
  Archimedean copulas (LSHAC): Theory and empirical tests}.
\newblock {\em J. Banking Finance}, 69:20--36.

\end{thebibliography}
\end{document}